\documentclass[12pt,onecolumn]{article}
\usepackage[left=2cm,right=2cm,top=3cm,bottom=3cm]{geometry}
\usepackage[utf8]{inputenc}
\usepackage[english]{babel}
\usepackage{hyperref}
\usepackage{paralist}
\usepackage{graphicx}
\usepackage{authblk}
\usepackage{natbib}
\usepackage{enumitem}
\usepackage{color}
\usepackage{multicol}
\usepackage{soul}

\newcommand{\dd}{\ensuremath{\textrm{d}}}
\newcommand{\Dp}[2]{\ensuremath{\frac{\partial#1}{\partial#2}}}
\newcommand{\DDp}[2]{\ensuremath{\frac{\partial^2 #1}{\partial #2 ^2}}}
\newcommand{\ddf}[2]{\ensuremath{\frac{\dd#1}{\dd#2}}}

\begin{document}
\title{Global climate modeling of Saturn's atmosphere.\\ Part II: multi-annual high-resolution dynamical simulations} 
\author[1,2]{Aymeric Spiga\thanks{Corresponding author: aymeric.spiga@sorbonne-universite.fr}}
\author[1]{Sandrine Guerlet}
\author[1]{Ehouarn Millour}
\author[1]{Mikel Indurain} 
\author[3]{Yann Meurdesoif}
\author[1]{Simon Cabanes}
\author[1]{Thomas Dubos}
\author[4]{J\'er\'emy Leconte}
\author[1]{Alexandre Boissinot}
\author[1]{Sébastien Lebonnois}
\author[1,5,6]{Mélody Sylvestre}
\author[6]{Thierry Fouchet}

\affil[1]{\footnotesize
Laboratoire de M\'et\'eorologie Dynamique / Institut Pierre-Simon Laplace (LMD/IPSL),
Sorbonne Université, 
Centre National de la Recherche Scientifique (CNRS),
\'Ecole Polytechnique,
\'Ecole Normale Supérieure (ENS),
\scriptsize \emph{address: Campus Pierre et Marie Curie BC99, 
4 place Jussieu 75005 Paris, France}
\normalsize}
\affil[2]{\footnotesize 
Institut Universitaire de France (IUF),
\scriptsize \emph{address: 1 rue Descartes,
75005 Paris, France}
\normalsize}
\affil[3]{\footnotesize 
Laboratoire des Sciences du Climat et de l'Environnement (LSCE/IPSL),
Commissariat à l’énergie atomique et aux énergies alternatives (CEA),
Centre National de la Recherche Scientifique,
Université Paris-Saclay,
\scriptsize \emph{address: Campus du CEA - Orme des Merisiers,
Saclay, France}
\normalsize}
\affil[4]{\footnotesize 
Laboratoire d'Astrophysique de Bordeaux (LAB), 
Univ. Bordeaux, 
Centre National de la Recherche Scientifique (CNRS),
\scriptsize \emph{address: B18N, all\'ee Geoffroy Saint-Hilaire, 33615 Pessac, France}
\normalsize}
\affil[5]{\footnotesize 
School of Earth Sciences, University of Bristol,
\scriptsize \emph{address: Wills Memorial Building, Queens Road, 
Bristol BS8 1RJ, UK}
\normalsize}
\affil[6]{\footnotesize 
Laboratoire d'{\'E}tudes Spatiales et d’Instrumentation en Astrophysique (LESIA), 
Observatoire de Paris, 
Université Paris Sciences et Lettres (PSL), 
Centre National de la Recherche Scientifique (CNRS),
Sorbonne Université, 
Univ. Paris Diderot,
\scriptsize \emph{address: 5 place Jules Janssen, 92195 Meudon, France}
\normalsize}

\date{Version: \today}
\maketitle
\newpage
\section*{Highlights}

\begin{itemize}
\item A new Global Climate Model for Saturn with radiative transfer
\item High-resolution numerical simulations on a duration of 15 Saturn years
\item Results on zonal jets, waves, eddies in Saturn's troposphere
\end{itemize}

\section*{Abstract}

\emph{The Cassini mission unveiled the intense and diverse activity
in Saturn's atmosphere: banded jets, waves, vortices,
equatorial oscillations.
To set the path towards a better understanding of 
those phenomena, 
we performed high-resolution multi-annual numerical simulations
of Saturn's atmospheric dynamics.
We built a new Global Climate Model [GCM] for Saturn,
named the Saturn DYNAMICO GCM,
by combining a radiative-seasonal model tailored for Saturn
to a hydrodynamical solver based on an icosahedral grid
suitable for massively-parallel architectures.
The impact of numerical dissipation, 
and the conservation of angular momentum,
are examined in the model
before a reference simulation
employing the Saturn DYNAMICO GCM
with a~$1/2^{\circ}$ latitude-longitude resolution
is considered for analysis.
Mid-latitude banded jets showing similarity with observations
are reproduced by our model.
Those jets are accelerated and maintained
by eddy momentum transfers
to the mean flow, with the magnitude of momentum fluxes
compliant with the observed values.
The eddy activity is not regularly distributed
with time, but appears as bursts; 
both barotropic and baroclinic instabilities
could play a role in the eddy activity.
The steady-state latitude of occurrence of jets
is controlled by poleward migration during the spin-up 
of our model.
At the equator, a weakly-superrotating tropospheric jet
and vertically-stacked alternating stratospheric jets
are obtained in our GCM simulations.
The model produces Yanai (Rossby-gravity), Rossby and Kelvin waves
at the equator, as well as extratropical Rossby waves,
and large-scale vortices in polar regions.
Challenges remain to reproduce
Saturn's 
powerful superrotating jet and
hexagon-shaped circumpolar jet in the troposphere,
and downward-propagating equatorial oscillation
in the stratosphere.}

\newpage\setcounter{tocdepth}{3}\tableofcontents
\newpage
\section{Introduction \label{sec:intro}}

It has been decades since Saturn's meteorological phenomena observed by Earth-based and space telescopes, and the pioneering Voyager missions, are challenging the fundamental knowledge of geophysical fluid mechanics \citep[e.g.,][]{Inge:90,Dowl:95}. Yet, a mission as richly instrumented as Cassini \citep{Porc:05}, offering from 2004 to 2017 an unprecedented spatial and seasonal coverage of Saturn's weather layer, brought a new impulse to the studies of giant planets' atmospheric dynamics \citep[e.g., review papers by][]{Delg:09,Show:18review}.

In Saturn's troposphere, the Cassini measurements confirmed the banded structure of alternating westward (retrograde) and eastward (prograde) jets, which features a $450$~m~s$^{-1}$ super-rotating equatorial jet \citep{Porc:05,Garc:10,Stud:18}. Furthermore, the Cassini instruments assessed the remarkable stability of the enigmatic hexagonal jet in the northern polar region \citep{Bain:09,Sanc:14hexagon,Antu:15,Flet:18hexagon}, with exquisite details on the structure of the turbulent polar vortex \citep{Saya:17,Bain:18}. They also offered a detailed record of mid-latitude convective storms \citep{Dyud:07,dRio:12} and vortices \citep{Vasa:06,Dyud:08,Tram:16,DelR:18}, including a chain of infrared bright spots named the ``String of Pearls'' \citep{Saya:14} and an exceptional coverage of Saturn's latest Great White Spot \citep{Fisc:11,Sanc:11,Saya:13}. Cassini observations of Saturn's cloud layer was also employed to demonstrate the high rate of conversion of energy from eddies to jets \citep{Delg:07,Delg:12}, to detail the structure of vorticity \citep{Read:09pv}, and to explore Jupiter's and Saturn's atmospheric energetic spectra across spatial scales \citep{Galp:14,Youn:17}, confirming pre-Cassini theoretical studies about geostrophic turbulence and the inverse energy cascade \citep{Suko:02}. All those observations strongly suggest that large-scale tropospheric banded jets emerge from forcing by smaller-scale eddies and waves arising from hydrodynamical instabilities.

In Saturn's stratosphere, not only the Cassini instruments led to key discoveries, but the longevity of the mission permitted a seasonal monitoring of the unveiled phenomena. Cassini's highlights in atmospheric science for the stratosphere include a spectacular stratospheric warming associated with the 2010 Great White Spot \citep{Flet:11,Flet:12beacon,Fouc:16}, an equatorial oscillation of temperature in Saturn's stratosphere \citep{Fouc:08,Guer:11,Li:11} with semi-annual periodicity \citep{Orto:08,Guer:18}, and a seasonal monitoring of the meridional distribution of Saturn's stratospheric hydrocarbons \citep{Guer:09,Guer:10,Sinc:13,Flet:15,Sylv:15,Guer:15}, hinting at a possible interhemispheric transport of chemical species. Cassini measurements even enabled to link a disruption in the downward propagation of the equatorial oscillation to the 2010 Great White Spot occurrence \citep{Flet:17}. Analogies can be drawn between Saturn's and the Earth's stratospheres \citep{Dowl:08}. Saturn's equatorial oscillation is reminiscent of Earth's Quasi-Biennal Oscillation and Semi-Annual Oscillation \citep{Andr:87,Bald:01,Lott:13,Guer:18}, driven by the propagation and breaking of Rossby, Kelvin and inertio-gravity waves. The interhemispheric transport of chemical species, which may affect the hydrocarbons distribution, might be analogous to the Earth's Brewer-Dobson circulation \citep{Butc:14}. 

In this stimulating observational context, new modeling efforts are needed to broaden the knowledge of Saturn's atmospheric dynamics by demonstrating the mechanisms underlying the above-mentioned observed phenomena. A great deal of past modeling work focused on the processes responsible for the banded tropospheric jets. 
A major difficulty with a giant planet is that the depth at which the atmosphere merges with the internal dynamo region and the strength at which the atmospheric circulations couple with magnetic disturbances have remained poorly constrained by observations \citep{Inge:90,Liu:08} until gravity measurements were recently performed on board Juno and Cassini \citep{Kasp:13gasgiants,Gala:17,Kasp:18,Gala:19}.
Two distinct modeling approaches have been adopted to account for Saturn's tropospheric jets: ``shallow-forcing'' climate models [see next paragraph for references] account for processes in the weather layer (baroclinic instability, moist convective storms), while ``deep-forcing'' dynamo-like models \citep{Heim:05,Yano:05,Kasp:09,Heim:11,Gast:14,Heim:16,Caba:17} resolve convection throughout gas giants' molecular envelopes. Contrary to deep models, shallow climate models have had difficulties reproducing gas giants' equatorial super-rotating jets. This has been overcome by including either bottom drag and intrinsic heat fluxes to simulate deep interior phenomena \citep{Lian:08,Schn:09,Liu:10jets}, or latent heating by moist convective storms \citep{Lian:10}, although the simulated equatorial jets are still about twice as less strong in simulations than in observations \citep[e.g.,][]{Garc:10}. The situation for off-equatorial jets is reversed, with better agreement with observations obtained by shallow models compared to deep models, although the latter can be modified to obtain more realistic results \citep{Heim:05}. The recent results from the Juno mission for Jupiter \citep{Kasp:18,Guil:18} and the Cassini mission for Saturn \citep{Gala:19} show that banded jets extend several thousand kilometers below the cloud layer, i.e. 
deeper than what shallow models consider 
and shallower than what deep models consider,
which probably indicates that shallow and deep models have both their virtues to represent part of the reality.

Here we adopt the approach of ``shallow-forcing'' climate modeling to study Saturn. In the last decade, the traditional approach using idealized modeling \citep{Cho:96pof,Will:03,Vasa:05} -- which still has great value to study how baroclinic and barotropic instabilities shape Saturn's jets \citep{Li:06,Kasp:07,Show:07}, including its polar hexagonal jet \citep{Rost:17saturn} and central vortex \citep{ONei:15} -- has been complemented by the development of complete three-dimensional Global Climate Models (GCMs) for Saturn and giant planets \citep{Dowl:98,Dowl:06,Lian:10,Liu:10jets,Youn:19partone,Youn:19parttwo}. A GCM is obtained by coupling a hydrodynamical solver of the Navier-Stokes equations for the atmospheric fluid on the sphere (the GCM's ``dynamical core'') with realistic models for physical processes operating at unresolved scales: radiation, turbulent mixing, phase changes, chemistry (the GCM's ``physical packages''). Most of those existing GCM studies for Saturn address the formation of tropospheric jets by angular momentum transfer through eddies and waves, often with either a theoretical approach aiming to address giant planets' atmospheric dynamics \citep{Schn:09,Lian:10,Liu:10jets,Liu:15} rather than a focused approach aiming to address Saturn specifically, or with a limited-domain approach using a latitudinal channel enclosing one specific jet to explain structures such as the Ribbon wave or the String of Pearls \citep{Saya:10,Saya:14}, to investigate the impact of convective outbursts \citep{Saya:07,Garc:13}, or to discuss the polar hexagonal jet \citep{Mora:11,Mora:15}. 
\newcommand{\bali}{The idealized GCM approach can also be employed to study equatorial oscillations in gas giants \citep{Show:18qboarxiv}. } \bali
All those existing studies use simple radiative forcing rather than computing a realistic ``physical package'' that includes seasonally-varying radiative transfer. The latter approach has been explored to study Saturn's stratosphere, either to constrain large-scale advection~/~eddy mixing in photochemical models \citep{Frie:12}, or to build a modeling framework applicable to extrasolar hot gas giants \citep{Medv:13}. Those studies of Saturn's stratosphere make use, however, of prescribed, \emph{ad hoc}, tropospheric jets.

The existing body of work on ``shallow-forcing'' modeling has thus paved the path towards a complete three-dimensional Global Climate Model (GCM) for giant planets. However, such a complete troposphere-to-stratosphere GCM for Saturn, capable to address the theoretical challenges opened by observations is yet to emerge. We propose that four challenges shall be overcome to develop a complete state-of-the-art GCM for Saturn and gas giants.
\begin{enumerate}[label=$\mathcal{C}_\arabic*$]
\item \label{ch:radi} The radiative transfer computations necessary to predict the evolution of atmospheric temperature, especially in the stratosphere, must be optimized for integration over decade-long giant planets' years, while still keeping robustness against observations.
\item \label{ch:dyna} Large-scale jets and vortices emerge from smaller-scale hydrodynamical eddies, through an inverse energy cascade driven by geostrophic turbulence. Relevant interaction scales (e.g. Rossby deformation radius) are only~$1^{\circ}$ latitude-longitude in gas giants vs.~$20^{\circ}$ on the Earth, making eddy-resolving global simulations over a full year four orders of magnitude more computationally expensive in gas giants.
\item \label{ch:strat} Terrestrial experience shows that models need to extend from the troposphere to the stratosphere with sufficient vertical resolution to resolve the vertical propagation of waves responsible for large-scale structures in both parts of the atmosphere. Moreover, a specific requirement of giant planets is to extend the model high enough in the stratosphere to model the photochemistry of key hydrocarbons impacting stratospheric temperatures \citep{Hue:16part2}.
\item \label{ch:bound} Climate models cannot extend neither deep enough to predict how tropospheric jets interact with interior convective fluxes and planetary magnetic field \citep{Kasp:09,Heim:11}, nor high enough to capture the coupling of stratospheric circulations with thermospheric and ionospheric processes \citep{Mull:12,Kosk:15}. A suitable approach to couple the weather layer with either the slowly-evolving convective interior, or the rapidly-evolving ionized external layers, remains elusive.
\end{enumerate}

Here we report the development and preliminary dynamical simulations of a new Saturn GCM at Laboratoire de M\'et\'eorologie Dynamique (LMD), which aims at understanding the seasonal variability, large-scale circulations, and eddy \& wave activity in Saturn's troposphere and stratosphere. It is a first step to further design a modeling platform dedicated to atmospheric circulations of Saturn and other solar system's giant planets. Challenge~\ref{ch:radi} about building fast and accurate radiative transfer for the Saturn GCM is addressed in \cite{Guer:14}. 
\newcommand{\paris}{In \citet{Guer:14}, which serves as Part I for the present study, a seasonal radiative–convective model of Saturn's upper troposphere and stratosphere is described and the sensitivity to composition, aerosols, internal heat flux and ring shadowing is assessed, with comparisons to the observed thermal structure by Cassini and ground-based telescopes. } \paris
In this Part II paper, we address Challenge~\ref{ch:dyna} by performing high-resolution dynamical simulations with our Saturn GCM. Our GCM is built by coupling the physical packages (notably, radiative transfer) of \cite{Guer:14} with DYNAMICO, a new dynamical core developed at LMD which uses an original icosahedral mapping of the planetary sphere to ensure excellent conservation and scalability properties in massively parallel resources \citep{Dubo:15}. 
\newcommand{\washington}{We describe here the insights gained from GCM simulations at high horizontal resolutions (reference at~$1/2^{\circ}$ latitude/longitude, and tests at~$1/4^{\circ}$ and~$1/8^{\circ}$) 
with two unprecedented characteristics at those horizontal resolutions: inclusion of realistic radiative transfer and long integration times up to fifteen simulated Saturn years.} \washington

The paper is organized as follows. Notations are defined in Table~\ref{tab:notations}. In section~\ref{sec:method}, we provide details on the characteristics of our Saturn DYNAMICO GCM, and the assumptions and settings adopted for the simulations discussed in subsequent sections, with appendix~\ref{sec:sens} featuring a necessary analysis of the impact of horizontal dissipation and the conservation of angular momentum in our Saturn GCM. In section~\ref{sec:reference}, we describe the results obtained with our reference 15-year-long $1/2^{\circ}$ Saturn DYNAMICO GCM simulation, with an emphasis on the driving and evolution of jets in section~\ref{sec:evolution}. In section~\ref{sec:discussion}, we summarize our conclusions and draw perspectives for future improvements of our Saturn DYNAMICO GCM needed to fully achieve challenges~\ref{ch:dyna}, \ref{ch:strat} and~\ref{ch:bound}, as it comes to no surprise that the present study is only a preliminary path towards fulfilling arguably ambitious scientific goals.

\begin{table}[p]
\begin{center}
\begin{tabular}{lll}
\hline
\multicolumn{3}{c}{\emph{Coordinates}} \\ \hline
$t$			& Time								& s \\ 
$(\lambda,\varphi)$	& Longitude, Latitude						& $^{\circ}$E, $^{\circ}$N  \\ 
$(x,y)$			& WE coordinate, SN coordinate (local frame)			& m \\ 
$z$			& Altitude							& m \\ 
$L_s$			& Saturn's heliocentric longitude ($0^{\circ}$ N spring)        & $^{\circ}$ \\
\hline 
\multicolumn{3}{c}{\emph{Meteorological variables}} \\ \hline
$p$			& Pressure							& Pa \\ 
$T$			& Temperature							& K \\ 
$u$			& Zonal wind component (W~$\rightarrow$~E)			& m~s$^{-1}$ \\ 
$v$			& Meridional wind component (S~$\rightarrow$~N)			& m~s$^{-1}$ \\ 
$w$			& Vertical wind component					& m~s$^{-1}$ \\ 
$m$			& Atmospheric mass						& kg \\ 
$\mathcal{M}$		& Axial Angular Momentum (AAM)					& kg~m$^2$~s$^{-1}$ \\ 
$\mu$			& Specific AAM (per unit mass, equation~\ref{eq:specificaam})	& m$^2$~s$^{-1}$ \\ 
$\zeta$			& Relative vorticity (vertical component~$\partial v / \partial x - \partial u / \partial y$) & m s$^{-2}$ \\
$q$			& Ertel potential vorticity					& kg m$^2$ s$^{-1}$ K$^{-1}$ \\
\hline 
\multicolumn{3}{c}{\emph{Planetary parameters and model settings}} \\ \hline
$\Omega$		& Rotation rate$^{\star}$					& $1.651 \times 10^{-4}$~s$^{-1}$ \\
$\omega$		& Obliquity							& $26.73^{\circ}$ \\
$g$			& Acceleration of gravity					& $10.44$~m~s$^{-2}$ \\
$a$ 			& Planetary radius						& $6.0268 \times 10^{7}$~m \\ 
$c_p$			& Specific heat capacity					& $11500$~J~kg$^{-1}$~K$^{-1}$ \\
$M$			& Molecular mass						& $2.34$~g~mol$^{-1}$ \\
$R$			& Ideal gas constant normalized with~$M$			& $0.309 \, c_p$ \\
$\Phi_i$		& Internal heat flux						& $2.6$~W~m$^{-2}$ \\
$\tau_R$ 		& Timescale for bottom Rayleigh drag				& 100 Earth days \\ 
$\varphi_R$		& Minimum latitude ($\pm \varphi_R$) for bottom drag 		& $33^{\circ}$ \\ 
\hline 
\multicolumn{3}{c}{\emph{Computations}} \\ \hline
$f=2\,\Omega\sin\varphi$& Coriolis parameter at latitude~$\varphi$ 			& s$^{-1}$ \\ 
$\beta = \Dp{f}{y} = \frac{2\,\Omega\,\cos\varphi}{a}$ & Beta parameter (meridional derivative of~$f$) & m$^{-1}$ s$^{-1}$ \\ 
$H = \frac{R\,T}{g}$	& Atmospheric scale height					& m \\
$N^2 = \frac{g}{\theta} \, \Dp{\theta}{z}$ & Brunt-V{\"a}is{\"a}l{\"a} frequency	& s$^{-2}$ \\
$\overline{\psi}$	& Axisymmetric component of variable~$\psi$			& Zonal average \\ 
$\psi^{\prime} = \psi - \overline{\psi}$	& Eddy$^{\dagger}$ component of variable~$\psi$ 	& Zonal anomaly \\
\end{tabular}
\end{center}
\caption{\label{tab:notations}\emph{Physical quantities used in the paper. 
The numerical values provided in the right column
corresponds to the values set in our Global Climate Model (section~\ref{sec:settings}). \newline
$^{\star}$~The rotation rate corresponds to Saturn ``days'' of~$38052$~s, 
according to the value obtained in \cite{Read:09rot} using an approach based on potential vorticity (denoted System IIIw). \newline
$^{\dagger}$~Eddies are defined as deviations (perturbations) from the zonal-mean flow 
and can represent the effects of turbulence, waves, and instabilities.
}}
\end{table}

\section{Characteristics of the Saturn DYNAMICO GCM \label{sec:method}} 

\subsection{Building the model}

As is reminded in the introduction, a GCM consists in coupling a dynamical core interfaced with physical packages (or parameterizations). Our project to develop a Saturn GCM started by the development of the latter: the physical packages used in our GCM are described in full detail in \cite{Guer:14}. Our model's radiative computations are based on a versatile correlated-$k$ method, suitable for any planetary composition \citep{Word:10gj581,Char:13,Leco:13} with $k$-coefficients derived from detailed line-by-line computations using the HITRAN spectroscopic database \citep{Roth:13}. Radiative contributions include the three main hydrocarbons (methane, ethane and acetylene), the broad H$_2$-H$_2$ and H$_2$-He collision-induced absorption \citep{Word:12}, and tropospheric and stratospheric aerosols layers. \newcommand{\newyork}{Our radiative computations also feature ring shadowing \citep[appendix A in][]{Guer:14} and account for internal heat flux independent with latitude \citep[section 2 in][]{Guer:14}.} \newyork

The spectral discretization of the correlated-$k$ model is optimized for Saturn, with a particular emphasis on accounting for absorption and emission bands of stratospheric methane CH$_4$ (the prominent driver of Saturn's stratospheric heating), and other hydrocarbons produced by its photodissociation (ethane C$_2$H$_6$ and acetylene C$_2$H$_2$, the prominent drivers of Saturn's stratospheric cooling). Compared to \citet{Guer:14}, the line list for methane has been updated beyond 9200~cm$^{-1}$ \citep[][in lieu of the \citet{Kark:10} band model]{Rey:18} and the two main isotopes~$^{13}$CH$_4$ and~CH$_3$D are now included. This improves the predicted temperatures in the middle stratosphere ($1-10$~mbar) by about~$2$~K. 
\newcommand{\montreal}{The vertical profiles of hydrocarbons' abundances are held constant with latitude and season, and set as is described in \citet{Guer:14} using a combination of Cassini observations \citep{Guer:09} and photochemical modeling \citep{Mose:00}. } \montreal
Variations up to~$100 \%$ of acetylene abundance are observed at high latitudes \citep[][note that \cite{Sylv:15} found weaker variations]{Flet:15} which would entail temperature variations of a couple K in the vicinity of the 1-mbar level \citep[][section 4.4]{Guer:14}; coupling our radiative model with a seasonal photochemical scheme is considered as a future development for dedicated middle-to-upper stratosphere GCM simulations \citep[see Challenge~\ref{ch:strat}, as well as][]{Hue:16part2,Hue:18}. 

\cite{Guer:14} showed that this seasonal model allowed for both efficiency and accuracy, with satisfactory comparisons with Cassini measurements -- including the observed temperature ``knee'' caused by heating at the top of the tropospheric aerosol layer, and the meridional gradient between the summer and winter stratosphere \citep{Flet:10season,Flet:15}. Temperatures predicted with our Saturn DYNAMICO GCM are compared with Cassini measurements in section~\ref{sec:thermal}.

The need to address specifically Challenge~\ref{ch:dyna} (i.e. to achieve fine-enough horizontal resolutions in order to predict the arising of smaller-scale eddies and the inverse cascade in the context of geostrophic turbulence) requires the use of a suitable dynamical core in our Saturn GCM. To that end, we chose to employ DYNAMICO, which is developed at LMD as the next state-of-the-art dynamical core for Earth and planetary climate studies \citep{Dubo:15}, and tailored for massively parallel High-Performance Computing resources (scalability tested up to~$10^5$ cores). 

Our dynamical core DYNAMICO solves the primitive hydrostatic equations assuming a shallow atmosphere, i.e. $z \ll a$ \citep[relaxing this assumption to solve the quasi-hydrostatic deep-atmosphere equations is considered for future developments of the model,][]{Tort:15}. The global horizontal mesh in DYNAMICO is set as a quasi-uniform icosahedral C-grid \citep{Dubo:15} obtained by subdivision of a regular icosahedron: the total number of hexagonal cells is~$10 \times N \times N$ corresponding to $N\times N$ sub-triangles subdividing each of the 10 main triangles of the icosahedron grid ($N$ is the parameter by which the horizontal resolution is set in DYNAMICO). Control volumes for mass, tracers and entropy/potential temperature are the hexagonal cells of the Voronoi mesh to avoid the fast numerical modes of the triangular C-grid. Vertical coordinates are mass-based coordinates: ``sigma'' levels defined as~$p / p_b$ where~$p_b$ is the pressure at the bottom of the model.

Spatial discretizations in DYNAMICO are formulated following an energy-conserving three-dimensional Hamiltonian approach \citep{Dubo:15}. Time integration is done by an explicit Runge-Kutta scheme (chosen for stability and accuracy). Subgrid-scale (unresolved) dissipation in the horizontal dimension is included as an additional hyperdiffusion term in the vorticity, divergence and temperature equations (see section~\ref{sec:dissip}). In the vertical dimension, subgrid-scale dissipation is handled in the physical packages through a combination of a \cite{Mell:82} diffusion scheme for small-scale turbulence, and a dry convective adjustment scheme for organized turbulence \citep[convective plumes, see section 2.4 of ][]{Hour:93}. In the case of our Saturn DYNAMICO GCM simulations, the adjustment scheme is the dominant term enabling a neutral profile in the troposphere.
\newcommand{\miami}{This simple adjustment scheme computes the temperature tendencies required to reach the entropy-conserving mixed layer of any convectively-unstable layer appearing in the model. Those characteristics entail that this scheme is not a source of small-scale eddies in the model, which was checked in practice in our Saturn DYNAMICO GCM.} \miami

The XIOS library \citep[XML Input/Output Server,][]{Meur:12,Meur:13} is employed to handle any input/output operations independently from the timeframe imposed by the numerical integrations: not only this improves the efficiency of the numerical integrations in massively-parallel computing clusters, but this also enables for complex operations on computed fields to be carried out during model runtime rather than as a post-processing operation. Notably, mapping the dynamical fields computed in the non-conformal icosahedral DYNAMICO grid towards a regular latitude-longitude grid, using finite-volume weighting functions, is performed by XIOS directly during our GCM runtime.

\subsection{Model settings and boundary conditions \label{sec:settings}}

The simulations discussed in this paper are obtained from integrations with our Saturn DYNAMICO GCM employing an horizontal icosahedral mesh with~$N=160$, corresponding to an approximate horizontal resolution of~$1/2^{\circ}$ in longitude/latitude (hereafter simply referred to as ``$1/2^{\circ}$ simulations''). Test simulations with~$N=301$ ($1/4^{\circ}$ simulations) and~$N=625$ ($1/8^{\circ}$ simulations), aimed at model testing rather than scientific exploration, are discussed in section~\ref{sec:discussion} to open perspectives for future work. The integration, dynamical timestep in the~$1/2^{\circ}$ Saturn simulations is~$118.9125$~s. Computations in physical packages are done every~160 dynamical timesteps (i.e. every half a Saturn day) with the exception of radiative computations, which are done every~40~physical timesteps (i.e. every~$6400$ dynamical timesteps, which is every~20 Saturn days). 
\newcommand{\lima}{This means that while the radiative tendency of temperature is added to the dynamical integrations at each physical timestep (every half a Saturn day), it is only updated by our radiative package every 20 Saturn days. } \lima
This is long compared to what is considered standard in GCMs for terrestrial planets, yet compliant with the comparatively long radiative timescales (or, equivalently, weak radiative forcing) in gas giants.
\newcommand{\rio}{As is indicated in \citet{Guer:14}, typical radiative timescales on Saturn are longer than a Saturn year below the 400-mbar pressure level and still about a third of a Saturn year at the 10-mbar level. This is much longer than the timescales of dynamical phenomena (most notably eddies) resolved in the model. } \rio 
We tested that simulations with smaller radiative timesteps yield similar results as reference simulations; we also checked that the diurnal cycle in radiative tendencies is negligible both in Saturn's troposphere and stratosphere.

Our~$1/2^{\circ}$ simulations feature~$32$ levels in the vertical dimension, ranging from $p_b \sim 3$~bars at the model bottom, to $1$~mbar at the model top. The Saturn DYNAMICO GCM simulations discussed in this paper thus extend from the lower troposphere to the middle stratosphere. Our model top is too low, and our vertical resolution too coarse in the stratosphere, to address Challenge~\ref{ch:strat}. This shall be improved in further studies dedicated specifically to Saturn's stratospheric phenomena (notably, the equatorial oscillation). Our DYNAMICO model features an optional absorbing (``sponge'') layer with a Rayleigh drag acting on the topmost model layers as a surrogate for gravity wave drag in the stratosphere, but we do not use it for the simulations presented in this paper, similarly to previous studies \citep{Schn:09,Liu:10jets}. Indeed, \cite{Shaw:07} showed that the inclusion of sponge-layer parameterizations that do not conserve angular momentum (which is the case for Rayleigh drag), or allow for momentum to escape to space, implies a sensitivity of the dynamical results (especially zonal wind speed) to the choice for model top or drag characteristic timescale, because of spurious downward influence when momentum conservation is violated.

Our bottom condition at the~$3$-bar level is similar to \cite{Liu:10jets}. We include a simple Rayleigh-like  drag~$\textrm{d}u / \textrm{d}t = - u / \tau_R$, with a timescale~$\tau_R = 100$~Earth~days. This drag plays the role devoted to surface friction on terrestrial planets, which allows to close the angular momentum budget through downward control \citep{Hayn:87,Hayn:91}. This could also be regarded as a zeroth-order parameterization for Magneto-HydroDynamic (MHD) drag as a result of Lorenz forces acting on jet streams putatively extending to the depths of Saturn's interior \citep{Liu:08,Gala:19}, much deeper than the shallow GCM's model bottom. Whether or not including a bottom drag at~$3$ bars is physically justified is out of the scope of the present paper, and improving on this admittedly simplistic bottom boundary condition is an entire research goal on its own (part of what we named Challenge~\ref{ch:bound}). In the present study, we take this bottom drag as an imperfect, yet unambiguous, means to close the angular momentum budget and accounting for deep-seated phenomena in shallow-forcing models for gas giants \citep{Schn:09,Liu:10jets,Liu:15}.

\newcommand{\london}{Similarly to \cite{Liu:10jets}, the bottom drag is not exerted at equatorial latitudes (i.e. $|\varphi| < \varphi_R = 33^{\circ}$) as it artificially suppresses the cylindrical barotropic circulation structures that develop along the rotational axis (Taylor columns). This approach mimics the so-called tangent cylinder, which is thought to cause the equatorial super-rotating jets in deep-convective models \citep{Heim:05,Kasp:09}. }
\london
The value~$\varphi_R = 33^{\circ}$ is obtained by the geometrical constraint~$r_R = a \cos \varphi_R$, with~$r_R = 0.84 \, a \simeq 50600$~km, ie. a depth of $\simeq 9600$~km below the 1-bar level. This corresponds to the depths at which electrical conductivity significantly increases and the Lorenz drag putatively slows down Saturn's deep jets \citep{Liu:08}. This value of $r_R$ is also consistent with the~$8000 - 10000$~km value of the depth of Saturn's jet streams determined recently from Cassini ``Grand Finale'' gravity measurements \citep{Gala:19}.

The initial temperature field in the three-dimensional Saturn DYNAMICO GCM consists in the same vertical profile being set in every grid point of the horizontal mesh. This profile is obtained from uni-dimensional (single-column) computations \citep[\emph{\`a la}][]{Guer:14} using the exact same physical parameterizations and vertical discretization than the full Saturn DYNAMICO GCM integrations. The single-column model is initialized with an isothermal profile and run for two Saturn decades to ensure that the annual-mean steady-state radiative-convective equilibrium is reached, especially at the deepest layers at~$3$ bars (a couple Saturn years is usually enough to reach equilibrium in the stratosphere where radiative timescales are shorter than in the troposphere, see Figure~\ref{fig:spinup} and section~\ref{sec:thermal}). The initial zonal and meridional wind fields in our reference~$1/2^{\circ}$ Saturn DYNAMICO GCM simulations are set to zero.

Discussions on the impact of numerical dissipation and on the conservation of angular momentum in our DYNAMICO-Saturn model are respectively detailed in appendices~\ref{sec:dissip} and~\ref{sec:aam}.

\section{Atmospheric dynamics in our reference Saturn GCM simulation \label{sec:reference}}

Hereafter are discussed the results of 15 complete Saturn years
simulated by our Saturn DYNAMICO GCM 
with~$1/2^{\circ}$ longitude/latitude resolution.

\subsection{Thermal structure \label{sec:thermal}}
 
\begin{figure}[h!]
\begin{center}
\includegraphics[width=0.49\textwidth]{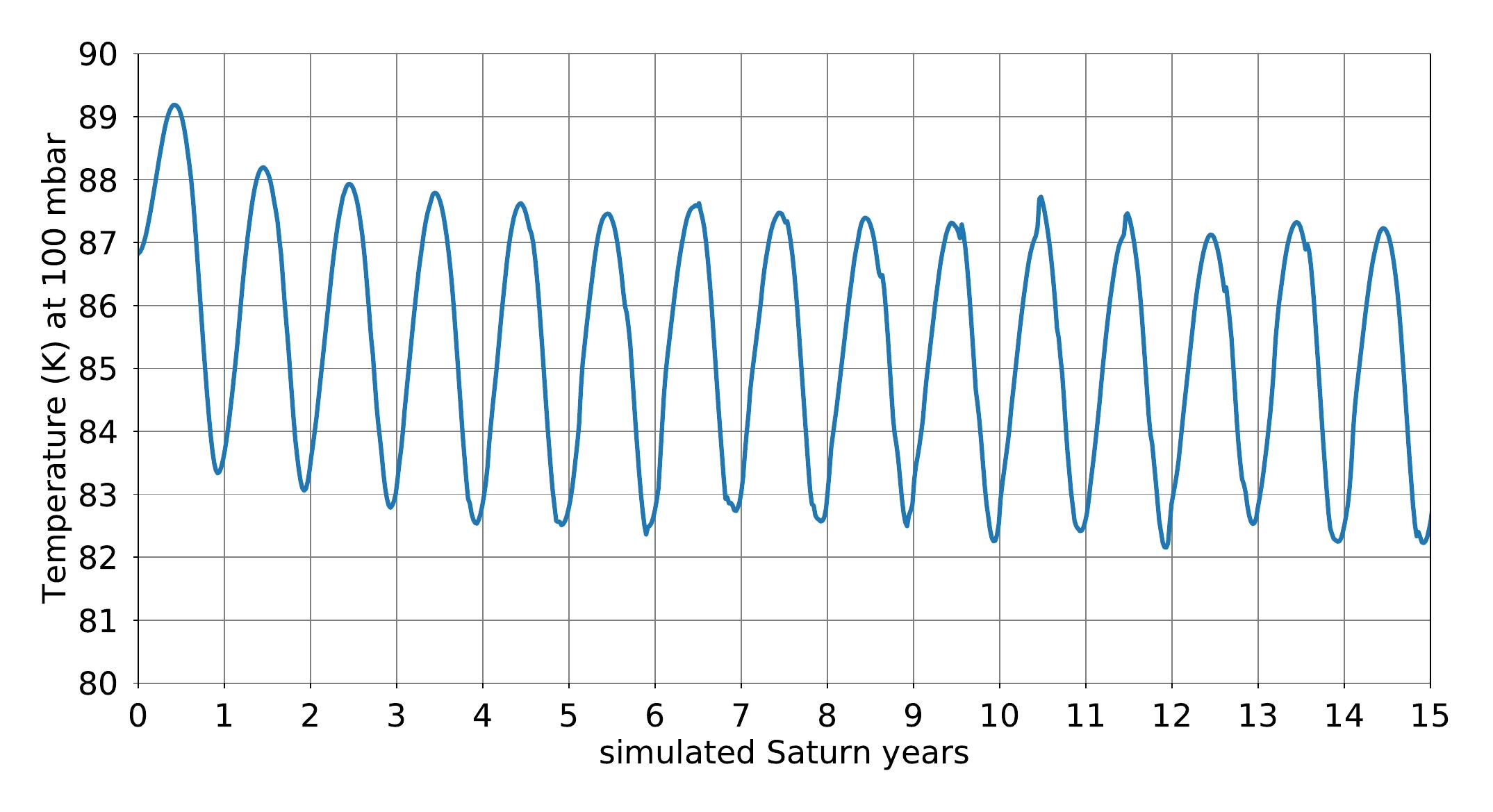}
\includegraphics[width=0.49\textwidth]{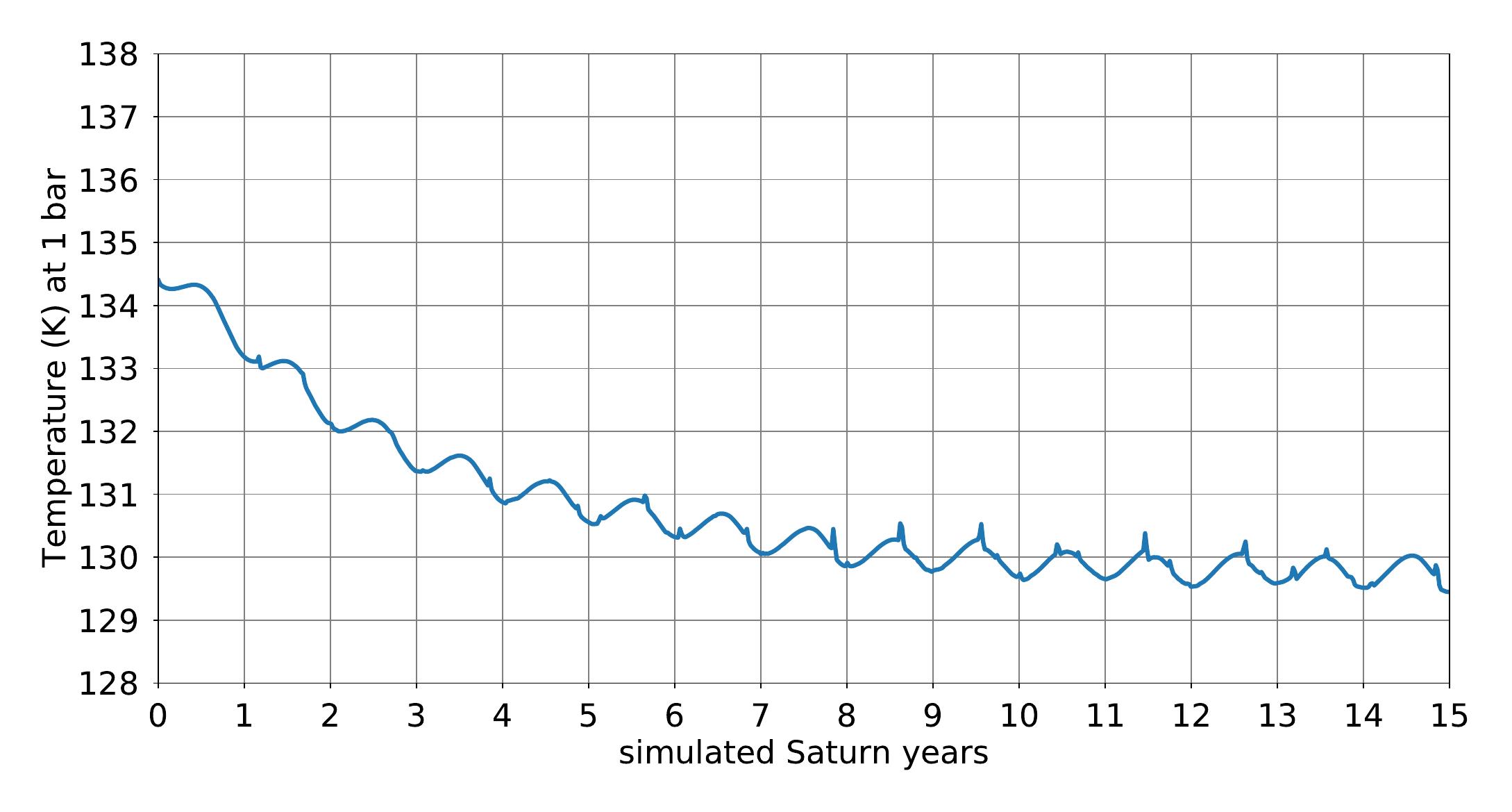}
\caption{\label{fig:spinup}
\emph{Evolution 
over the whole 15-year duration of 
the reference Saturn GCM simulation 
of zonal-mean temperature~$\overline{T}$, 
averaged over latitudes~$\varphi = 20-60^{\circ}$N, 
in the tropopause region ($p = 100$~mbar, left)
and in the troposphere ($p = 1$~bar, right).}}
\end{center}
\end{figure}

The analysis of angular momentum
in appendix~\ref{sec:aam}
shows that 
the 15-year duration of simulation ensures 
dynamical spin-up
and that the dynamical fields are 
in quasi-steady state.
Full radiative spin-up must also be
ensured, along with dynamical spin-up.
In Figure~\ref{fig:spinup}
the evolution of the mean
temperature in the northern hemisphere
is shown:
as is expected from differences in radiative timescales,
the troposphere takes longer to reach steady-state
seasonal cycle (about eight Saturn years) than
the tropopause level does (about three Saturn years).
\newcommand{\sydney}{This shows that 
satisfactory spin-up, 
both dynamical and radiative, is ensured 
starting from the ninth simulated year.} \sydney

The comparison of our seasonal radiative-convective model 
with observations, 
both from instruments on board the Cassini spacecraft
and ground-based telescopes,
is discussed at length in \citet{Guer:14}.
Yet a sanity check is necessary, given
that we now use this model interactively
with a three-dimensional dynamical core.

\newcommand{\canberra}{Model results are similar 
should any of the simulated Saturn year
starting from year eight be considered.}
\begin{figure}[p!]
\begin{center}
\includegraphics[width=0.65\textwidth]{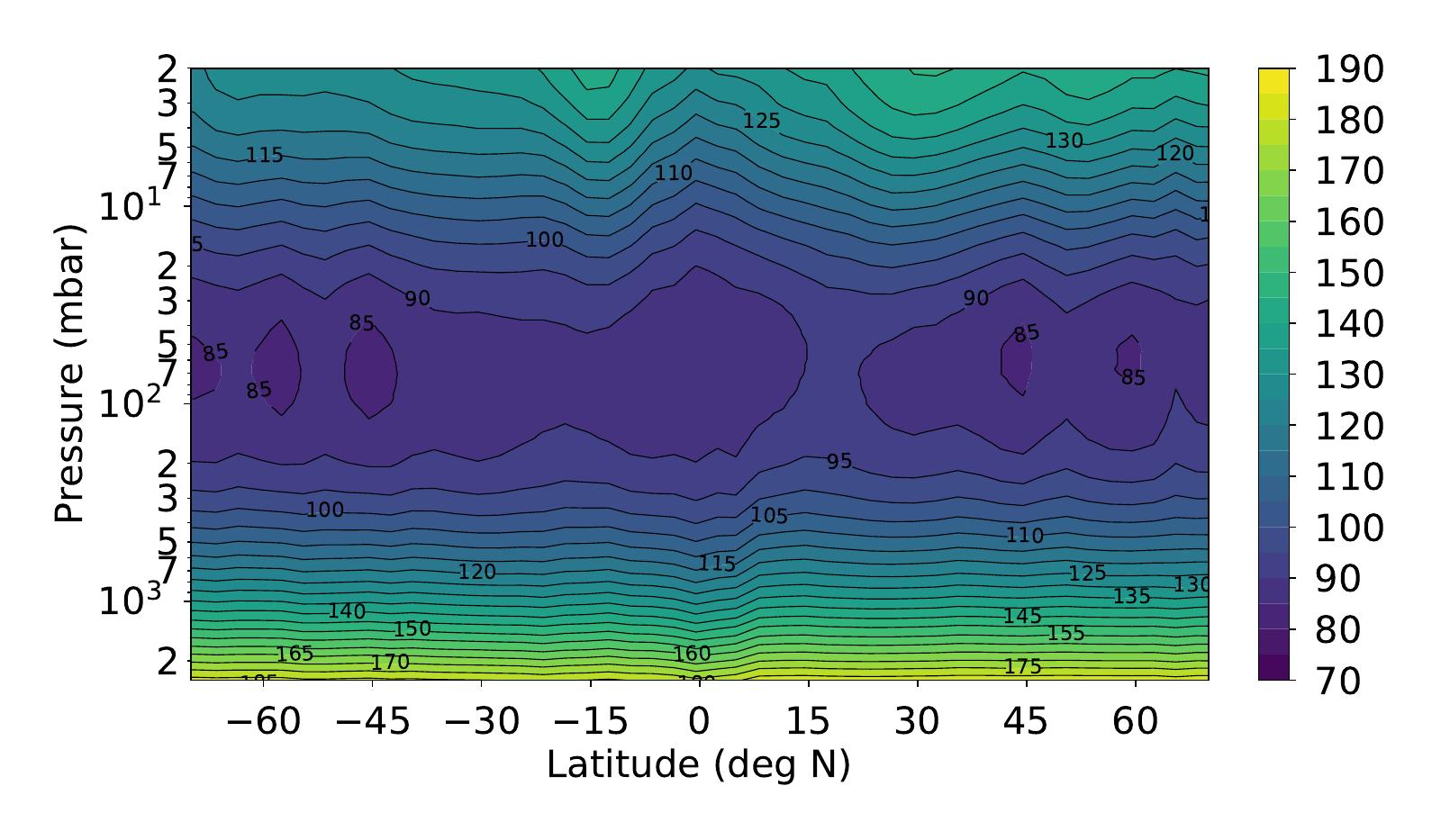}
\includegraphics[width=0.65\textwidth]{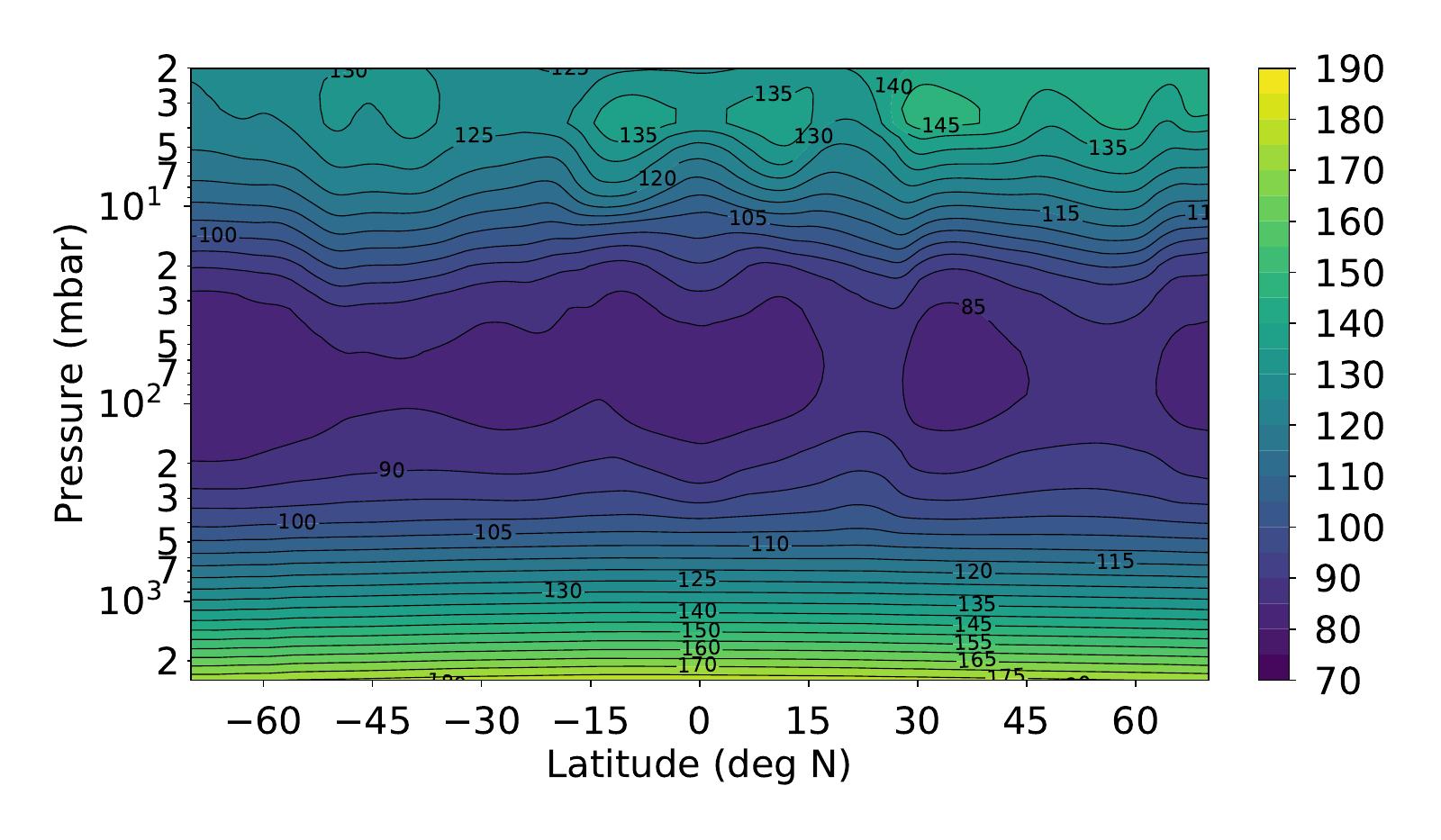}
\caption{\label{fig:tsections}
\emph{Latitude-pressure section of zonal-mean temperature~$\overline{T}$
(top) observed by Cassini/CIRS in 2015 \citep{Guer:15dps} vs.
(bottom) produced by our Saturn DYNAMICO GCM
in the fourteenth simulated Saturn year at~$L_s \sim 70^{\circ}$ 
(see Table~\ref{tab:notations} for a definition of~$L_s$).
The Cassini/CIRS observations shown here 
are nadir retrievals, with optimal sensitivity
in the~$500-70$~mbar and~$5-0.5$~mbar ranges.
\canberra}}
\end{center}
\end{figure}

\begin{figure}[ht]
\begin{center}
\includegraphics[width=0.65\textwidth]{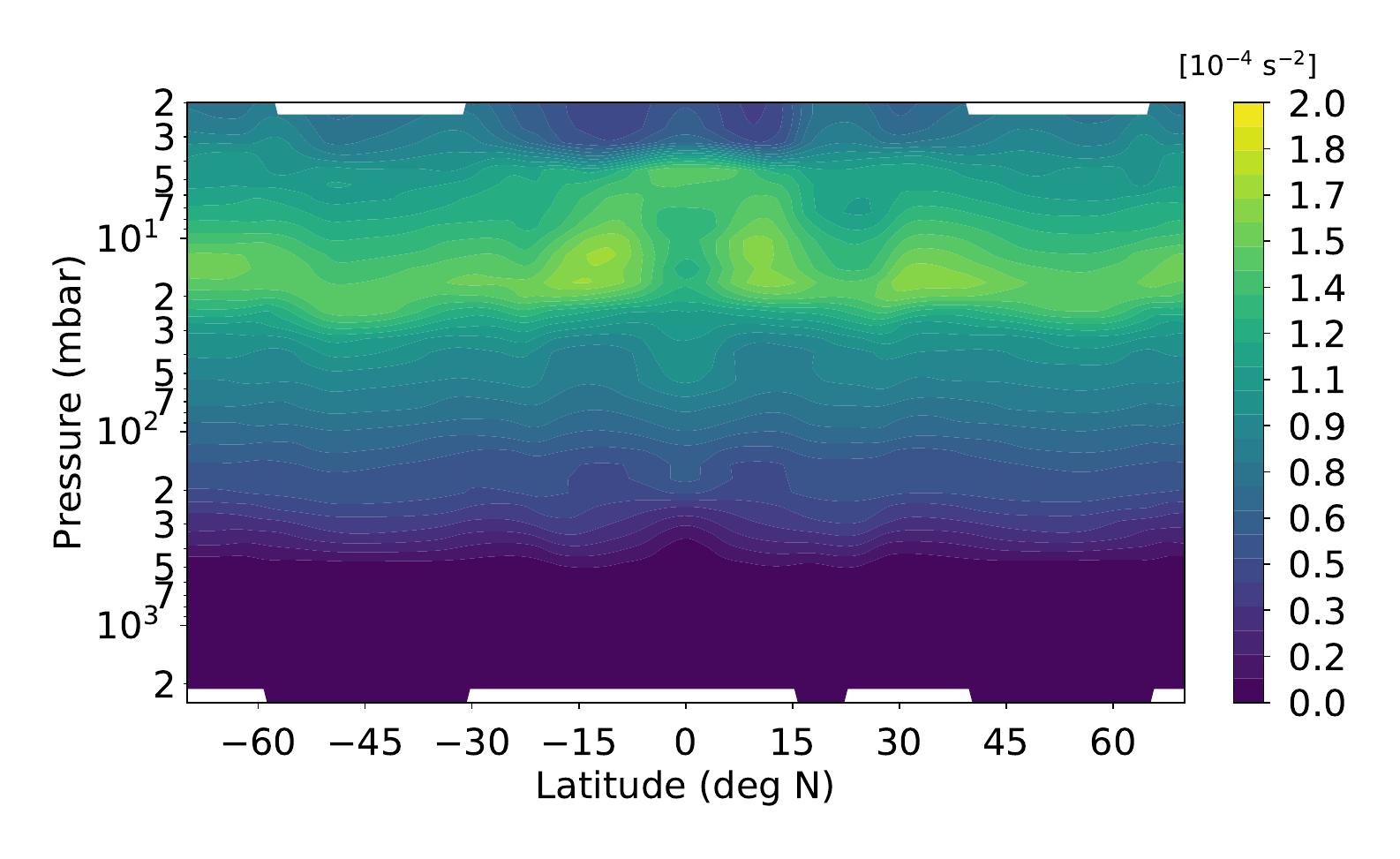}
\includegraphics[width=0.65\textwidth]{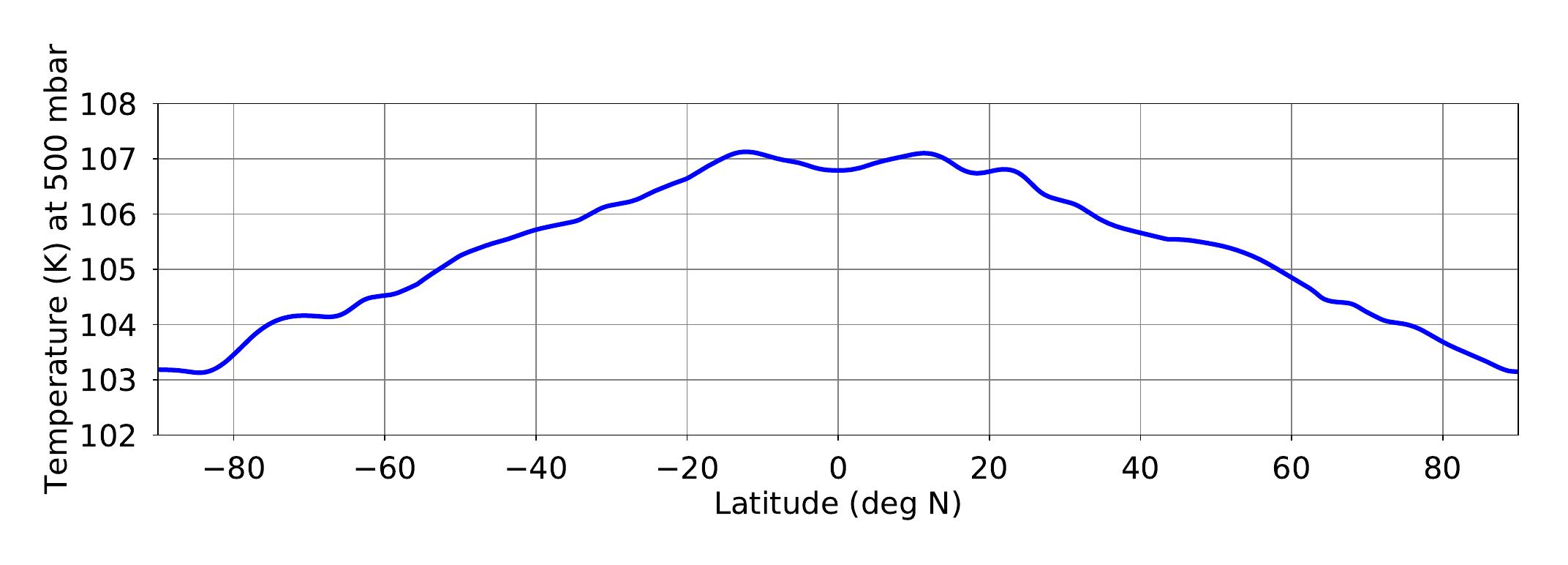}
\caption{\label{fig:bv}
\emph{(Top) Latitude-pressure section of zonal-mean 
Brunt-V{\"a}is{\"a}l{\"a} frequency~$\overline{N^2}$
and (bottom) meridional profile of zonal-mean temperature~$\overline{T}$
at 500 mbar, 
averaged over the fourteenth simulated Saturn year
with our Saturn DYNAMICO GCM.}}
\end{center}
\end{figure}

Figure~\ref{fig:tsections} shows meridional-vertical
sections of zonal-mean temperatures, both
simulated by our Saturn DYNAMICO GCM
and
observed by the Cassini / 
Composite InfraRed Spectrometer (CIRS)
instrument in 2015. 
The model successfully reproduces 
the vertical transition
from troposphere to stratosphere,
and the rather flat meridional gradients of temperature at this season ($L_s \sim 70^{\circ}$).
In Figure~\ref{fig:bv} (top), the simulated
meridional-vertical section of the 
Brunt-V{\"a}is{\"a}l{\"a} frequency~$\overline{N^2}$
indicates that the radiative-convective transition,
between the neutral profile ($N^2 \sim 0$) 
in the bulk of the troposphere
and the stable profile ($N^2 > 0$) 
in the upper troposphere and lower stratosphere,
occurs around~$500-600$~mbar, which is
in agreement with observations
\citep{Pere:06,Flet:07}.

\newcommand{\moscow}{Nevertheless, 
while a $1-2$~K contrast between low latitudes and
the pole is observed by Cassini 
at 500 mbar \citep[][Figure 3 bottom]{Flet:10season},
the simulated
pole-to-equator meridional gradient
of temperature in our Saturn DYNAMICO GCM
at the 500 mbar level is about~$4$~K (Figure~\ref{fig:bv} bottom).
This is half of the meridional gradient
simulated by \citet{Liu:10jets},
but still at least twice
the observed gradient.
Most of the incident solar flux 
on Saturn is absorbed in the 
upper troposphere haze
\citep{Pere:06,Flet:07},
but our radiative model \citep{Guer:14}
shows that the solar flux is
not zero at~$500$~mbar
and enough to cause
a hemispheric meridional 
gradient of temperature.} \moscow

\begin{figure}[p!]
\begin{center}
\includegraphics[width=0.75\textwidth]{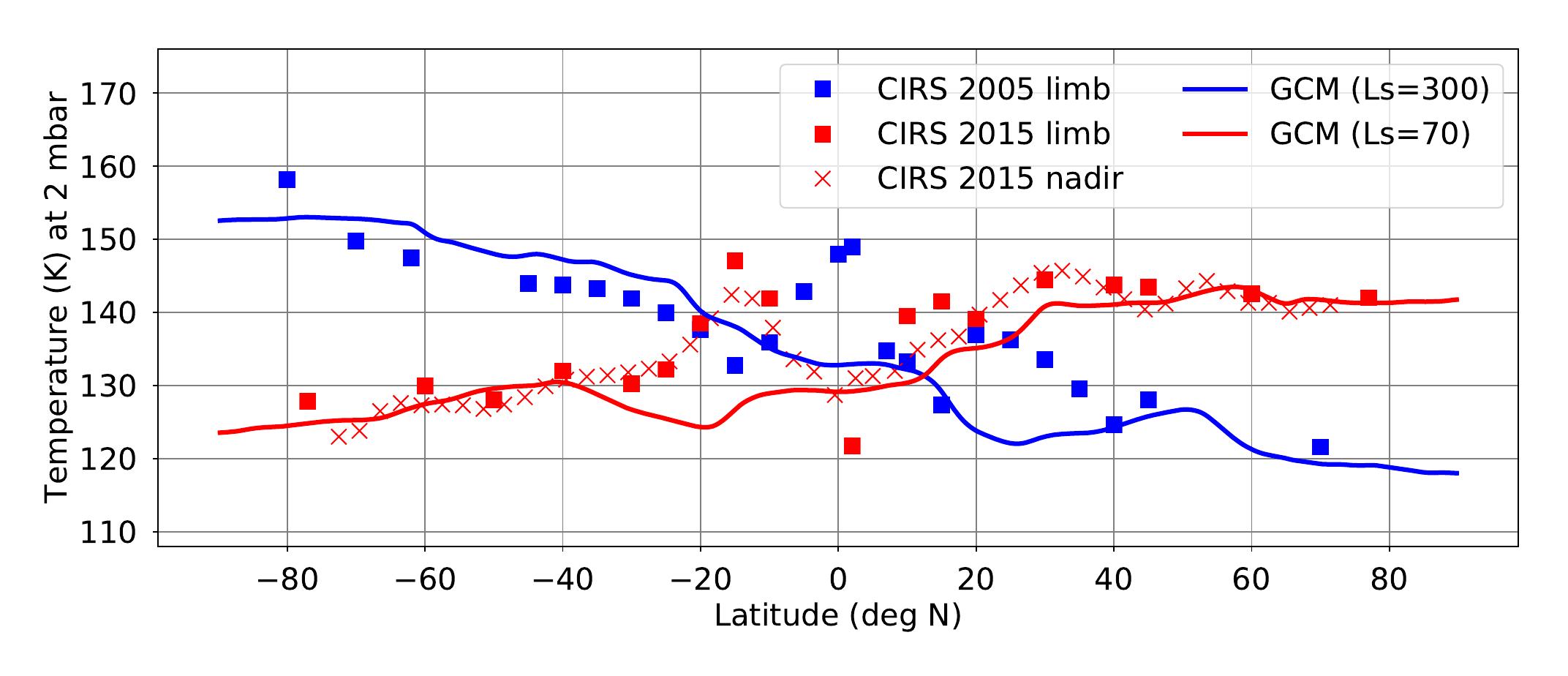}
\includegraphics[width=0.75\textwidth]{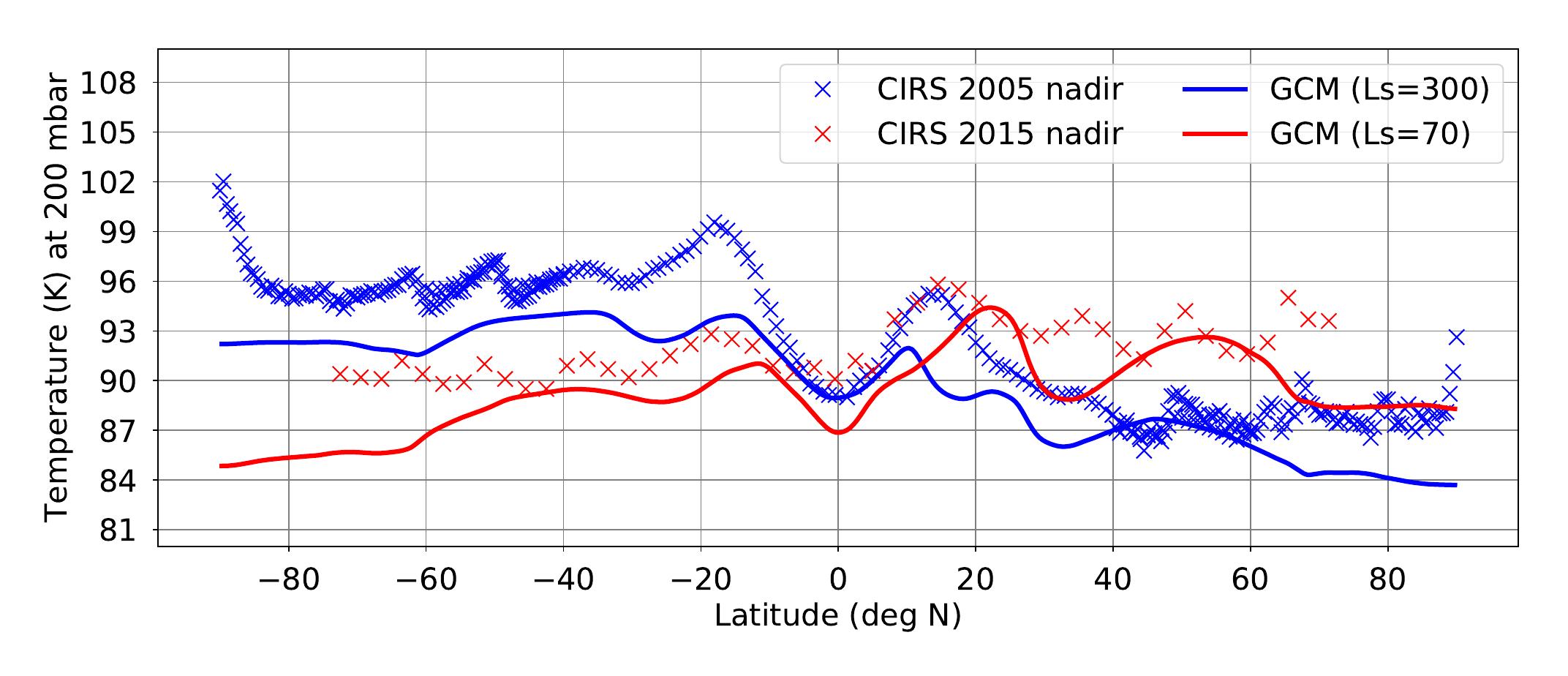}
\caption{\label{fig:compcirs}
\emph{Meridional profiles of zonal-mean temperature~$\overline{T}$
at the two opposite solstices explored by the Cassini spacecraft
(blue: year 2005, $L_s = 300^{\circ}$;
red: year 2015, $L_s = 70^{\circ}$).
Squares correspond to Cassini / CIRS limb retrievals \citep{Guer:09,Guer:15dps},
crosses correspond to Cassini / CIRS nadir retrievals \citep{Flet:07,Guer:15dps},
lines correspond to Saturn DYNAMICO GCM simulations
(fourteenth simulated year). 
Middle stratospheric conditions (2~mbar)
are shown in the top plot,
where both CIRS limb and nadir retrievals are valid;
upper tropospheric conditions (200~mbar)
are shown in the bottom plot,
where CIRS nadir retrievals are valid.
\canberra}}
\end{center}
\end{figure}

A more detailed comparison than Figure~\ref{fig:tsections}
is displayed in Figure~\ref{fig:compcirs},
at two typical pressure levels where
Cassini/CIRS is the most sensitive,
close to the two solstices within which Cassini was operating.
The meridional gradient of temperature, 
and the seasonal variability thereof,
is correctly represented in our model.
The fact that summer stratospheric temperatures are
$5-10$~K too warm compared to CIRS observations
was also noted with 
one-column radiative-convective 
modeling \citep{Guer:14,Sylv:15}
and is not a feature introduced by our dynamical simulations.
Putative dynamical effects (e.g. Brewer-Dobson seasonal circulations)
had been proposed to explain this discrepancy between
radiative models and CIRS observations; however, the adopted
setting for our Saturn DYNAMICO GCM simulations does
not allow us to address this question that would
require to raise the model top above the 1-mbar level.

The zonal jets produced by our dynamical
model (discussed at length in what follows)
are associated with distinctive temperature signatures,
i.e. localized meridional gradients of temperature
(see Figure~\ref{fig:tsections} and~\ref{fig:compcirs}),
according to the thermal wind equilibrium
which links $\partial \overline{u} / \partial z$ 
to $\partial \overline{T} / \partial y$.
These thermal signatures associated with jets are 
of similar amplitude between modeling and observations,
although the localization (i.e. latitude) of those 
thermal signatures is not compliant
between models and observations, 
echoing the discrepancies in latitude between
the observed and modeled jet structures
(see section~\ref{sec:jets}).

\subsection{Tropospheric and stratospheric jets \label{sec:jets}}

\subsubsection{Mid-latitude jets \label{sec:midlatjets}} 

\begin{figure}[ht]
\begin{center}
\includegraphics[width=0.65\textwidth]{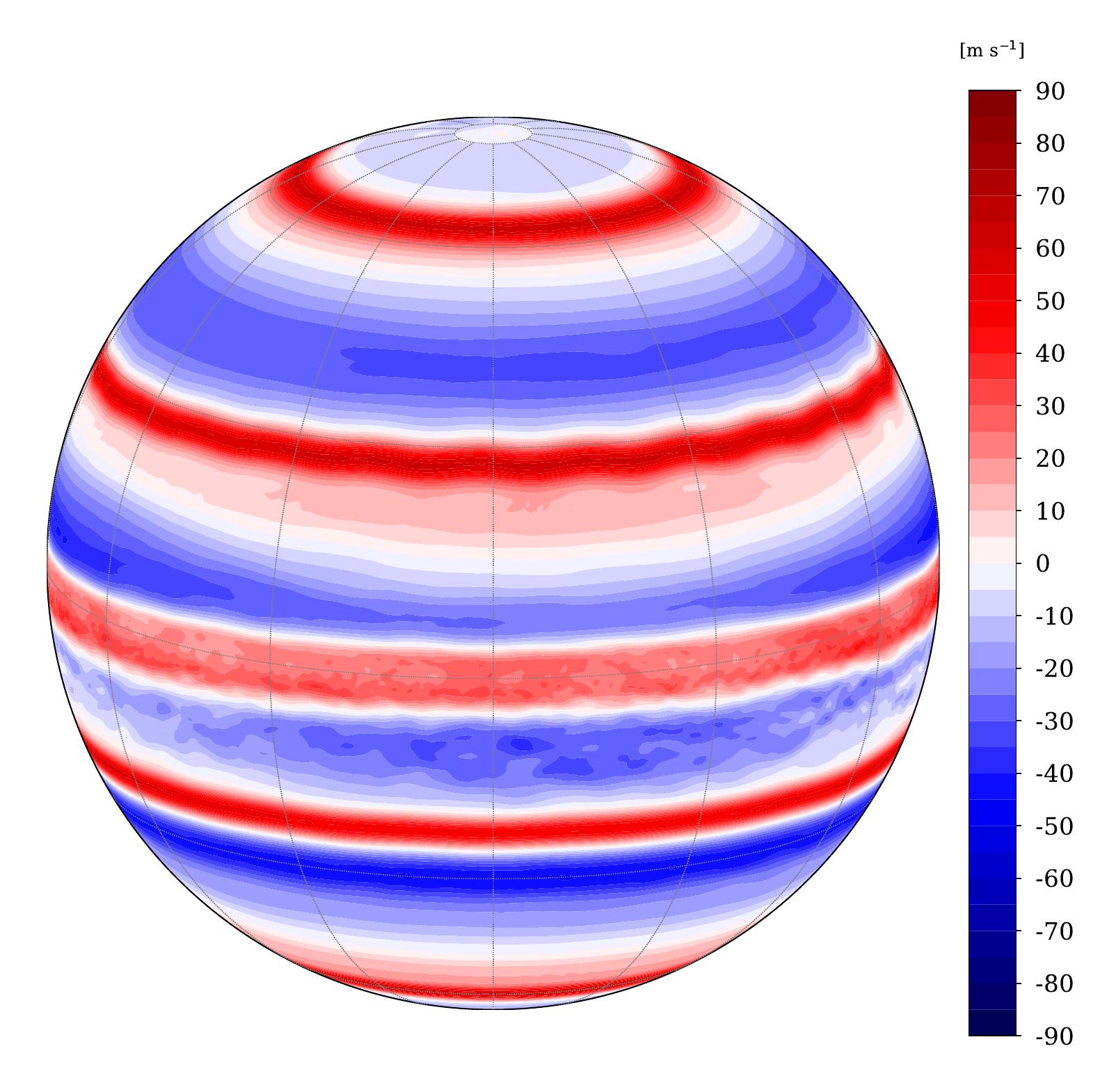}
\caption{\label{fig:sphereu}
\emph{Instantaneous zonal wind~$u$ 
in the beginning ($L_s \sim 0^{\circ}$) 
of the twelfth simulated year 
(after 270 thousands simulated Saturn days),
on the fifth 
sigma level of the model 
(pressure level~$\sim 1.5$~bar,
corresponding to Saturn's visible cloud deck 
and tropospheric conditions).}}
\end{center}
\end{figure}

Figure~\ref{fig:sphereu} is a snapshot of the steady-state zonal flow
of our $1/2^{\circ}$ Saturn DYNAMICO GCM simulation.
Our model produces mid-latitude zonal jets,
both eastward and westward
(i.e. prograde and retrograde),
which average intensities over a year
reach 
about~$45 - 50$~m~s$^{-1}$ for westward jets
and~$60 - 65$~m~s$^{-1}$ for eastward jets
at the visible cloud deck ($0.8-1.5$~bar).
Thus the strength of mid-latitude zonal jets
modeled in our Saturn DYNAMICO GCM
are, to first order, consistent with
the observed winds \citep{Porc:05,Garc:10}
recast in the System IIIw rotating frame 
following \citet{Read:09rot} (see their Figure 2a). 
The quantitative match between our GCM and
the observations is not perfect, 
for modeled mid-latitude jets are underestimated
by about~$25\%$ compared to observations.
Although the number of mid-latitude prograde zonal jets produced in our Saturn DYNAMICO GCM 
is compliant with observations ($2-3$ per hemisphere),
the latitude of occurrence of the modeled zonal jets
do not exactly match the observations, where the mid-latitude jets are more closely grouped.
Our GCM predictions for mid-latitude zonal jets 
are broadly consistent with 
the previously-published body of work
employing Saturn GCMs \citep[e.g.,][]{Liu:10jets,Lian:10}.

\begin{figure}[p!]
\begin{center}
\includegraphics[width=0.75\textwidth]{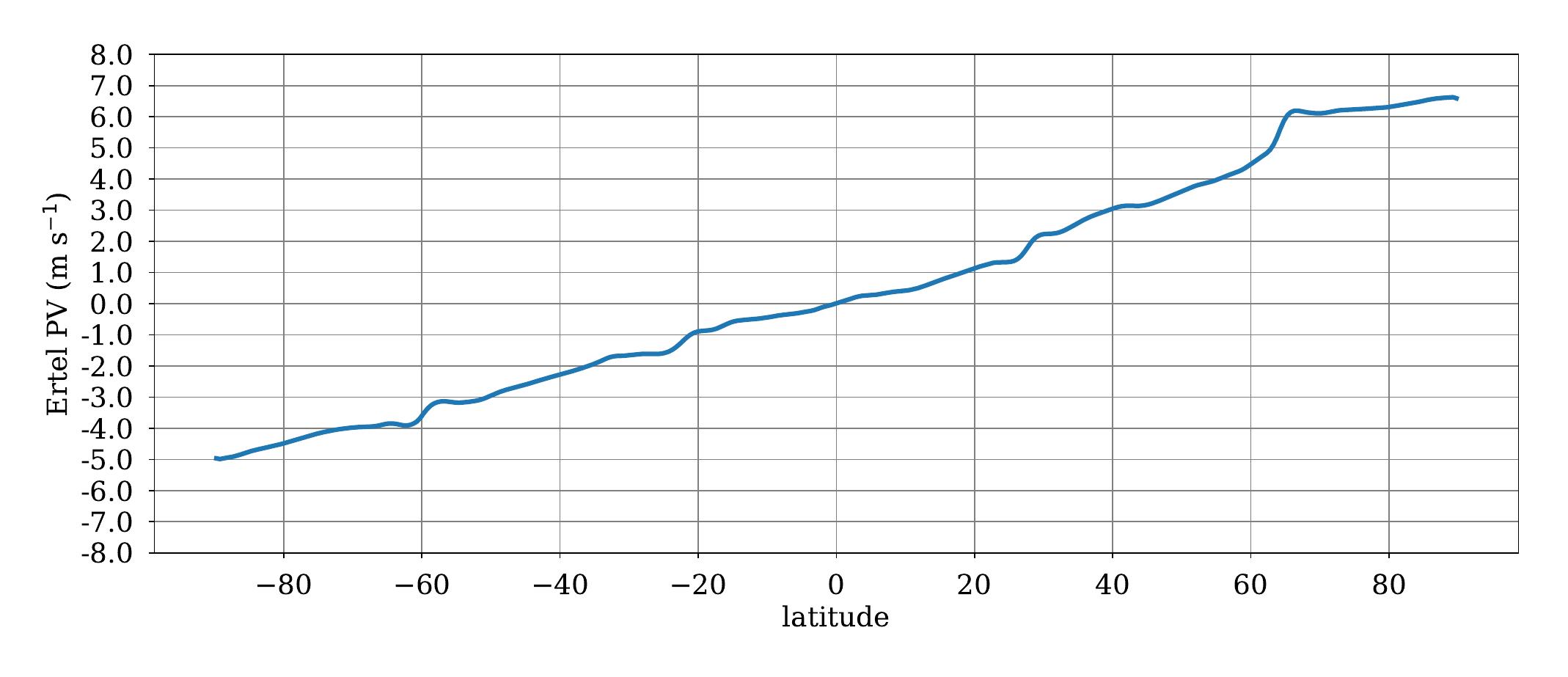}
\includegraphics[width=0.75\textwidth]{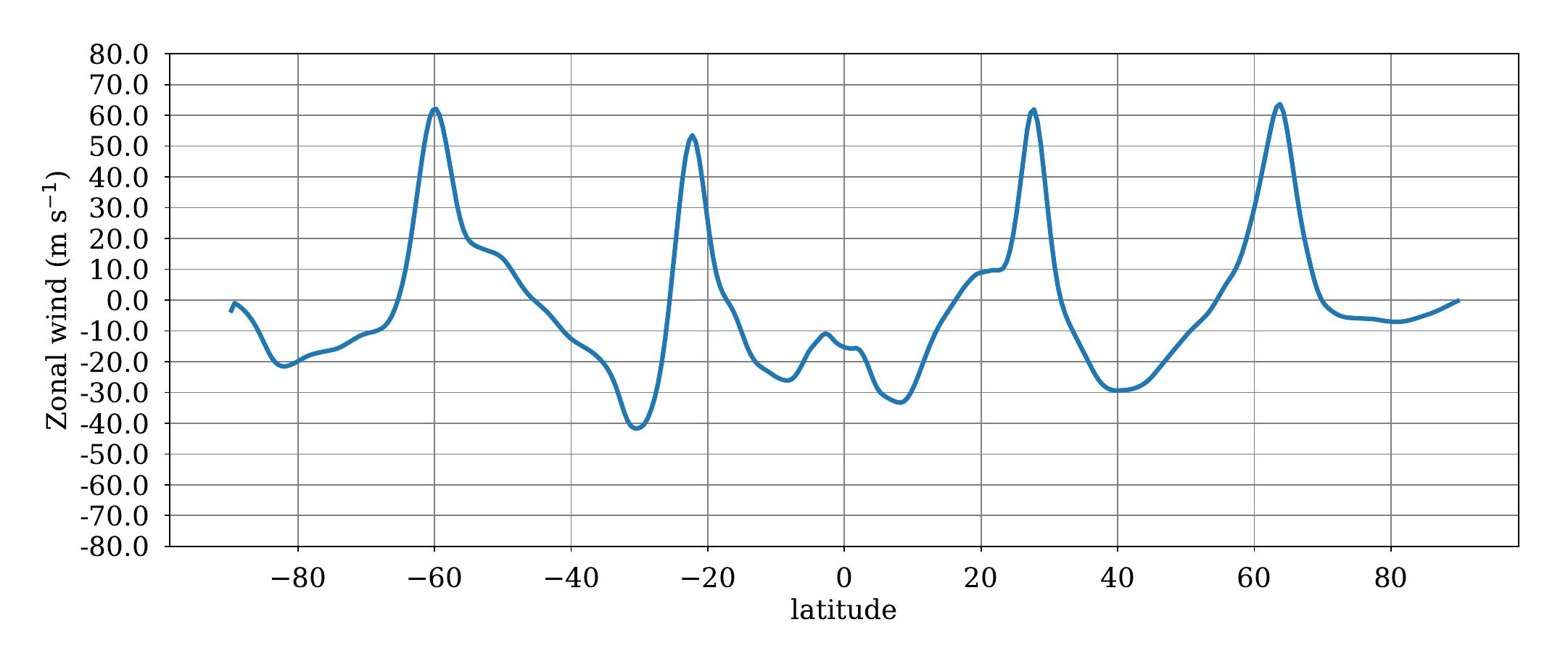}
\caption{\label{fig:pvprof}
\emph{Instantaneous
zonal-mean Ertel potential vorticity (top)
and zonal wind (bottom)
projected on a surface of equal potential temperature~$205$~K,
corresponding to tropospheric conditions.
The time adopted for this plot
is similar to Figure~\ref{fig:sphereu}.
The $205$~K value of potential temperature
corresponds to upper tropospheric conditions,
higher in altitude than Figure~\ref{fig:sphereu};
lower tropospheric conditions are not accessible
through this diagnostic due to the
difficulty of defining the Ertel potential vorticity
on isentropes in neutral conditions ($N^2 \sim 0$).
Our Python code used to calculate 
potential vorticity contains excerpts
from the code by \cite{Barl:17code}.}}
\end{center}
\end{figure}

\newcommand{\dakar}{The Ertel potential vorticity 
(PV)~$q_{\theta}$ calculated on
isentropic surfaces 
-- a conserved quantity
(i.e. a flow ``tracer'') for
adiabatic motions \citep{Vall:06} --
is defined under hydrostatic approximation
following \citet{Read:09pv} equation~3
\begin{equation}\label{ipv}
q_{\theta} = -g \, \left( f + \zeta_{\theta} \right) \, \Dp{\theta}{p}
\end{equation}
\noindent where 
$f$ and $\zeta_{\theta}$ are defined as 
in Table~\ref{tab:notations}, 
with the $\theta$ subscript denoting evaluation 
across a surface of uniform potential temperature~$\theta$. } \dakar 
The meridional profile of PV associated
with the tropospheric jet structure simulated by our
Saturn DYNAMICO GCM is shown in
Figure~\ref{fig:pvprof}.
Characteristic PV ``staircases''
(i.e. sharp PV gradients) are found 
within the core of each mid-latitude eastward jet,
surrounded by areas with
``mixed PV''  (i.e. uniform PV with latitude)
on the flanks of the jets.
The Ertel PV field obtained in Figure~\ref{fig:pvprof}
with our model
is similar to the one obtained through 
Cassini measurements by \cite{Read:09pv}.
This result, reminiscent
of those obtained with idealized models
of rapidly-rotating flows
\citep{Dunk:08,Drit:11,Marc:11},
shows that the emergence
and sharpening
of mid-latitude eastward jets
is associated with 
PV mixing.
This homogeneization of PV 
on the flanks of the jets
is associated with the breaking
of Rossby waves emitted 
at the core of the jet \citep[e.g.,][]{Drit:08}.
This creates a convergence
of eastward momentum towards the
regions of wave emission,
i.e. the core of the eastward jets
\citep{Vall:06,Schn:09,Show:11polvani},
helping to maintain the jet
structure against dissipation
\citep{OGor:08}. 

\begin{figure}[ht]
\begin{center}
\includegraphics[width=0.85\textwidth]{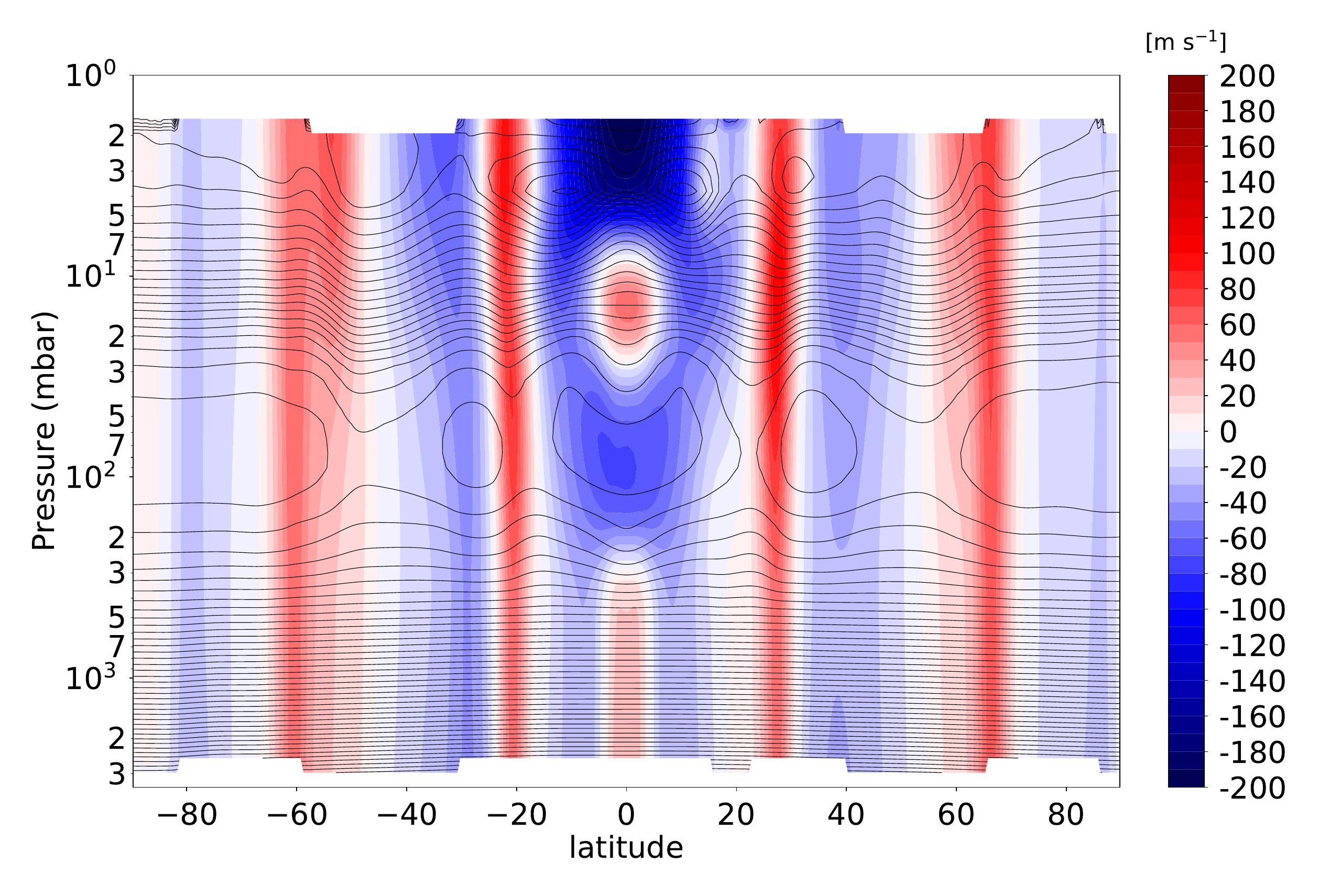}
\caption{\label{fig:section}
\emph{Instantaneous latitude-pressure cross-section of
zonal-mean zonal wind~$\overline{u}$ (colors)
and zonal-mean temperature~$\overline{T}$ (contours)
in the beginning ($L_s \sim 0^{\circ}$) 
of the fifteenth simulated year 
(after about 342 thousands simulated Saturn days).
\canberra}}
\end{center}
\end{figure}

The vertical structure of the zonal-mean zonal jet system
simulated by our Saturn DYNAMICO GCM is displayed in
Figure~\ref{fig:section}. The mid-latitude 
eastward and westward jets
exhibit a barotropic structure (i.e. weak vertical shear)
in the deep troposphere,
and a baroclinic structure 
(i.e. significant vertical shear)
in the upper troposphere / stratosphere.
The latter is in balance with the meridional temperature
variations observed and modeled in the 
temperature structure in Figure~\ref{fig:compcirs}.
Interestingly, eastward mid-latitude jets 
simulated in our Saturn DYNAMICO GCM
does not weaken from the cloud level around 1 bar
to the upper troposphere, 
as is observed \citep{Garc:10,Delg:12};
the jet intensity actually tends to slightly
increase upwards from the troposphere
to the stratosphere in our simulations.
Using Cassini VIMS images,
\cite{Stud:18} found that the intensities
of mid-latitude jets were generally increasing from
the 2-bar level to the 300-500~hPa level,
which tends to confirm our Saturn DYNAMICO GCM results
at and below the cloud level.

Accounting for the preferential zonal wavenumber~$n=6$ (hexagonal) mode 
in the circumpolar jet structure on Saturn is still an open question \citep{Mora:11,Mora:15}, 
given the narrow parameter space which allows for this mode 
to predominate over other modes \citep{Barb:10,Rost:17saturn}. 
The polar jet in our Saturn DYNAMICO GCM simulation 
(Figures~\ref{fig:sphereu} and~\ref{fig:instpv}) 
has a different morphology than the other mid-latitudes jets,
exhibiting meandering with time, 
which cause it to undergo latitudinal 
deformation and temporal variability.
However, the meandering of our simulated polar jet 
is intense and very variable with time,
with neither a~$n=6$ nor any mode~$n$ predominance.
This is clearly at odds with the observed stable 
slowly-moving hexagonal jet 
in Saturn's northern polar regions
\citep{Sanc:14hexagon,Antu:15}. 
The polar jet's zonal wind speed, latitudinal position and width are, however, 
key factors to account for Saturn's northern polar hexagon \citep{Mora:11}. 
The polar jet simulated by our Saturn GCM is 
both too weak and too equatorward (as in Figure~\ref{fig:sphereu})
to possibly lead to a predominance of the~$n=6$ mode. 
Furthermore, the polar jet's 
temporal evolution influenced by 
poleward migration
causes it to break under 
intensified meandering
by barotropic and baroclinic instability
(see section~\ref{sec:evolution}).
This obviously prevents
any high-latitude jet
to settle as a stable, wavenumber-6 hexagon-shaped, structure. 
Either the baroclinicity in the polar regions is not realistic 
enough in our Saturn DYNAMICO GCM 
\citep[this influence of vertical shear is discussed in][]{Mora:15};
and/or the central polar vortex 
is insufficiently resolved
in our simulations, hence
too weak to stabilize the hexagonal shape of the polar jet against meandering
\citep[the influence of the central polar vortex is discussed in][]{Rost:17saturn}.

\subsubsection{Equatorial jets \label{sec:eqjets}} 

A prograde equatorial jet is produced in the troposphere
by our~$1/2^{\circ}$ Saturn GCM simulation. 
It is, however, severely underestimated by one order of magnitude
compared to the observed value by Cassini 
\citep[$\sim 350-400$~m~s$^{-1}$ in System IIIw,][]{Read:09rot,Garc:10}.
The local super-rotation index~$s$ associated with this equatorial jet writes
\citep[e.g.,][]{Read:18}
\begin{equation}
s = \frac{\mu^m + \overline{\mu^w}}{\mu^m_{\varphi=0}} - 1
= \frac{a \, \cos\varphi \, (\Omega \, a \, \cos\varphi + \overline{u})}{\Omega \, a^2} - 1
\end{equation}
\noindent The 
equatorial zonal jet simulated by our GCM is only weakly super-rotating~($s \sim 0.3 \%$)
in a very limited area across the equator ($\varphi = \pm 3^{\circ}$),
while the observed superrotating index is 
an order of magnitude larger \citep[][their table 1]{Read:18}
and the observed equatorial jet extends towards latitudes $\pm 15-20^{\circ}$.

\newcommand{\tunis}{This temporal averaging is carried out 
because, contrary to other jets,
the equatorial jet does not migrate with time
and the equatorial eddies develop continuously 
rather than through ``bursts''
(see Figure~\ref{fig:jetevolution} in section~\ref{sec:evolution}).}
\begin{figure}[ht]
\begin{center}
\includegraphics[width=0.7\textwidth]{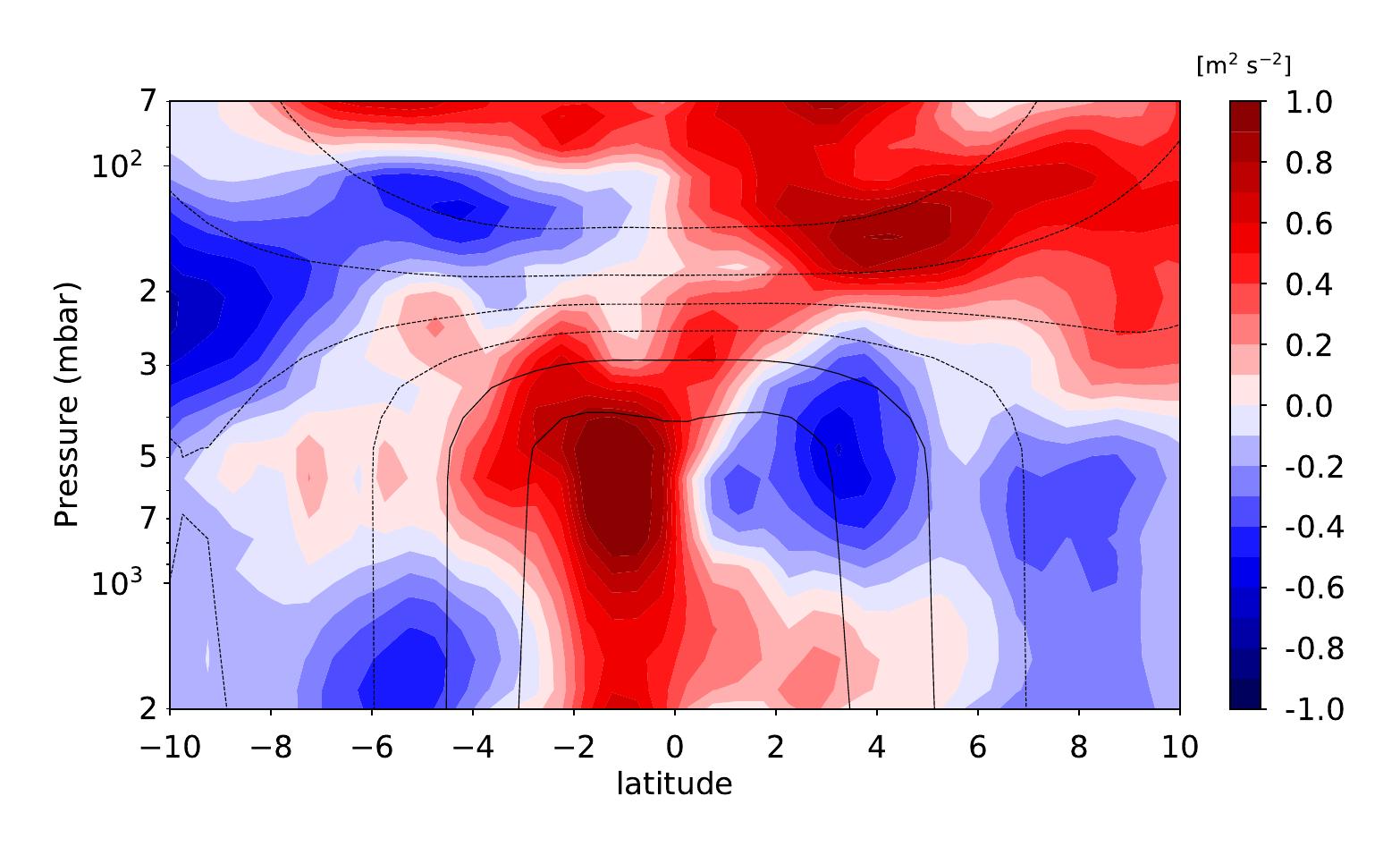}
\caption{\label{fig:section_eqsuper}
\emph{Latitude-pressure cross-section of
zonal-mean eddy momentum transport $\overline{u^{\prime}v^{\prime}}$
(colors) and zonal wind~$\overline{u}$ (contours)
averaged over the whole 15-year duration of the reference simulation.
\tunis
}}
\end{center}
\end{figure}

The prograde, eastward, equatorial jet in the Saturn DYNAMICO GCM
arises from acceleration caused by
convergence of eddy momentum towards the equator
(see equation~\ref{eq:eddyacc} in section~\ref{sec:evolution}).
Figure~\ref{fig:section_eqsuper} shows
that, within the equatorial super-rotating jet,
the eddy momentum transport $\overline{u^{\prime}v^{\prime}}$
is positive south of the equator
and negative north of the equator,
meaning that waves and eddies cause
a convergence of eastward momentum at the equator.
Yet this equatorial acceleration by waves and eddies
is probably underestimated by our Saturn DYNAMICO GCM,
given the resulting modeled jet being ten times less strong
than the observed equatorial jet \citep{Garc:10}. 
This is consistent with the latitudinal profile of Ertel PV
shown in Figure~\ref{fig:pvprof}, where PV mixing
by Rossby waves in equatorial regions appears
not sufficient to yield a truly PV-mixed area,
as is the case for mid-latitude jets.

Based on the existing literature \citep[e.g.,][]{Gier:00,Lian:10}, 
a possible source of this underestimate 
of Saturn's equatorial superrotation could be 
the lack of a parameterization for moist convection 
in our model \citep[see, e.g., the work on Jupiter by][]{Zuch:09clouds,Youn:19parttwo}. 
Our GCM results are actually in contrast with the simulations by~\cite{Liu:10jets} 
which did not include an additional (moist) convective source, 
apart from the combination of internal heat flux and convective adjustment. 
The fact that the convective adjustment scheme in~\cite{Liu:10jets} 
has a non-zero relaxation time \citep[cf.][appendix B, section d]{Schn:09}, 
while ours is instantaneously adjusting, 
might be an element of explanation for this discrepancy. 
\newcommand{\oslo}{Using a non-zero relaxation time might emulate
the convective overturning time of dry and moist convective structures,
confirming the need to add, in our Saturn DYNAMICO GCM, 
a (dry and moist) convective parameterization
more sophisticated than our simple convective adjustment scheme
\citep[e.g. thermal plume modeling like ][]{Hour:02,Rio:08,Cola:13}
that will better represent the local dynamics 
underlying convective mixing
and the impact thereof on the generation
of waves and eddies.} \oslo

\begin{figure}[p!]
\begin{center}
\includegraphics[width=0.65\textwidth]{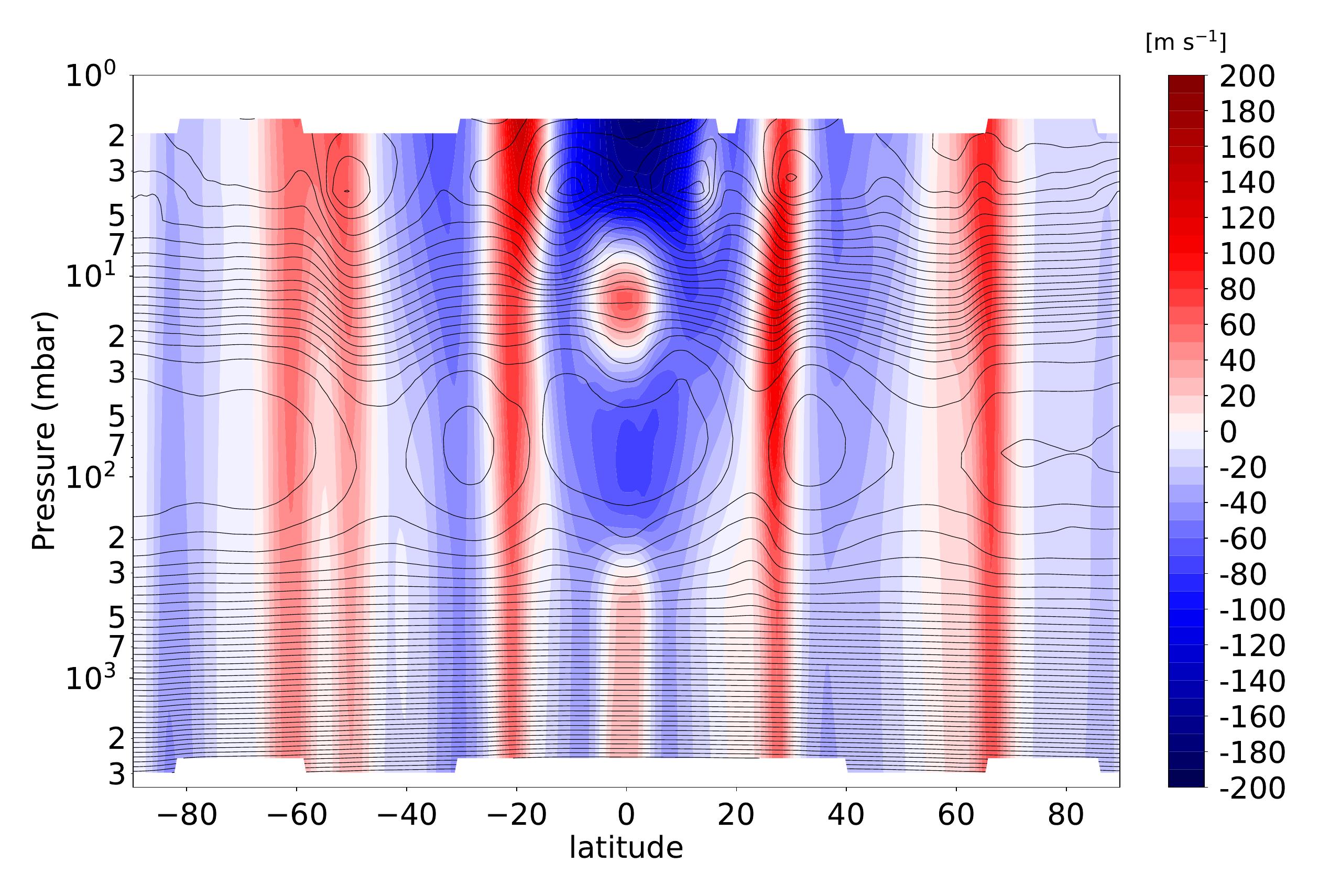}
\includegraphics[width=0.65\textwidth]{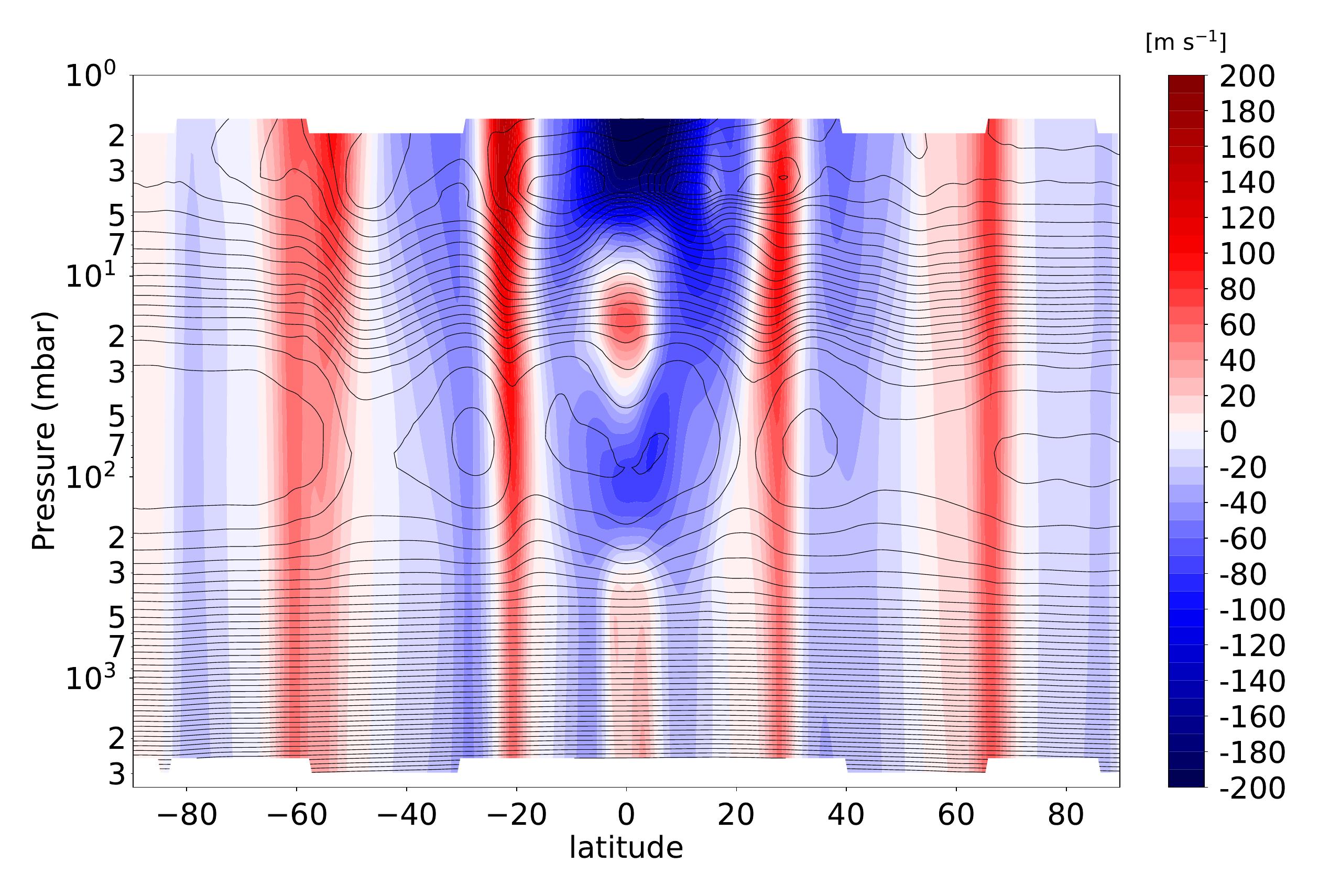}
\caption{\label{fig:section_season}
\emph{Same as Figure~\ref{fig:section}
for northern summer solstice ($L_s \sim 90^{\circ}$, top panel)
and fall equinox ($L_s \sim 180^{\circ}$, bottom panel).}}
\end{center}
\end{figure}

While the strengths of mid-latitude jets
increase with altitude in our Saturn DYNAMICO GCM simulations
(see section~\ref{sec:midlatjets}),
the intensity of the simulated equatorial jet decreases with altitude, 
which is also in line with the Cassini observations reported in \cite{Stud:18}.
Although a quantitative comparison with observations
is prevented by the severely underestimated
equatorial wind speed in our GCM,
the fact that the equatorial jet decays from
the cloud level to the tropopause
is also observed by 
\cite{Flas:05},
\cite{Li:08}, and \cite{Sanc:16}.
Figure~\ref{fig:section_eqsuper}
indicates that this decay is caused in
our Saturn DYNAMICO GCM by 
divergence of eastward eddy momentum 
at the tropopause,
in contrast to the 
convergence of eastward eddy momentum
causing the super-rotating jet 
in the troposphere.
\newcommand{\leicester}{Interestingly, 
to interpret the Cassini VIMS observations of tracers,
\citet{Flet:11vims} proposed that two stacked, reversed cells
are present in Saturn's troposphere,
one resulting from ``jet damping'' in the upper troposphere
and
one resulting from ``jet pumping'' in the mid-troposphere
(their section 6).
This is compliant with Figure~\ref{fig:section_eqsuper} 
showing in equatorial regions an
area of eddy divergence / westward jet
sitting on top of an 
area of eddy convergence / eastward jet.} \leicester

Saturn's equatorial stratosphere 
exhibits a downward propagation of
(supposedly zonally-symmetric)
alternating positive and negative
temperature perturbations 
with respect to the 
radiative-equilibrium temperature field
\citep{Fouc:08,Guer:11,Li:11,Flet:17,Guer:18}.
Those temperature signatures are
thought to be associated with
eastward and westward jets
alternating along the vertical,
similarly to the Earth's 
Quasi-Biennal Oscillation \citep{Bald:01}.
The typical period of this
equatorial stratospheric oscillation is~$15$~Earth years,
half a Saturn year \citep{Orto:08}.
Modeling could help to identify
the mechanisms responsible for 
Saturn's equatorial oscillation,
putatively the interaction of planetary-scale 
and small-scale waves with the mean zonal flow 
as it is the case of the terrestrial 
Quasi-Biennal Oscillation \citep{Bald:01}.

In Figure~\ref{fig:section},
above the eastward equatorial tropospheric jet,
we note the presence 
in our reference simulation
of stacked,
alternatively eastward and westward, 
stratospheric jets.
An eastward jet is
centered at pressure level~$15$~mbar
in the latitudinal range~$-5^{\circ} \rightarrow 5^{\circ}$N,
surrounded above and below
by westward jets.
This structure is reminiscent of the 
stacked jet signature derived
through thermal wind balance from
Cassini observations of equatorial oscillation
of temperature \citep[e.g.,][]{Guer:18}.
Yet, 
contrary to Cassini observations,
no downward propagation of
this jet signature is
reproduced by our Saturn DYNAMICO GCM
between northern summer 
and fall (Figure~\ref{fig:section_season}).

Section~\ref{sec:eqwave}
features a discussion of the
equatorial waves being most probably responsible
for the stacked jet structure in
the stratosphere,
and possible explanations
for the lack of vertical propagation
of the structure.
A key point, not related to the wave analysis,
that shall be emphasized here is that
the model top is too low (at 1~mbar), 
and the vertical resolution
of the model in the stratosphere
is too coarse, to correctly study
the observed equatorial oscillations.
For instance, the equatorial westward
jet above 10~mbar is compliant
with the thermal wind field 
derived by \cite{Fouc:08},
but our too-low model top
prevents us from discussing
the bulk of the observed oscillation above the 10-mbar pressure level.

\subsection{Waves and vortices}

\subsubsection{Equatorial waves \label{sec:eqwave}}

\begin{figure}[ht]
\begin{center}
\includegraphics[width=0.99\textwidth]{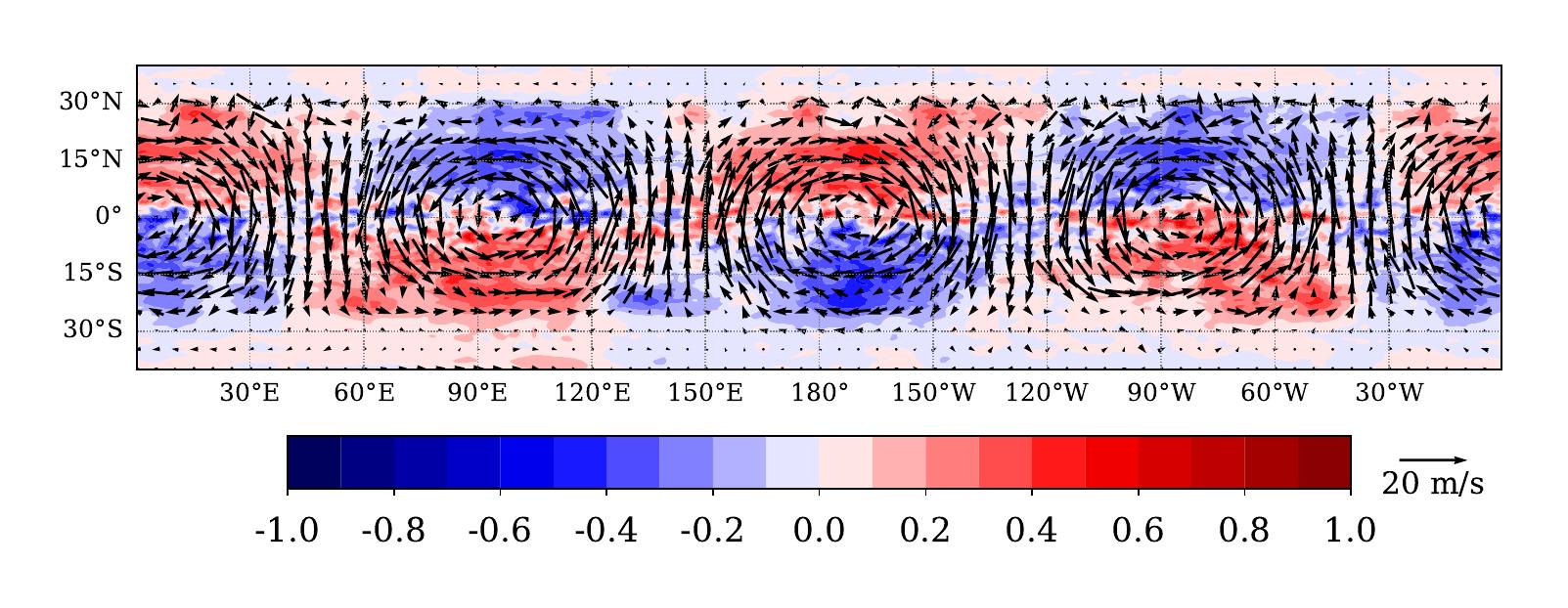}
\caption{\label{fig:eqwave}
\emph{Instantaneous view on the 
zonal perturbations of temperature~$T^{\prime}$ (shaded contours) and 
horizontal winds~$[u^{\prime},v^{\prime}]$ (wind vectors)
in the beginning ($L_s \sim 0^{\circ}$) 
of the twelfth simulated year 
(after 270 thousands simulated Saturn days),
on the fifteenth 
sigma level of the model 
(pressure level~$\sim 130$~mbar,
corresponding to Saturn's tropopause,
transition between troposphere and stratosphere).
Only latitudes below~$40^{\circ}$N/S are shown (tropical channel).
A similar figure is obtained at Saturn's cloud level
($\sim 1.5$ bar, as in Figure~\ref{fig:sphereu}),
with an even stronger prominence of the wavenumber-2 signal.}}
\end{center}
\end{figure}

A close examination of the tropical structure of zonal wind 
in Figure~\ref{fig:sphereu} hints at planetary wave activity 
-- notably a prominent wavenumber-2 signal.
This is confirmed by~Figure~\ref{fig:eqwave} 
in which the eddy (non-axisymmetric) components~$T^{\prime},u^{\prime},v^{\prime}$ 
are shown in the equatorial region at the tropopause level.
The prominent wavenumber-2 signal features 
zonal wind and temperature perturbations
(about~$0.5$~K) 
which are anti-symmetric about the equator, 
while meridional wind perturbations are symmetric about the equator.
The pattern shown in Figure~\ref{fig:eqwave}
is strongly reminiscent of 
an equatorial Yanai wave
\citep[also named mixed Rossby-gravity wave,][their Figure 3]{Kila:09},
although an interpretation as an
eastward inertio-gravity wave is also possible 
at this stage of the analysis.

To further characterize the wavenumber-2 signal,
and offer a more complete wave analysis 
(Figure~\ref{fig:eqwave} hints at other signals being
present besides the prominent wavenumber-2 signal),
we follow the method of \citet{Whee:99}
commonly employed to study equatorial waves 
in the terrestrial tropical atmosphere \citep{Kila:09,Maur:14}.
We perform a two-dimensional Fourier transform, 
from the longitude~$\lambda$ / time~$t$ space 
to the zonal wavenumber~$s$ / frequency~$\sigma$ space,
of the symmetric ($T_{\mathcal{S}}$) and antisymmetric ($T_{\mathcal{A}}$) 
components of the temperature field about the equator
\begin{equation}\label{eq:refwheeler}
T_{\mathcal{S}} = \frac{1}{2} \left( T_{10^{\circ}N} + T_{10^{\circ}S} \right) 
\quad
T_{\mathcal{A}} = \frac{1}{2} \left( T_{10^{\circ}N} - T_{10^{\circ}S} \right)
\end{equation}
\noindent (similar computations are also performed for
zonal wind~$u$ and meridional wind~$v$).
Our code
uses the Fast Fourier Transform package included in the \emph{scipy} Python library.
We validated independently our spectral analysis code on well-defined
(semi-)diurnal tides and Kelvin waves simulated 
in the Martian atmosphere \citep{Wils:96,Lewi:05,Guze:16msl}.

\begin{figure}[p!]
\begin{center}
\includegraphics[width=0.49\textwidth]{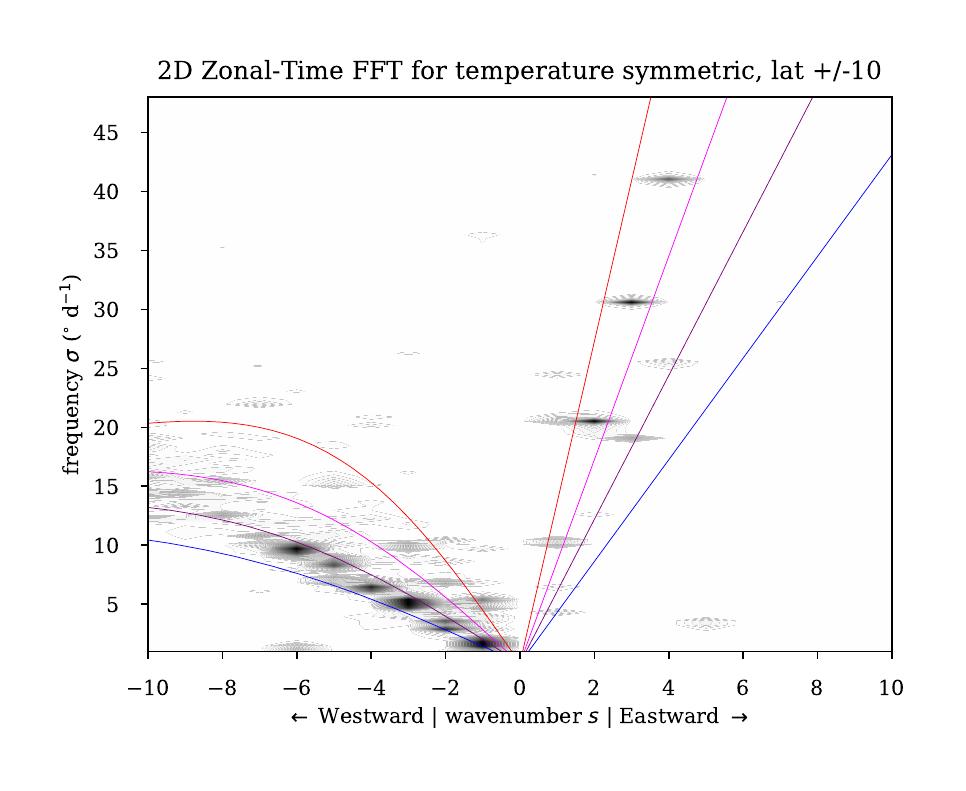}
\includegraphics[width=0.49\textwidth]{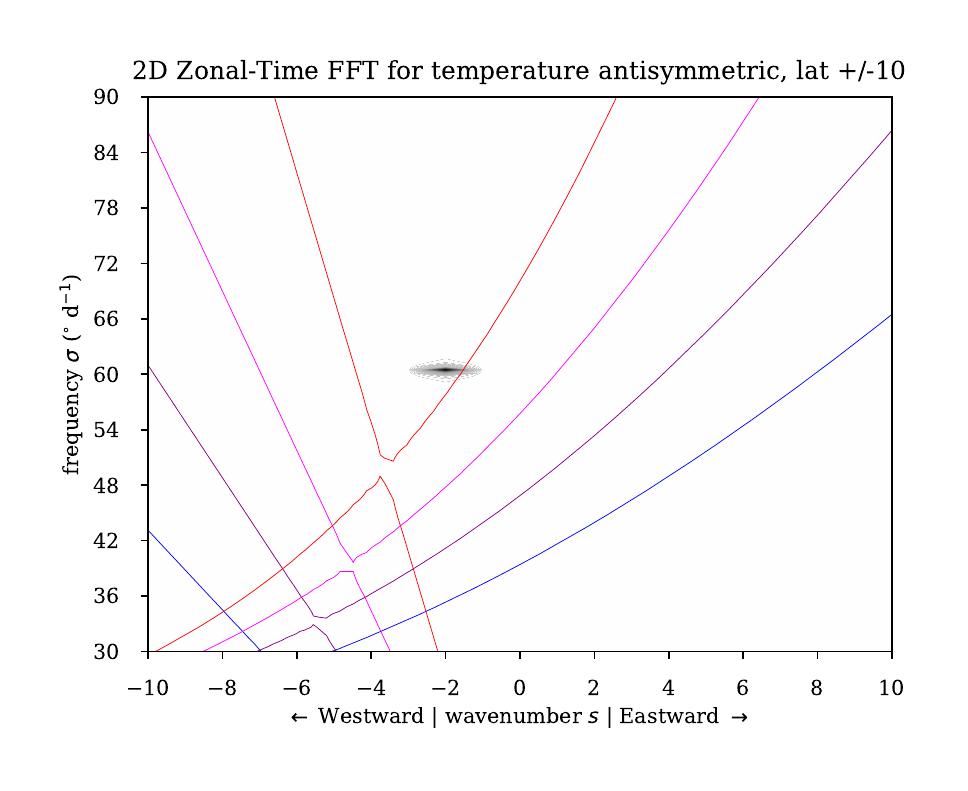}
\caption{\label{fig:wheelerspectra} \emph{Spectral analysis of the equatorial waves produced
in our Saturn DYNAMICO GCM (in dynamical steady-state)
at the same vertical level as Figure~{fig:eqwave}.
This spectral power mapping is obtained 
by a two-dimensional Fourier transform
from the longitude~$\lambda$ / time~$t$ space
to the zonal wavenumber~$s$ / frequency~$\sigma$ space \citep{Whee:99}
of the symmetric component of temperature~$T_{\mathcal{S}}$ about the equator (left, see equation~\ref{eq:refwheeler})
and antisymmetric compoment of temperature~$T_{\mathcal{A}}$ about the equator (right).
We follow the common assumption of a positive sign convention
for frequency~$\sigma$, meaning
that~$s>0$ modes are eastward-propagating
and~$s<0$ modes are westward-propagating.
The dispersion relation for equatorial waves 
(equation~\ref{eq:reldis}) 
is superimposed
for meridional mode number~$\nu=-1,+1$
(left panel, Rossby waves on the~$s<0$ side
and Kelvin waves on the~$s>0$ side)
and meridional mode number~$\nu=0$
(right panel, Yanai waves
on the part of the curves increasing with~$s$).
Four values of equivalent depths (equation~\ref{eq:eqdepth})
are included:
$h = 5$~km (blue), $h = 10$~km (purple),
$h = 20$~km (magenta), $h = 50$~km (red).}}
\end{center}
\end{figure}

We perform the Fourier analysis 
on a specific 1000-day-long~$1/2^{\circ}$ Saturn DYNAMICO GCM 
run with frequent (daily) output,
restarted from the GCM state 
after~$270$~thousands simulated Saturn days 
(about 11 simulated Saturn years).
The spectral mapping in the~$(s,\sigma)$ space
enables to evidence Rossby and Kelvin waves
in the symmetric component~$T_{\mathcal{S}}$
and Yanai waves
in the antisymmetric component~$T_{\mathcal{A}}$
\citep[the former waves can also be detected in $u_{\mathcal{S}}$
and the latter waves in $u_{\mathcal{A}}$ and~$v_{\mathcal{S}}$, see][]{Whee:99,Kila:09}.
Results for the temperature field simulated at the tropopause
are shown in Figure~\ref{fig:wheelerspectra},
along with the dispersion relation for equatorial waves \citep{Maur:14}
\begin{equation}\label{eq:reldis}
\sqrt{\gamma} \, (2\,\nu+1) = \gamma \, \sigma^2 - s^2 - s/\sigma 
\qquad
\gamma = \frac{4\,a^2\,\Omega^2}{g\,h}
\end{equation}
\noindent where~$\nu$ is the meridional mode number 
and
defines the considered wave
(Rossby: $\nu = +1, +2, \ldots$
Yanai: $\nu = 0$, 
Kelvin: $\nu = -1, -2, \ldots$),
$\gamma$ is named the Lamb parameter, 
and~$h$ is an equivalent depth associated with
the vertical wavenumber~$m$
\begin{equation}\label{eq:eqdepth}
m^2 = \frac{N^2}{g \, h} - \frac{1}{4 \, H^2}
\end{equation}
Dominant modes in the symmetric and antisymmetric
components of the temperature and wind fields
are detailed in Table~\ref{tab:spectra}.

\newcommand{\lyon}{The frequencies and periods are intrinsic,
i.e. with respect to a frame fixed on the zonally-averaged zonal flow
$\overline{u} \sim -40 m/s$ at latitudes~$10^{\circ}$N/S.}
\begin{table}[p]
\begin{center}
\begin{tabular}{cccc}
\multicolumn{4}{c}{Dominant modes in $T_{\mathcal{S}}$} \\ \hline
 $s$ & $\sigma \, (^{\circ}$/d) & period (d) &  log(SP) \\ \hline 
  +2 &     22.0 &     16.3 &      7.9 \\ 
  +3 &     32.1 &     11.2 &      7.6 \\ 
  -3 &      3.5 &    102.0 &      7.5 \\ 
  -6 &      8.2 &     43.8 &      7.4 \\ 
  -4 &      5.0 &     72.4 &      7.3 \\ 
  -2 &      1.4 &    262.7 &      7.3 \\ 
  -5 &      6.8 &     53.2 &      7.2 \\ 
  +4 &     42.5 &      8.5 &      7.2 \\ 
\multicolumn{4}{c}{~} \\
\multicolumn{4}{c}{Dominant modes in $u_{\mathcal{S}}$} \\ \hline
 $s$ & $\sigma \, (^{\circ}$/d) & period (d) &  log(SP) \\ \hline 
  -6 &      8.2 &     43.8 &      9.8 \\ 
  -2 &      3.5 &    102.0 &      9.7 \\ 
  -3 &      3.2 &    113.6 &      9.6 \\ 
  -4 &      4.3 &     84.7 &      9.6 \\ 
  -5 &      5.3 &     67.5 &      9.4 \\ 
  -7 &      8.9 &     40.3 &      9.4 \\ 
  +2 &     22.0 &     16.3 &      9.3 \\ 
\multicolumn{4}{c}{~} \\
\multicolumn{4}{c}{Dominant modes in $T_{\mathcal{A}}$, $u_{\mathcal{A}}$, $v_{\mathcal{S}}$} \\ \hline
 $s$ & $\sigma \, (^{\circ}$/d) & period (d) &  log(SP) \\ \hline 
  -2 &     59.0 &      6.1 &     10.1 \\
\end{tabular}
\end{center}
\caption{\label{tab:spectra} \emph{Spectral modes detected 
by Fourier analysis and depicted in Figure~\ref{fig:wheelerspectra}.
The spectral mapping for wind components
is not shown, as it is similar to the spectral
mapping for temperature shown in Figure~\ref{fig:wheelerspectra}.
\emph{SP} stands for spectral power, \emph{d} for Saturn days.
\lyon}}
\end{table}

The spectral analysis shows that (consistently in the three analyzed fields) 
the prominent wavenumber-2 signal is a westward-propagating Yanai wave 
with a period~$6$~days (frequency~$60^{\circ}$ longitude per day).
Our analysis also evidences,
both in the temperature and zonal wind fields, 
westward-propagating Rossby waves
with wavenumbers~$s=-2$ to~$s=-6$,
exhibiting long periods of hundreds Saturn days
and frequencies of a couple degrees longitude per day
(the wavenumber-$s=-1$ cannot be evidenced unambiguously 
since its intrinsic phase speed is too close to zero).
The longer-period Rossby waves are
modulating
the temperature variability 
caused by the wavenumber-2 Yanai wave
with a shorter 6-day period.
The temperature field also features
eastward-propagating Kelvin waves
with wavenumbers~$s=+2,+3,+4$,
periods $10-20$~Saturn days,
and frequencies a couple tens of degrees longitude per day;
the Kelvin wave signal is much fainter in the
zonal wind component.
This Kelvin wave signal is absent from the
temperature field lower in the troposphere, at the cloud level.

Elaborating from Voyager observations, \cite{Acht:96}
detected a Rossby wave\-number-2 signal
in the tropics and mid-latitudes 
of Saturn's tropopause (130~mbar),
confined by 
vertical variations of static stability.
\newcommand{\marseille}{A similar signal is present in our
our Saturn GCM DYNAMICO simulations:
1 degree longitude per day 
corresponds to $\sim 26.3$~m~s$^{-1}$,
hence the simulated Rossby wavenumber-2 signal
has a phase speed of~$\sim 37-92$~m~s$^{-1}$,
which is compatible with
the phase speed of the order~$100$~m~s$^{-1}$
discussed by \cite{Acht:96}. } \marseille
Our Saturn DYNAMICO GCM results indicate that other
tropical Rossby modes 
(wavenumbers 3, 4, 5, \ldots)
are likely to be significant 
within Saturn's tropics.
This is compliant with the recent
analysis by \cite{Guer:18},
based on Cassini CIRS observations,
which shows a complex structure
at a pressure level of 150~mbar;
interestingly, what appears
as a wavenumber-3 Rossby mode 
dominate in the upper stratosphere (0.5~to 5~mbar),
possibly indicating conditions for
breaking at this level or below
for the other modes.
Unless the wavenumber-2 signal 
found by \cite{Acht:96}
is actually eastward-propagating 
at phase speeds about~$1600$~m~s$^{-1}$
(the fast planetary modes were
discarded by this study in favour
of slower, more plausible, Rossby modes),
the prominent wavenumber-2 
westward-propagating Yanai wave
in our Saturn DYNAMICO GCM simulation
remains to be evidenced in observations.
A westward-propagating 
wavenumber-9 Yanai wave mode 
has been, however, detected by Cassini CIRS in
the upper stratosphere (1~mbar) by
\cite{Li:08}, but their observed 
temperature signature
being symmetric about the equator
(with a maximum at the equator)
would be more compliant
with a westward inertia-gravity 
wave \citep{Guer:18}.

The presence of equatorial 
vertically-propagating
eastward Kelvin waves
and
westward Rossby and Yanai waves 
in our Saturn DYNAMICO GCM simulations
at the 130-mbar level
means that both eastward and westward
momentum is transferred in the
stratosphere where 
vertically-stacked
westward/eastward jets
are found
(Figure~\ref{fig:section} and section~\ref{sec:eqjets}). 
This ``stacked jets'' equatorial signature is similar to
the jet structure putatively associated with
the equatorial oscillation of temperature \citep{Fouc:08,Guer:11}.
\newcommand{\perth}{Nevertheless, our model does not reproduce
the downward propagation of the observed
equatorial oscillation in Saturn's stratosphere \citep{Guer:18},
also obtained by idealized 
simulations \citep{Show:18qboarxiv}. } \perth
Furthermore, the modeled temperature 
contrasts between the equator and latitudes~$\pm 15^{\circ}$
associated with the stacked jets
($\sim \pm 5-10$~K, figure not shown)
are much lower than the contrasts
obtained by Cassini thermal infrared measurements
\citep[$\pm 15-20$~K,][]{Fouc:08,Guer:18}.
Those discrepancies with observations are 
probably related to 
a weak transfer of momentum to the mean flow
by the resolved waves in our model:
\begin{enumerate}[label=$(\alph*)$]
\item while our spectral analysis reveals a Kelvin-wave signal
at the tropopause, moist convection in the deep troposphere
of Saturn, not accounted for in the current version of our model,
could cause convectively-coupled Kelvin Waves,
which are an important component
to explain the Quasi-Biennal Oscillation on Earth \citep{Kila:09};
\item mesoscale (inertia-)gravity waves are not resolved
by our model and are known to contribute
to the momentum flux responsible for
equatorial oscillations on Earth \citep{Lind:68,Lott:13,Maur:14}
and this possibility has also been explored 
in Jupiter's stratosphere \citep{Cose:17};
\item the absence of a strong equatorial super-rotating jet
in our simulations means that the vertical propagation
(and possible filtering) of equatorial waves towards the stratosphere
is different between our simulations and the actual Saturn's atmosphere.
\end{enumerate}
According to terrestrial 
modeling studies 
\citep[e.g.][]{Taka:96,Niss:00,Wata:08,Lott:14},
our modeling setting is ultimately lacking
two key elements 
to reproduce Saturn's equatorial oscillation:
the model top must be raised to cover the stratospheric levels
($0.01-10$~mbar) where the stratospheric oscillation is observed,
and the vertical resolution must be refined in the stratosphere.
Those improvements, related to Challenge~\ref{ch:strat}, 
are deferred to a dedicated future study 
of Saturn's stratospheric circulations
using our Saturn DYNAMICO GCM
(Bardet D. et al., Part IV in preparation).

\subsubsection{Extratropical eddies \label{sec:extraeddies}}

\begin{figure}[p!]
\begin{center}
\includegraphics[width=0.48\textwidth]{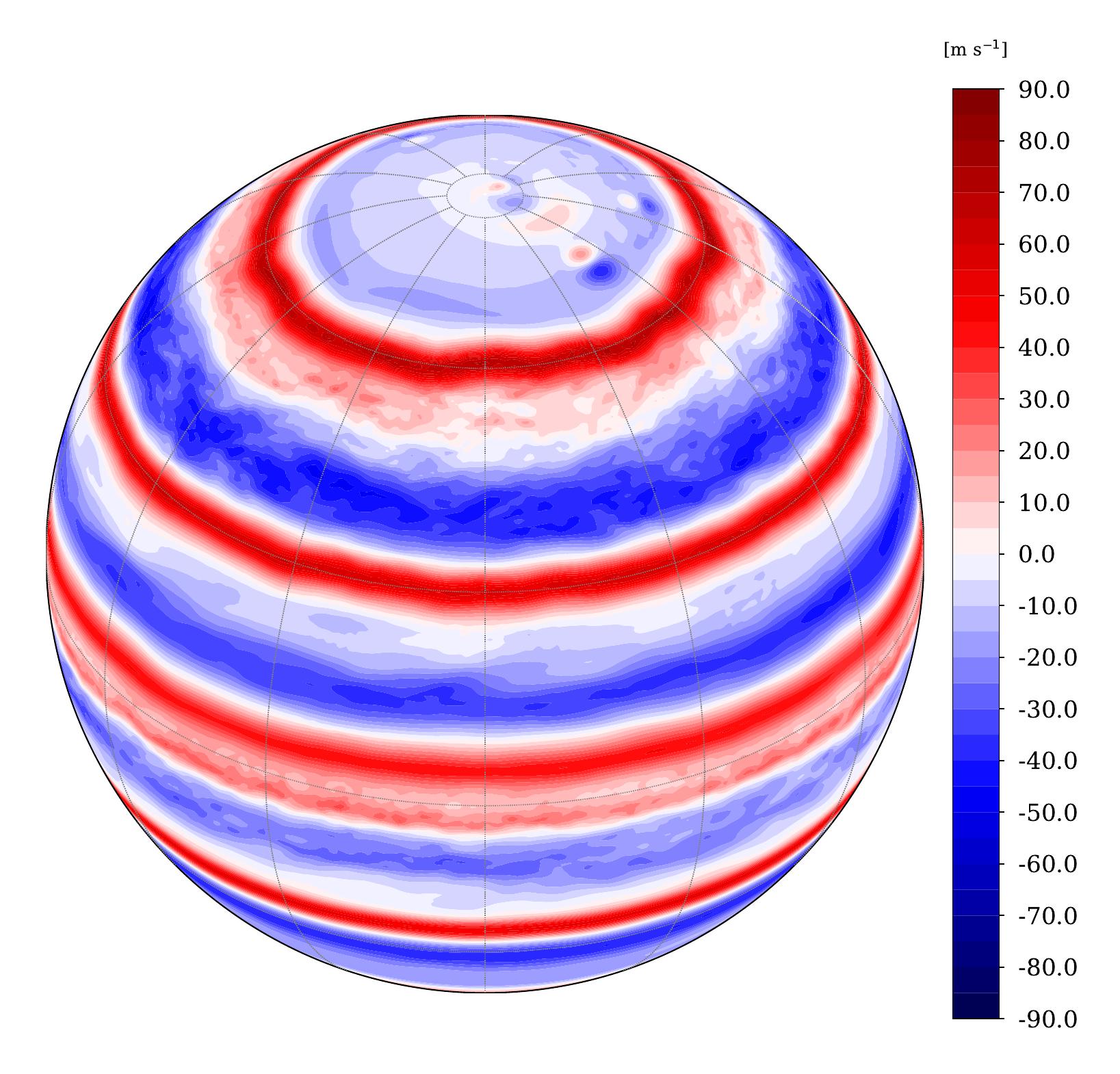}
\includegraphics[width=0.48\textwidth]{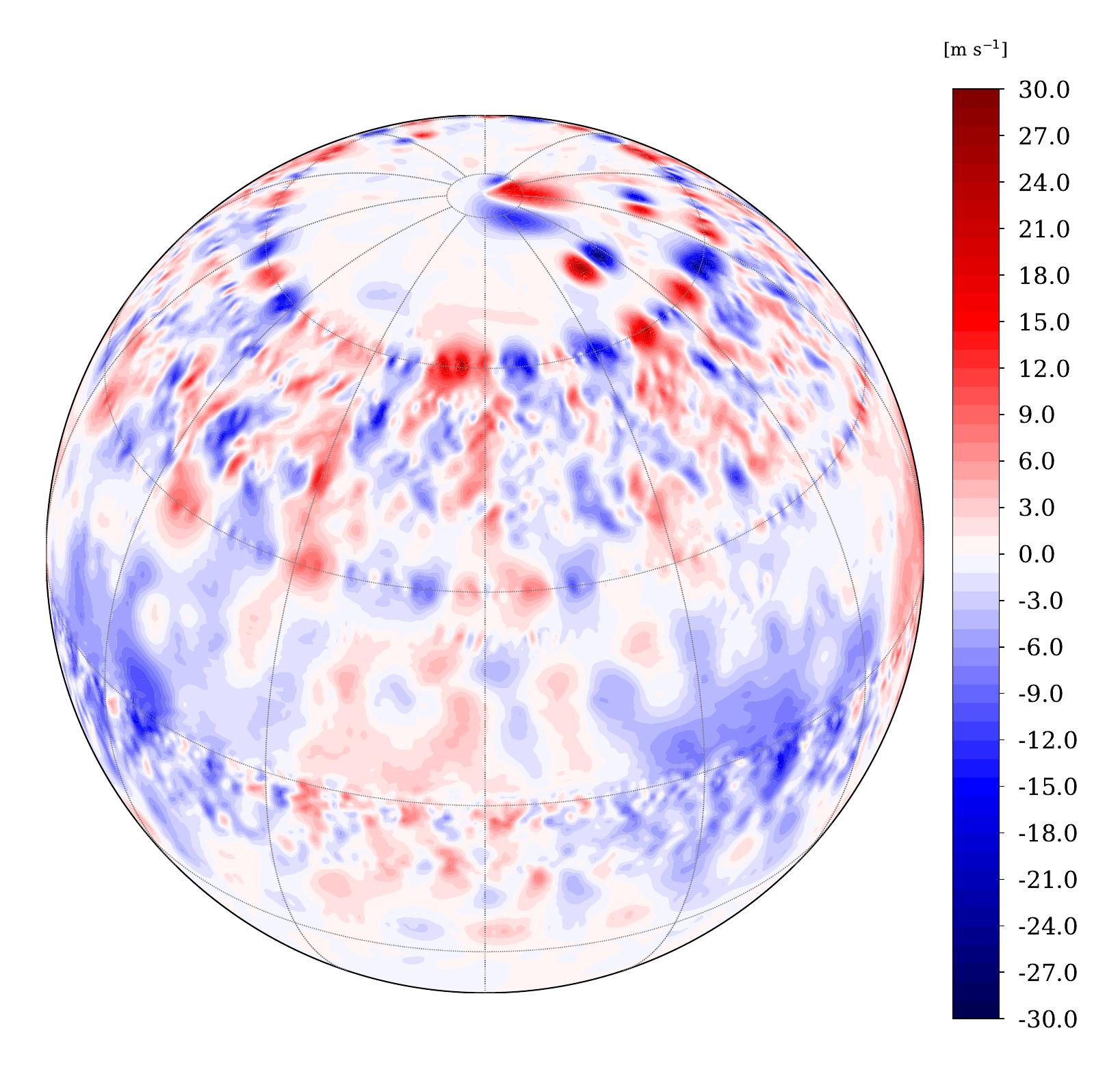}
\includegraphics[width=0.6\textwidth]{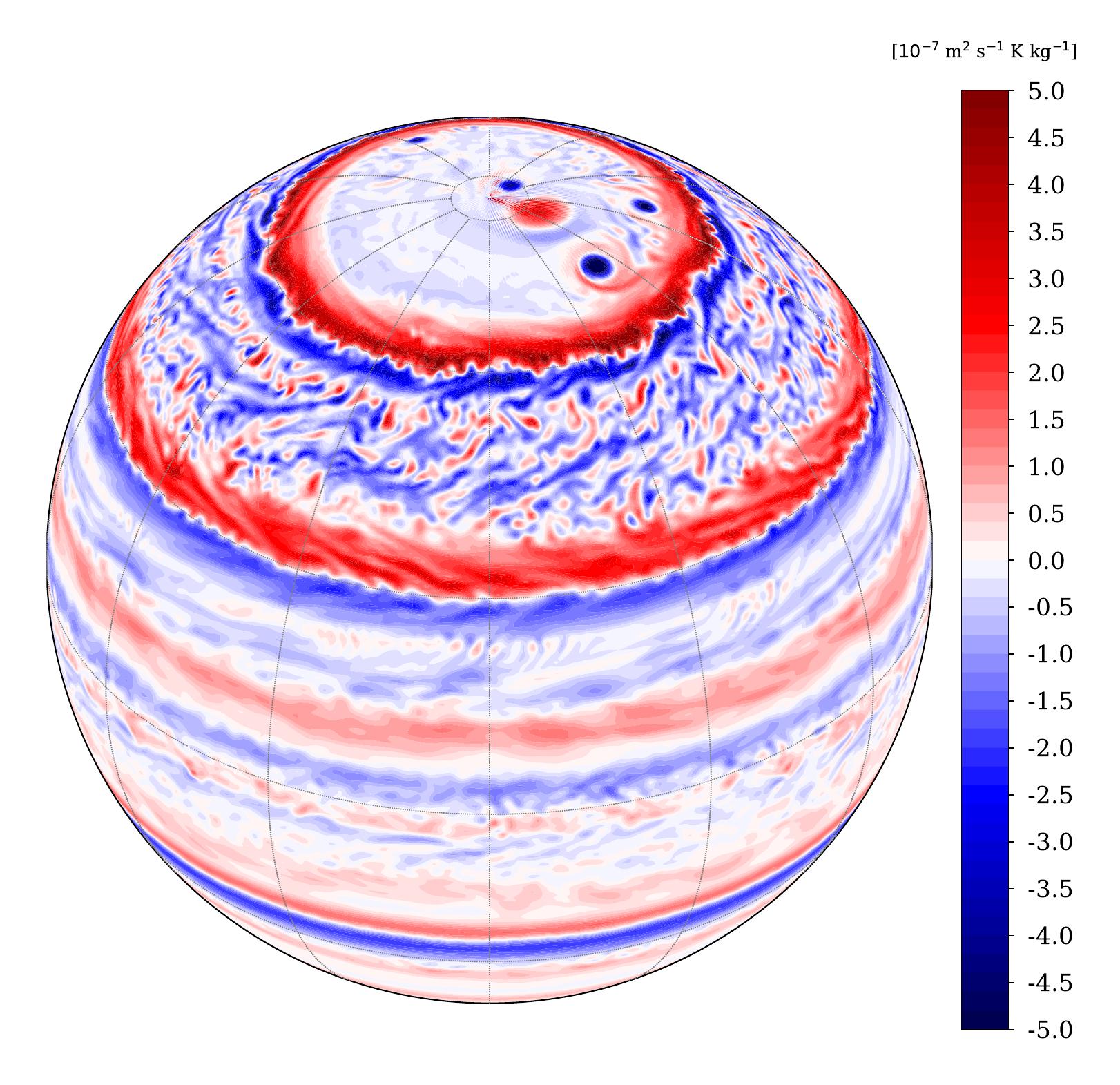}
\caption{\label{fig:instpv}
\emph{Instantaneous view on 
(top left panel) the zonal wind~$u$ and
(top right panel) the meridional wind~$v$
on the fifth 
sigma level of the model 
(pressure level~$\sim 1.5$~bar),
as in Figure~\ref{fig:sphereu}
and
(bottom panel)
relative Ertel potential vorticity (PV)
on tropospheric isentrope~$\theta=205$~K
after 171 thousands simulated Saturn days
(beginning of the seventh simulated year).
Ertel PV computations are described
in equation~\ref{ipv} and section~\ref{sec:midlatjets}.}}
\end{center}
\end{figure}

Our reference simulation with the Saturn DYNAMICO GCM
exhibits a variety of extratropical eddies, as is
evidenced in Figure~\ref{fig:instpv}.

The simulated tropospheric fields 
in Figure~\ref{fig:instpv} indicate that
the mid-latitude eastward jets 
at latitude~$30^{\circ}$N and~$60^{\circ}$N
are prone to meandering 
caused by high-wavenumber waves.
Those waves are found at the center
of the eastward jets,
featuring a strong inversion
of the meridional gradient
of potential vorticity
(Rayleigh-Kuo 
necessary condition for barotropic instability,
see section~\ref{sec:baroinstab}).
\newcommand{\mexico}{A spectral analysis on a 2000-day sample
of the temperature and wind fields,
performed similarly to the analysis 
in section~\ref{sec:eqwave}
(except for a Doppler-shift correction
considering the ambient eastward zonal 
jet $\overline{u} = 60$~m~s$^{-1}$)
indicates that
the~30$^{\circ}$N perturbations correspond 
to a westward-propagating wavenumber-19 Rossby wave 
with a period of $435$~days 
(frequency~$0.8^{\circ}$ longitude per day,
i.e. about~23~m~s$^{-1}$). } \mexico
The characteristics of this wave 
(wavenumber, phase speed, and
occurrence at the center of 
a mid-latitude eastward jet)
are very similar to the 
idealized modeling results 
obtained by \citet{Saya:10}
to explain the ``Ribbon wave'',
a Rossby wave propagating
in the extratropical latitudes
of Saturn as a result of
barotropic and baroclinic instability
\citep{Godf:86,Sanc:02,Gunn:18}.
\newcommand{\lapaz}{Furthermore, 
the meandering phase speed 
and wavenumber reproduced by our Saturn DYNAMICO GCM
in the 30$^{\circ}$N eastward jet
are compliant with the slow 
ribbon waves identified by 
\citet{Gunn:18}
with Cassini imaging 
(their Figure 3d).} \lapaz

It is worthy of notice that our Saturn 
DYNAMICO GCM simulations
exhibit 
a chain of about ten cyclonic
vortices with an horizontal extent
of a couple thousand kilometers
(the successive red spots
of positive vorticity in 
Figure~\ref{fig:instpv} bottom)
in the equatorward (southern) edge 
of the~$30^{\circ}$N eastward jet.
This signature shares
the characteristics of the
``String of Pearls''
observed by Cassini 
through infrared mapping \citep{Saya:14}.
Nevertheless, the simulated vortices
are more short-lived (typically a ten-day duration)
than the observed cyclonic vortices,
so the analogy between the modeled structures
and the ``String of Pearls'' is not complete.

The most distinctive and large-scale
vortices are found in polar regions
in our Saturn DYNAMICO GCM simulations.
Figure~\ref{fig:instpv} demonstrates
a clear and sharp transition
between 
the mid-latitudes where the vorticity field is dominated by eddies,
and 
the polar regions where the vorticity field is dominated by large-scale vortices.
This is somewhat compliant with 
the picture drawn by observations,
where no vortex activity 
was observed in mid-latitudes
before the appearance of the
2010 giant storm \citep{Tram:16}.
Both anticyclones and cyclones
are produced in our model:
the excerpt shown in Figure~\ref{fig:instpv}
comprises four anticyclones
and one larger cyclone.
Those simulated large-scale vortices
exhibit a strong temporal variability,
with merging phenomenon
combined with beta-drift effect
\citep[poleward for anti-cyclones, see][]{Saya:13}, 
which causes their typical duration to be
no more than several hundreds Saturn days.
The picture drawn by Figure~\ref{fig:instpv}
is typical of most of our Saturn DYNAMICO GCM simulation:
anticyclones appear favored against cyclones,
which is in agreement with 
the putative longer stability of 
anticyclones
compared to cyclones,
but at odds with the statistics 
derived from Cassini imagery
over seven Earth years by \cite{Tram:16}.
A more in-depth analysis of the
large-scale vortices is out of the scope of this paper;
furthermore, accounting for
moist convection and the
associated release of latent heat
appears to be a crucial addition to
carry out this analysis \citep{ONei:15}.

The last class of eddies are the remainder of
non-axisymmetric disturbances that are 
neither waves nor vortices.
Such ``non-organized'' eddies can be seen
in Figure~\ref{fig:instpv} between
latitudes~$30^{\circ}$N and~$60^{\circ}$N.
Their activity can strongly vary with time:
Figure~\ref{fig:instpv} shows a case with
a ``burst'' of eddy activity at those latitudes,
while the eddy activity can be close to none
at other times in the simulation
(Figure~\ref{fig:sphereu}).
\newcommand{\toronto}{The presence of intermittent bursts
of eddy activity is reminiscent of the results
of \citet{Pane:93} and is further discussed 
in section~\ref{sec:phenomevolution}.} \toronto

\section{Evolution of the tropospheric jet structure \label{sec:evolution}}

\subsection{Jets and eddies in the 15-year simulation \label{sec:phenomevolution}}

\begin{figure}[p]
\begin{center}
\includegraphics[width=0.7\textwidth]{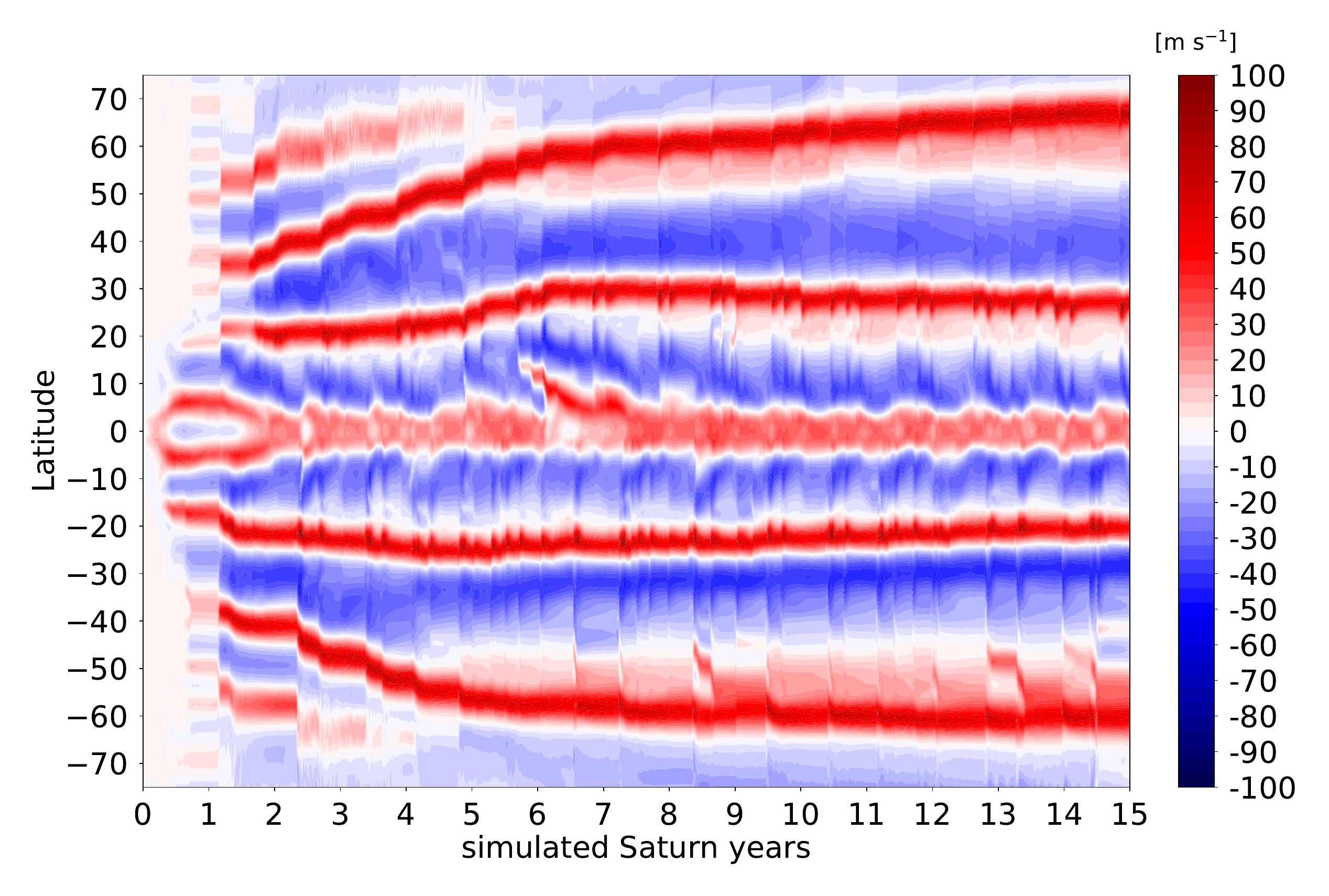}
\includegraphics[width=0.7\textwidth]{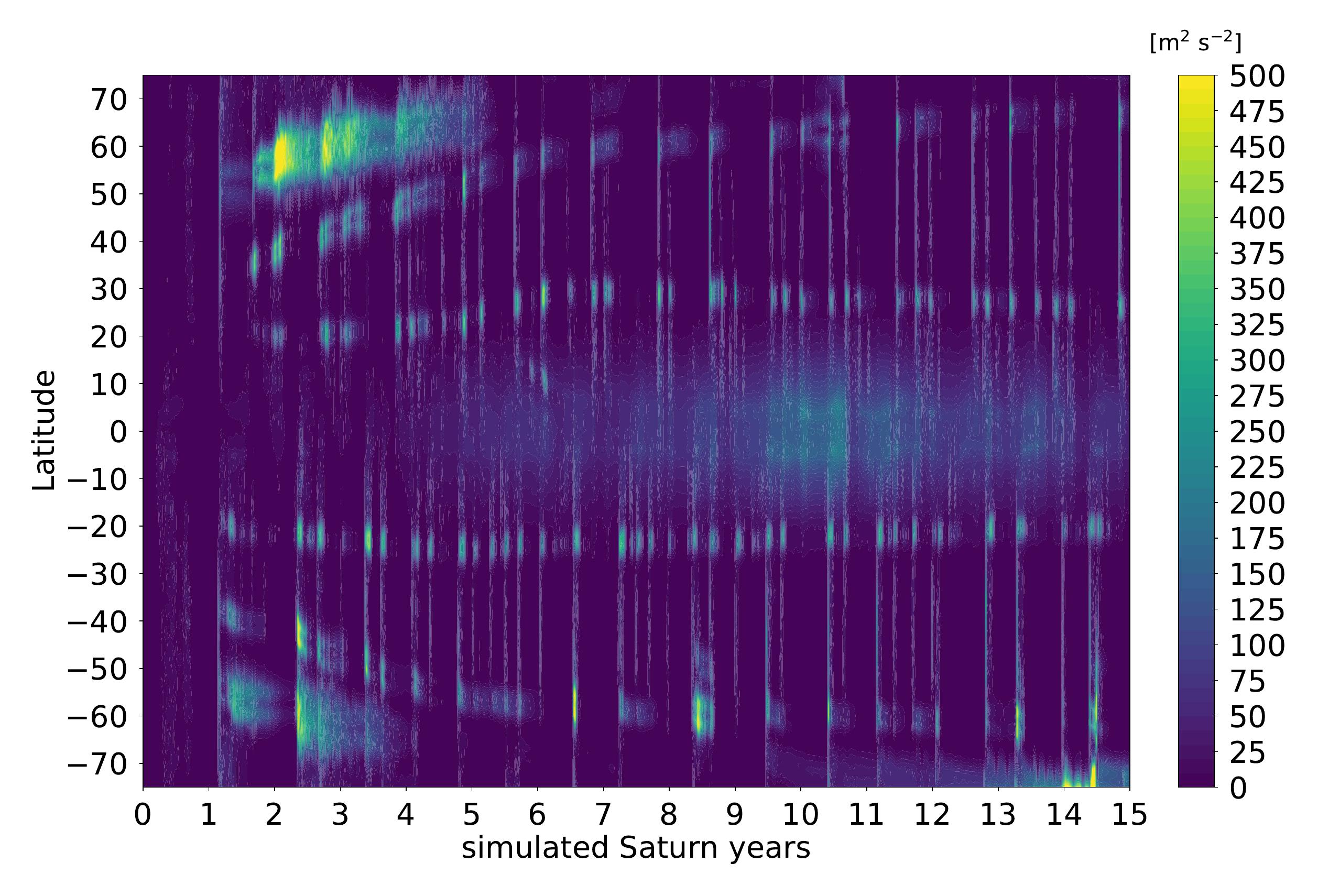}
\caption{\label{fig:jetevolution}
\emph{Evolution of the zonal-mean zonal wind~$\overline{u}$ (top plot)
and the zonal mean of the
Eddy Kinetic Energy (EKE)~$1/2 (u^{\prime 2} + v^{\prime 2})$ (bottom plot)
in Saturn troposphere (800 mbar)
within the whole 15-year duration of 
our Saturn DYNAMICO GCM simulation.}}
\end{center}
\end{figure}

The evolution of tropospheric jets with time in
the 15-year duration of our reference $1/2^{\circ}$ 
Saturn DYNAMICO simulation is summarized in
Figure~\ref{fig:jetevolution}.
It takes about about 6-7 simulated Saturn years for the jet system
to reach what most closely resembles a
steady-state equilibrium; 
a similar conclusion was drawn from
the analysis of the temporal evolution of AAM
(Figure~\ref{fig:aam} in section~\ref{sec:aam}).
The zonal mean of the Eddy Kinetic Energy (EKE)
\begin{equation}
\overline{e} = \frac{1}{2} \, \left( \overline{{u^{\prime}}^2} + \overline{{v^{\prime}}^2} \right)
\end{equation}
is also shown in Figure~\ref{fig:jetevolution}
to diagnose eddy activity.

\begin{figure}[p]
\begin{center}
\includegraphics[width=0.7\textwidth]{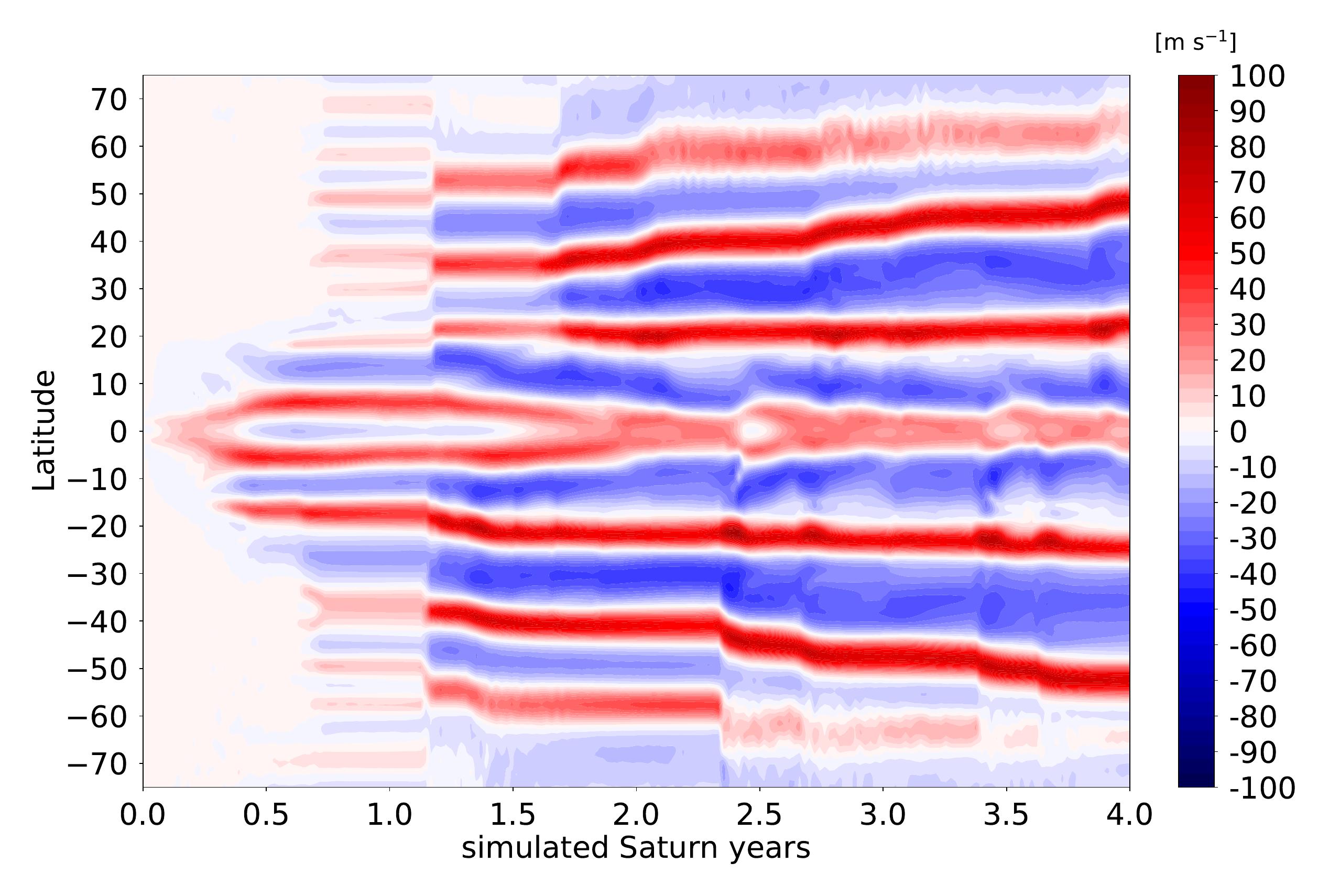}
\includegraphics[width=0.7\textwidth]{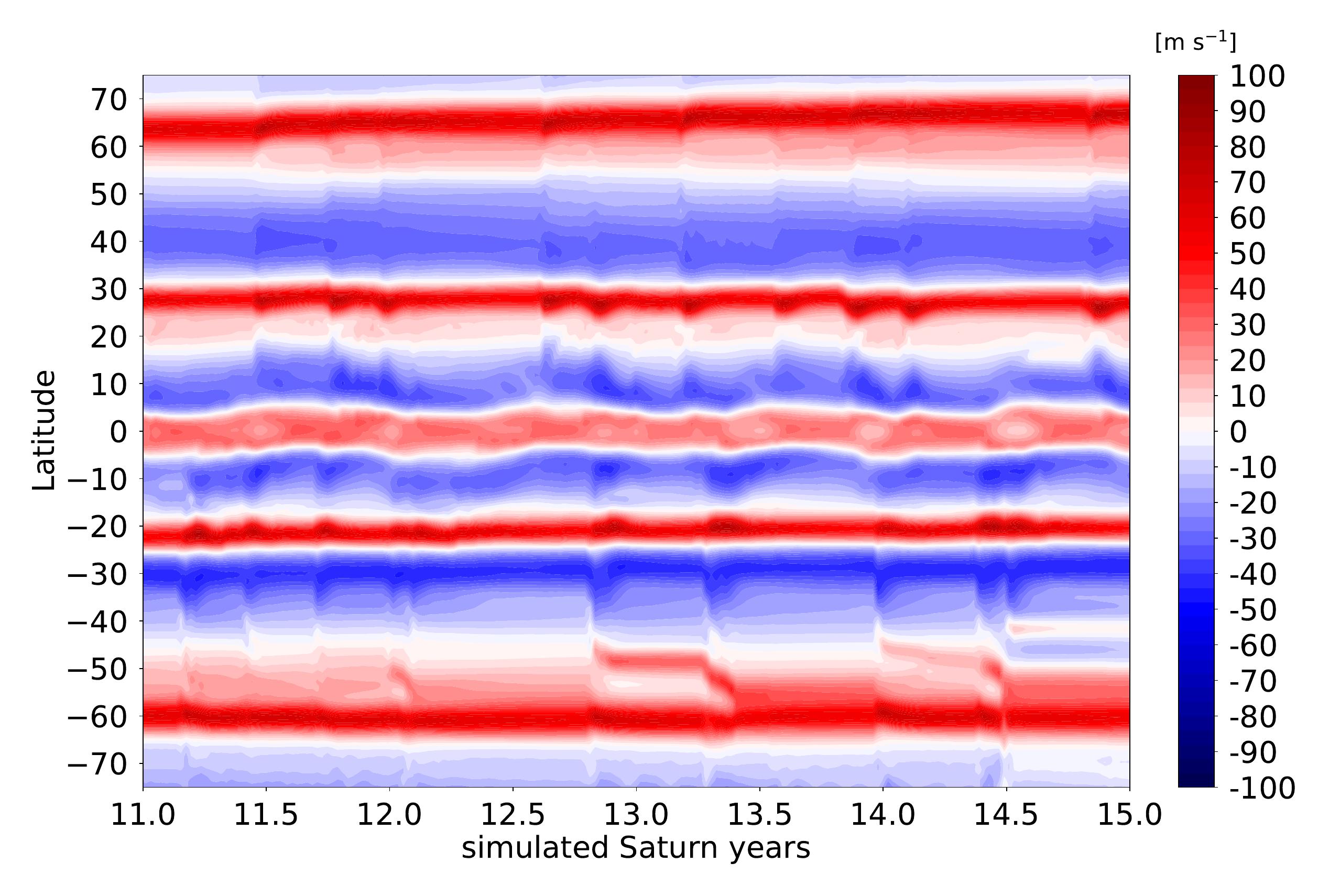}
\caption{\label{fig:jetevolutiondetail}
\emph{Same as Figure~\ref{fig:jetevolution}
with an emphasis on 
(top) the first four and 
(bottom) the last four 
simulated Saturn years
of the full 15-year duration
of our Saturn DYNAMICO GCM simulation.}}
\end{center}
\end{figure}

\begin{figure}[ht]
\begin{center}
\includegraphics[width=0.45\textwidth]{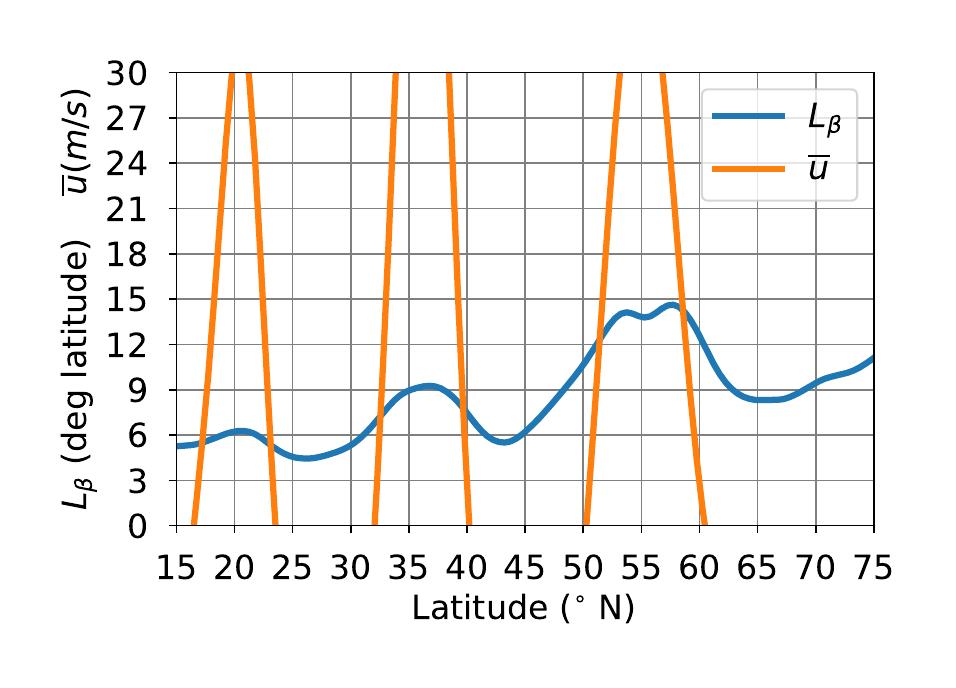}
\includegraphics[width=0.45\textwidth]{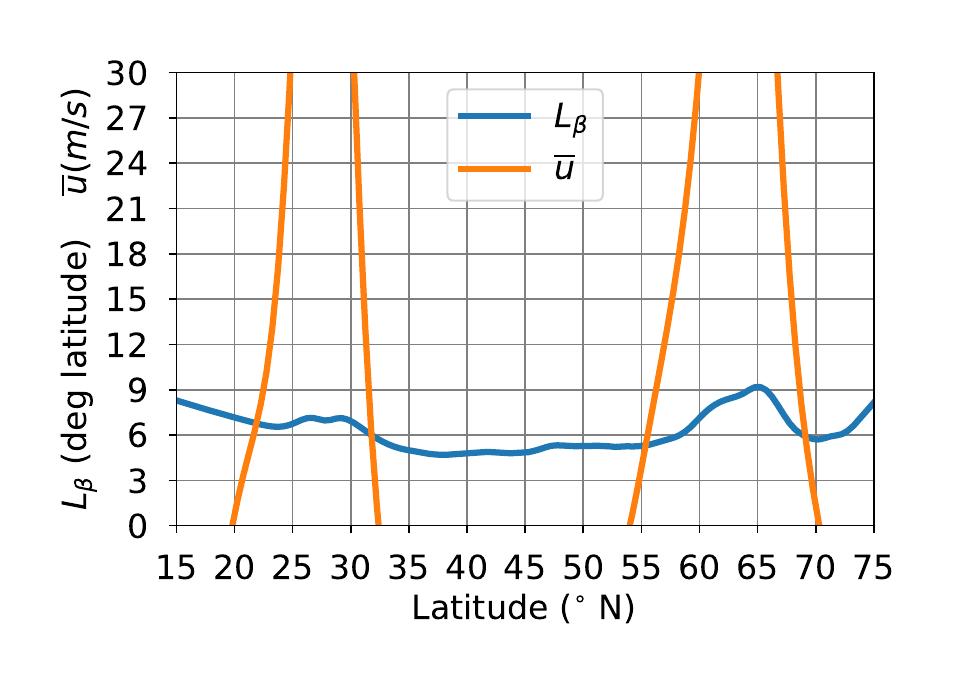}
\caption{\label{fig:rhines}
\emph{Meridional profiles of the Rhines scale~$L_{\beta}$
defined as in equation~\ref{eq:rhines} (orange line)
and eastward jet speed~$\overline{u}$ (blue line).
Those quantities are vertically-averaged 
in the pressure range~$[10^5,10^3]$~Pa.
The left plot shows a temporal average 
within half of the second simulated year
(during dynamical spin-up and strong jet migration)
and the right plot within half of the twelfth
simulated year (quasi-steady state).
The correlation between the Rhines scale
and the jet width / spacing is more obvious
in the former than in the latter.
The left plot is computed in conditions of 
strong poleward migration of the eastward jets,
a temporal evolution resembling the simulations
in \citet{Chem:15jas}.}}
\end{center}
\end{figure}

The first years of our Saturn DYNAMICO GCM simulations
follow the evolution of zonal jets typically obtained with
nonlinear analytical models prone to barotropic
and baroclinic instability. 
The evolution of our Saturn DYNAMICO GCM displayed 
in the top panel of Figure~\ref{fig:jetevolutiondetail}
is similar to the evolution described, 
e.g., in Figure 10 of \cite{Kasp:07}.
The first simulated half-year exhibits no particular
zonal organization (except at the equator).
In the second half of the first simulated year,
the growth of the fastest unstable mode
leads to the emergence of numerous weak zonal jets,
which subsequently reorganize, as additional growing eddy modes are present,
to lead to a system with less, and wider, jets.
The abrupt transition around 1.15 simulated
years in Figure~\ref{fig:jetevolution} is associated 
with significant eddy activity.
The merging of numerous weak jets
into a final jet structure with lesser and stronger jets
is typical of the inverse energy cascade
by geostrophic turbulence which shapes 
the jet structure \citep{Caba:17}.
\newcommand{\sofia}{Another typical feature of the
inverse energy cascade 
shown in Figure~\ref{fig:rhines}
is the overall correlation
between the Rhines scale \citep{Rhin:75}
\begin{equation} \label{eq:rhines}
L_{\beta} = 2 \, \pi \, \sqrt{ \frac{U}{\beta} } 
\qquad \textrm{with} \qquad
U = \sqrt{2 \, \overline{e}}
\end{equation}
\noindent and the eastward 
jets' width/spacing \citep{Chem:15jas}, with a tendency for 
broader jets and increased spacing between jets
towards higher latitudes \citep{Kids:10}. } \sofia
A full exploration of the dynamical regimes
(e.g., zonostrophy) and the inverse energy cascade 
in our Saturn DYNAMICO GCM requires
detailed spectral analysis 
of the flow energetics \citep{Suko:02,Galp:14,Youn:17}
that will be developed in a follow-up paper
(Cabanes S. et al., Part III submitted).

The most prominent feature of Figure~\ref{fig:jetevolution}
between the simulated years~1 and~6
is the poleward migration of mid-latitude jets.
This echoes the idealized simulations  
detailed in \cite{Chem:15} 
\citep[see also jovian simulations by][]{Will:03}.
The migration is gradual, but the jet migration
is much stronger in short-lived episodes characterized
by a burst of eddy activity,
which implies momentum transfer to the jet
altering its meridional structure.
A detailed analysis of a typical poleward migration
episode in Figure~\ref{fig:jetevolution}
is proposed in section~\ref{sec:local}.
The tropical jets undergo a much weaker
migration than the mid-latitude jets.
\newcommand{\lecap}{Most of the migration events are poleward,
but there is at least one clear equatorward migration event
of a jet appearing at latitude~$15^{\circ}$N
in the beginning of year 7. This equatorward
migration contributes to accelerate the 
equatorial jet.} \lecap
\newcommand{\tripoli}{This was also noticed by \citet{Youn:19partone}
in their Jupiter GCM, although their simulations
exhibit a global tendency for equatorward jet migration,
contrary to the global tendency for poleward jet migration
found in our GCM simulations and in \citet{Chem:15}.} \tripoli

The poleward migration of the mid-latitude jets 
continues until the migration rate
slows down and the zonal jets reach
their final latitude of occurrence
in our Saturn DYNAMICO GCM simulation.
\newcommand{\jerusalem}{Starting from the seventh
simulated year to the fifteenth simulated year
(Figures~\ref{fig:jetevolution} top 
and~\ref{fig:jetevolutiondetail} bottom),
we notice a continuing, very slow 
poleward migration of the mid-latitude eastward jets and 
equatorward migration of the tropical jets.
The 15-year duration of our Saturn DYNAMICO GCM simulation
allows us to reach robust conclusions about
the overall steady-state jets' structure and intensity;
the long duration of our simulations also allows us
to conclude that this jet structure is still
impacted by a slowly-evolving transient state
over timescales of tens of Saturn years. } \jerusalem
It is important to note here that 
we are not interpreting our Saturn GCM simulations
as indicative of a current migration
of Saturn zonal jets, 
which is not supported by observations.
\newcommand{\lisbon}{Our GCM simulations start with a zero-wind state
which is not encountered 
in the actual Saturn's atmosphere,
and may have never been encountered in
the past history of Saturn's atmosphere.
Thus, we can only speculate 
that jet migration could have been an important
process in the past evolution of Saturn's atmosphere
and might explain the present 
latitudes of Saturn's zonal jets.} \lisbon

The bursts of eddy activities are also associated
with acceleration of the zonal jets,
would it be a case of a migration episode or not
(once the jets have migrated, the impact
of the eddies is actually solely jet acceleration and
no longer migration).
This suggests that eddy forcing plays a
great role in shaping the zonal jets
in Saturn's atmosphere, as argued by
existing modeling studies \citep{Show:07,Lian:10,Liu:10jets}
and Cassini observations \citep{Delg:07,Delg:12}.
We discuss this matter in 
section~\ref{sec:global} for a global analysis
and
section~\ref{sec:local} for a local analysis.
Despite our Saturn DYNAMICO GCM resolving 
the seasonal evolution
of Saturn's troposphere and stratosphere, 
there is no clear seasonal trend associated
with the bursts of mid-latitude eddy activity
(although the typical timescale
between the bursts is close to one year).
This is in line with theoretical studies
supporting abrupt stochastic transitions
in the zonal jet structure prone to
barotropic and baroclinic instabilities
\citep{Bouc:09,Bouc:13}.
The typical timescale between bursts
is not so much set by the seasonal cycle,
but by the typical life cycle of
instabilities \citep{Pane:93}.

\subsection{Kinetic energy conversion rate \label{sec:global}}

To investigate the mechanism by which zonal banded jets
arise in our Saturn DYNAMICO GCM simulation,
we first consider a global diagnostic,
the conversion rate~$\mathcal{C}$ 
of eddy-to-mean kinetic energy. 
Cloud tracking with Cassini Imaging Science Subsystem (ISS) images
has been employed 
to address the driving of Saturn's zonal jets 
by eddy momentum fluxes \citep{Delg:07,Delg:12}.
The conversion rate~$\mathcal{C}$ in m$^2$~s$^{-3}$ (or W~kg$^{-1}$),
estimating the conversion per unit mass of eddy kinetic energy to zonal-mean kinetic energy,
can be obtained by multiplying 
eddy momentum transport~$\overline{u^\prime v^\prime}$ 
by the meridional curvature of the zonal flow~$\Dp{\overline{u}}{y}$.
\begin{equation}\label{eq:convrate}
\mathcal{C} = \overline{u^\prime v^\prime} \, \Dp{\overline{u}}{y}
\end{equation}
Wind observations by cloud tracking exhibit a globally positive conversion rate~$\mathcal{C}>0$,
both at the middle troposphere ammonia cloud at~$1$~bar and the upper troposphere haze at~$100$~mbar,
which suggests that Saturn's zonal jets are eddy-driven \citep{Delg:12}.
The fact that~$\mathcal{C}$ is positive means that 
the eddy flux is, on average, equatorward in cyclonic shear regions
and poleward in anticyclonic shear regions \citep{Delg:12},
hence eastward jets are accelerated by the convergence of eddy flux
while westward jets are decelerated by the divergence of eddy flux
(see also PV discussions in section~\ref{sec:midlatjets}).
The values of kinetic energy conversion rate observed by Cassini
by \citet{Delg:12} (their Figure~11) are large:
$\mathcal{C} \sim 1 \times 10^{-5}$~W~kg$^{-1}$ at 100~mbar,
and four times larger in the troposphere at 1 bar.
Those estimates of the energy conversion rates 
support the idea
that eddy momentum transfers are able
to maintain jets against dissipation.
Similar conclusions were reached for 
Jupiter's weather layer by \citet{Saly:06}.

\begin{figure}[ht]
\begin{center}
\includegraphics[width=0.75\textwidth]{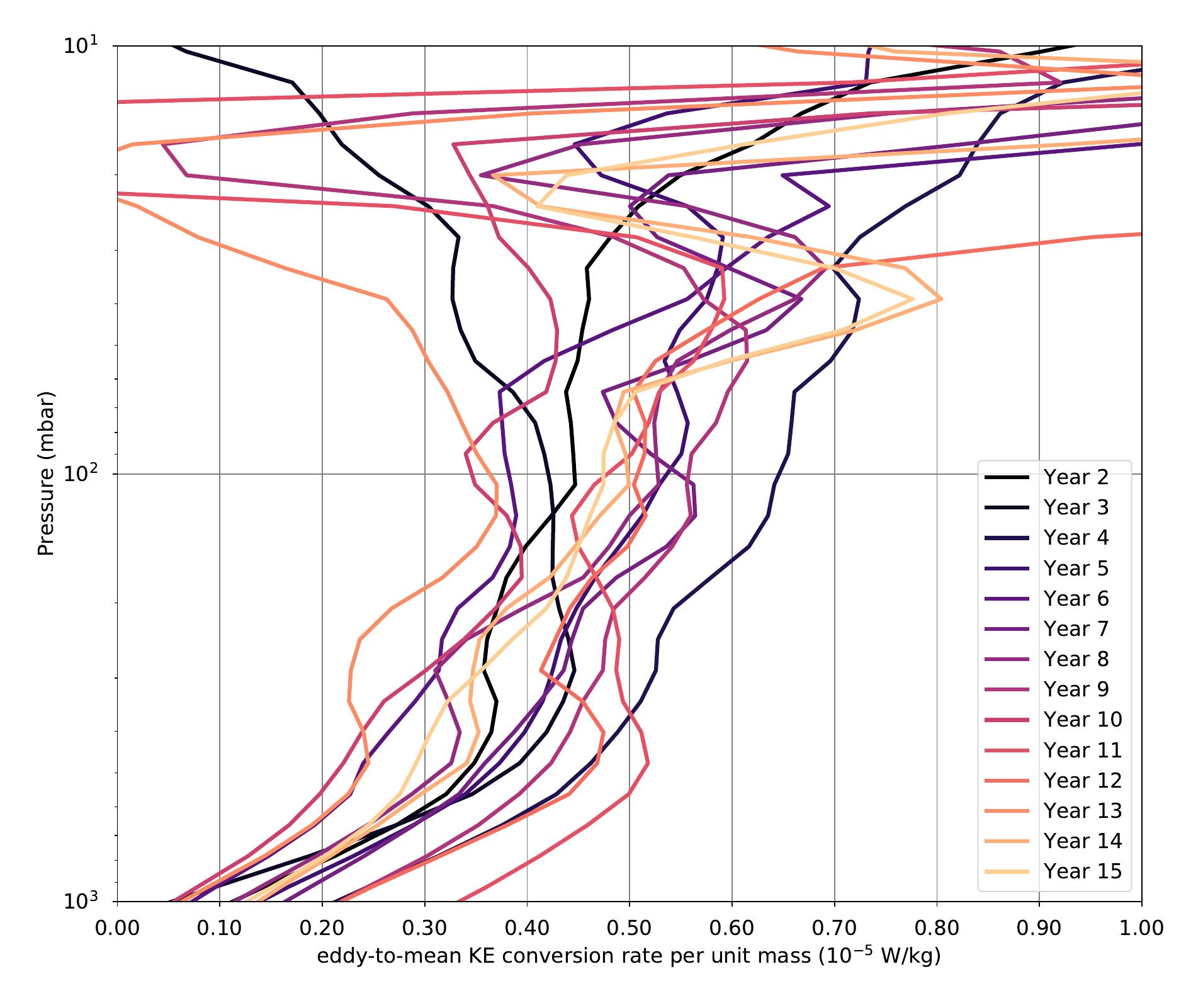}
\caption{\label{fig:globalforcingcomp}
\emph{Rate per unit mass~$\mathcal{C}$ in m$^2$~s$^{-3}$ (or W~kg$^{-1}$)
estimating the conversion of eddy kinetic energy to zonal-mean kinetic energy.
Vertical profiles of~$\mathcal{C}$ are shown, 
anually-averaged 
(one line per simulated year, from year~2 to~15)
and globally-averaged 
($60^{\circ}$S to~$60^{\circ}$N latitudes)
This spatial and seasonal averaging is chosen to allow
a direct comparison with Cassini estimates of~$\mathcal{C}$ in \citet{Delg:12}.}}
\end{center}
\end{figure}

The positive conversion rates~$\mathcal{C}$ 
simulated by our Saturn DYNAMICO GCM,
shown in Figure~\ref{fig:globalforcingcomp},
indicate that our model supports the conclusion 
of \citet{Delg:12} that Saturn's zonal banded jets
are, for a significant part, driven and maintained by eddies in the weather layer
\citep[see also][]{Read:09pv}.
This also confirms the diagnostics 
obtained from previous GCM studies \citep{Lian:08,Liu:10jets}.
In the upper troposphere haze layer at~$100$~hPa, 
our model predicts~$\mathcal{C} \sim 0.5 \times 10^{-5}$~m$^2$~s$^{-3}$, 
which matches to the order-of-magnitude
the quantitative estimates obtained from Cassini by \citet{Delg:12},
implying a typical timescale of replenishing 
the jets of less than a Saturn year.
This shows that our Saturn DYNAMICO GCM
resolves a satisfactory conversion rate from eddies 
to zonal jets in the tropopause, providing support 
for a ``downward control'' of jets at deeper levels \citep{Hayn:91}
by eddy forcing in the radiatively-driven upper troposphere,
as proposed by \citet{Schn:09} and \citet{Liu:10jets}.
Nevertheless, our modeled values for~$\mathcal{C}$
are half those obtained by cloud tracking on board Cassini,
indicating room for improvement in predicting the eddy activity
and jet curvature resolved by our GCM in the upper troposphere,
suggesting the need for either more accurate radiative computations,
or an additional physical process causing eddies.

The conversion rate~$\mathcal{C}$ increases 
with altitude in our Saturn GCM simulation, 
whereas it decreases with altitude in the Cassini observations of \citet{Delg:12}.
In other words, if our simulations match the observations at 100 mbar,
the conversion rate at the cloud layer
is one order of magnitude
lower in the Saturn GCM simulations 
than it is in the observations.
\citet{Delg:12} already noticed this discrepancy 
by comparing their data to the GCM results of \citet{Liu:10jets}. 
We speculate that our simulated eddy forcing of jets 
being compliant with observations in the radiatively-driven tropopause, 
but not in the deeper troposphere,
indicates that a source of tropospheric eddy forcing  
\citep[e.g. latent heat release and 
convective motions associated 
with moist processes][]{Zuch:09clouds,Lian:10}
is missing in our Saturn GCM.
This is also consistent with our equatorial jet super-rotating 
too weakly for a possible lack 
of convectively-generated Rossby waves \citep{Schn:09}.

\subsection{Local analysis of eddy-induced jets \label{sec:local}}

The approach using kinetic 
energy conversion rate in section~\ref{sec:global}
is to be understood on a global sense;
here we present diagnostics 
for eddy-induced jets that bear
a more local sense.

\begin{figure}[p]
\begin{center}
\includegraphics[width=0.65\textwidth]{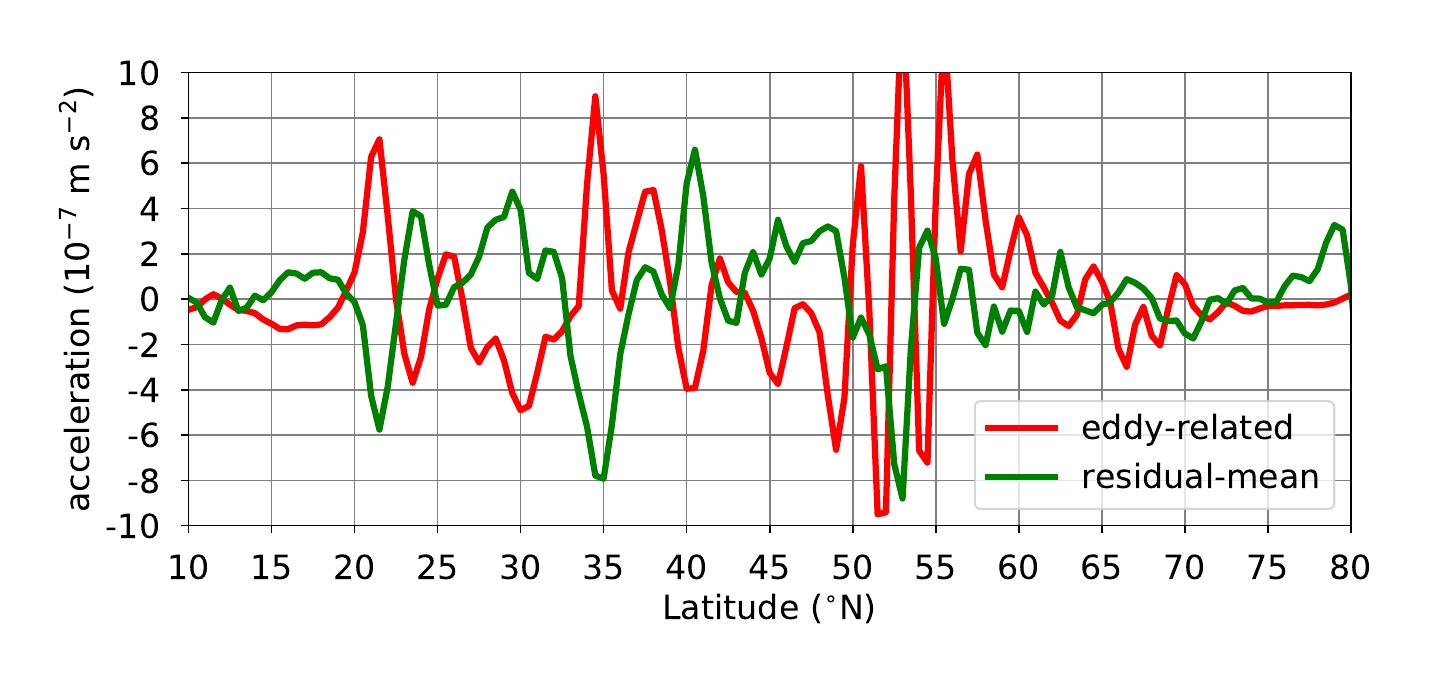}
\includegraphics[width=0.65\textwidth]{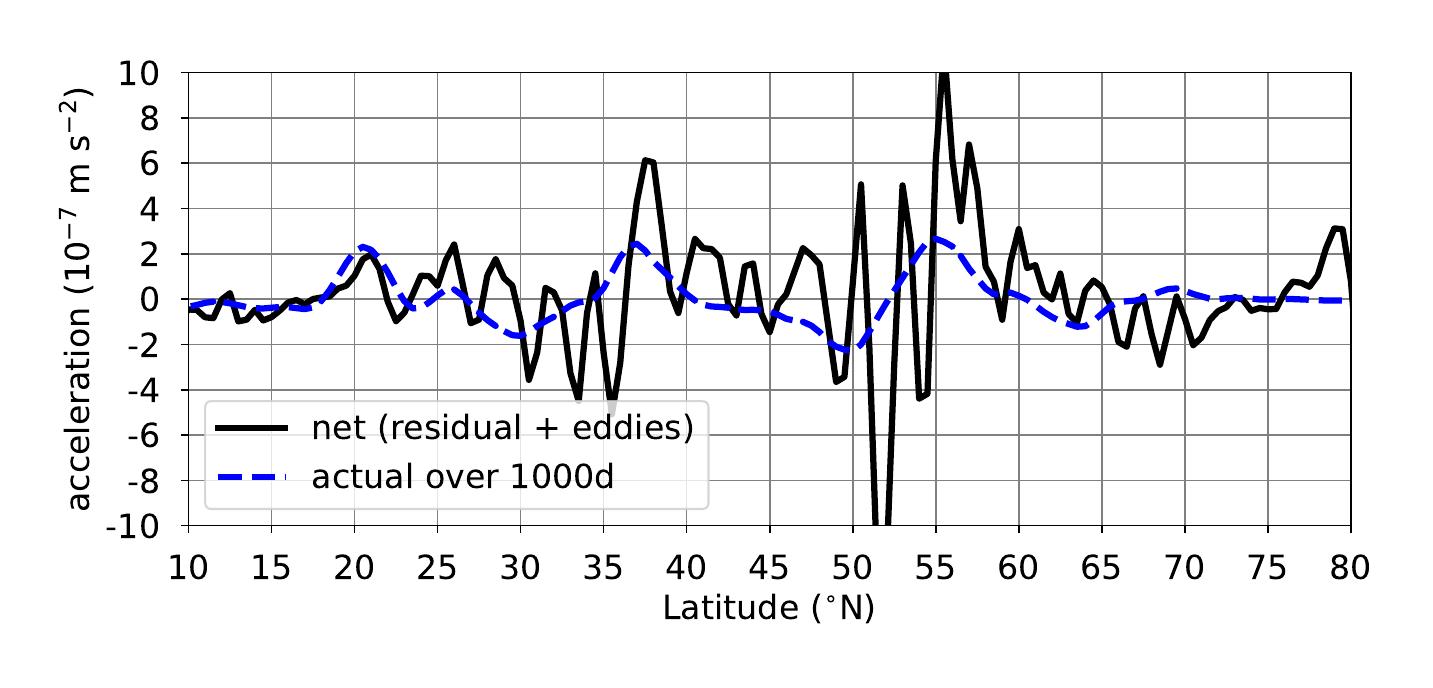}
\includegraphics[width=0.65\textwidth]{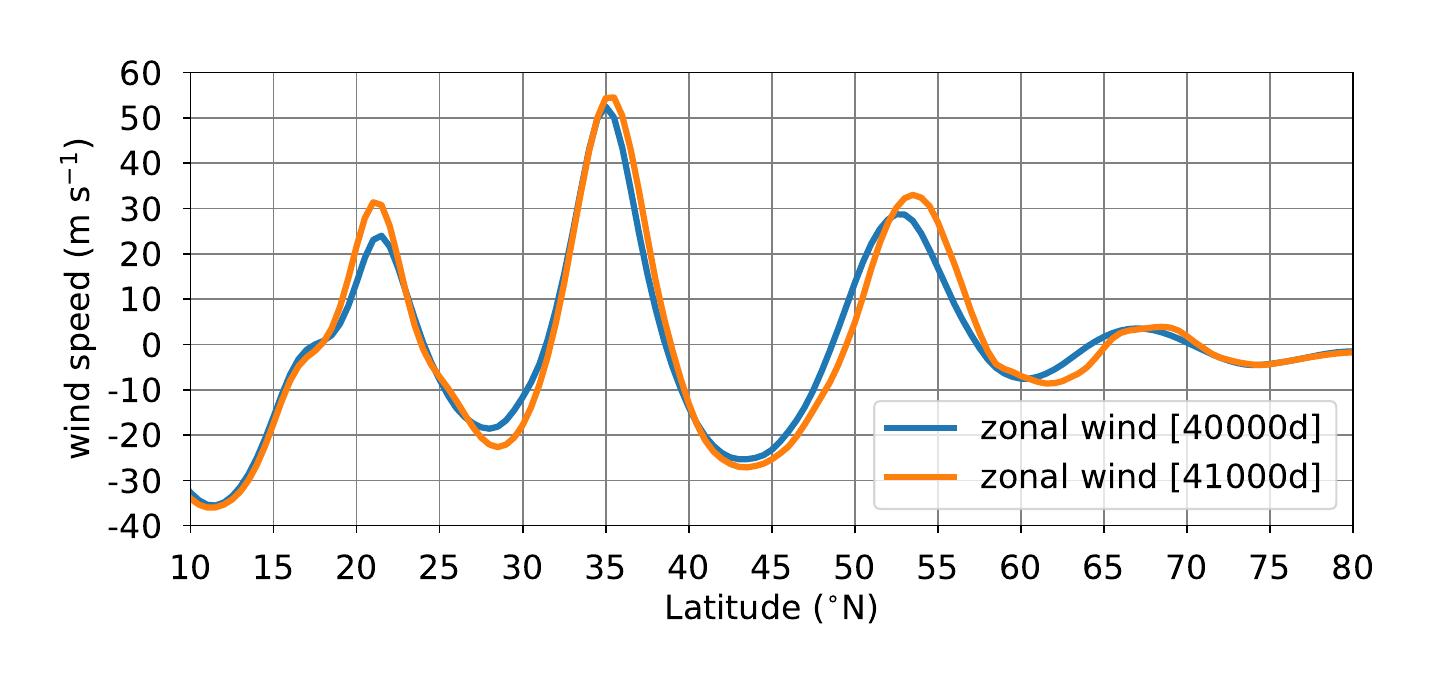}
\caption{\label{fig:dynbudg}
\emph{Meridional profiles of the zonal-mean acceleration terms
in the Eulerian-mean formalism described by equation~\ref{eq:zmzmeq}.
The analysis is carried out after
41 thousands simulated Saturn days
(around~$1.6$~simulated year),
corresponding to 
a typical burst of eddies in the northern 
hemisphere evidenced in Figure~\ref{fig:jetevolution}.
The plots are obtained after averaging over 1000 Saturn days.
The residual-mean term (equation~\ref{eq:accrmc}) is shown in the red line
and the eddy-related term (equation~\ref{eq:acceddh}) is shown in the green line.
The net resulting acceleration is shown in the black full line
and the actual acceleration obtained within the considered 1000 days
is shown in the dashed blue line.
The comparison of the latitudinal profiles of 
zonal-mean zonal winds at the beginning and in the
end of the 1000-day sequence are shown 
respectively in light blue and orange in the bottom plot.}}
\end{center}
\end{figure}

We can consider, as a typical 
and particularly illustrative
example,
the impact of the strong burst of eddy activity
taking place in the northern hemisphere
within 1000 Saturn days
between~$1.6$ and~$1.7$ simulated years
(Figure~\ref{fig:jetevolution}).
In order to study the 
contributions to the zonal acceleration,
the Eulerian-mean form of the zonal-mean zonal 
momentum equation of the atmospheric flow motion 
can be written as follows
\begin{equation}\label{eq:zmzmeq}
\Dp{\overline{u}}{t}
=
\left[ \Dp{\overline{u}}{t} \right]_{\textrm{R}}
+
\left[ \Dp{\overline{u}}{t} \right]_{\textrm{E}}
+ 
\overline{\mathcal{X}}
\end{equation}
\noindent where~$\overline{\mathcal{X}}$ is a mean nonconservative force such as e.g. diffusion, 
and the respectively ``residual-mean'' and ``eddy-related'' terms are written \citep[e.g. equation 2.5 in][]{Andr:83}
\begin{equation}\label{eq:accrmc} 
\left[ \Dp{\overline{u}}{t} \right]_{\textrm{R}}
=
- \left[ \frac{1}{a\cos\varphi} \Dp{\overline{u}\cos\varphi}{\varphi} - f \right] \overline{v}
- \Dp{\overline{u}}{p} \overline{\omega} 
\end{equation}
\begin{equation}\label{eq:acceddh}  
\left[ \Dp{\overline{u}}{t} \right]_{\textrm{E}}
=
- \frac{1}{a \cos^2\varphi} \Dp{\overline{u^\prime v^\prime} \cos^2\varphi}{\varphi} 
- \Dp{\overline{u^\prime \omega^\prime}}{p} 
\end{equation}
The contributions of each of those terms,
within the considered 1000 Saturn days prone
to significant eddy activity, are provided
in Figure~\ref{fig:dynbudg}.
As was also noticed by 
\citet[][their Figure 8]{Lian:08},
the two terms described in
equations~\ref{eq:accrmc}
and~\ref{eq:acceddh}
are generally anti-correlated,
which indicates that a significant
part of eddy-related acceleration
(which might 
reach~$2 \times 10^{-6}$~m~s$^{-2}$
on average
over 1000 Saturn days)
contributes to maintaining
an associated meridional circulation,
in addition to contributing
to the zonal jets.
Under the assumption of steady-state zonal jets
($\partial u / \partial t \simeq 0$),
the quasi-equilibrium 
between
eddy-related 
and meridional-circulation
terms
is used in studies 
exploring the circulations underlying
observed temperature and aerosol fields
in gas giants \citep[e.g.,][]{West:92}.

The three northern zonal jets
featured in Figure~\ref{fig:dynbudg}
illustrate the possible distinct outcome
of local eddy forcing:
the~$21^{\circ}$N eastward jet 
is slightly accelerated,
in a situation where
the residual-mean and eddy-related terms almost compensate;
the~$35^{\circ}$N eastward jet
does not accelerate at its core, 
but is migrating as a result of
an eddy-induced acceleration on its poleward flank;
the~$52^{\circ}$N eastward jet
undergoes strong eddy perturbations,
especially on its poleward flank,
not compensated by an evolution of
the residual-mean circulation,
which results in both
a poleward migration and
an acceleration of this eastward jet. 
Note that, concomitantly with
this overall acceleration of
the eastward jets, 
westward jets are decelerating.
This supports the interpretation proposed
in sections~\ref{sec:midlatjets} 
and~\ref{sec:global},
and in the literature \citep[e.g.,][]{Schn:09},
that there is a net transfer of
momentum from the westward jets
towards the eastward jets.

\begin{figure}[htbp]
\begin{center}
\includegraphics[width=0.48\textwidth]{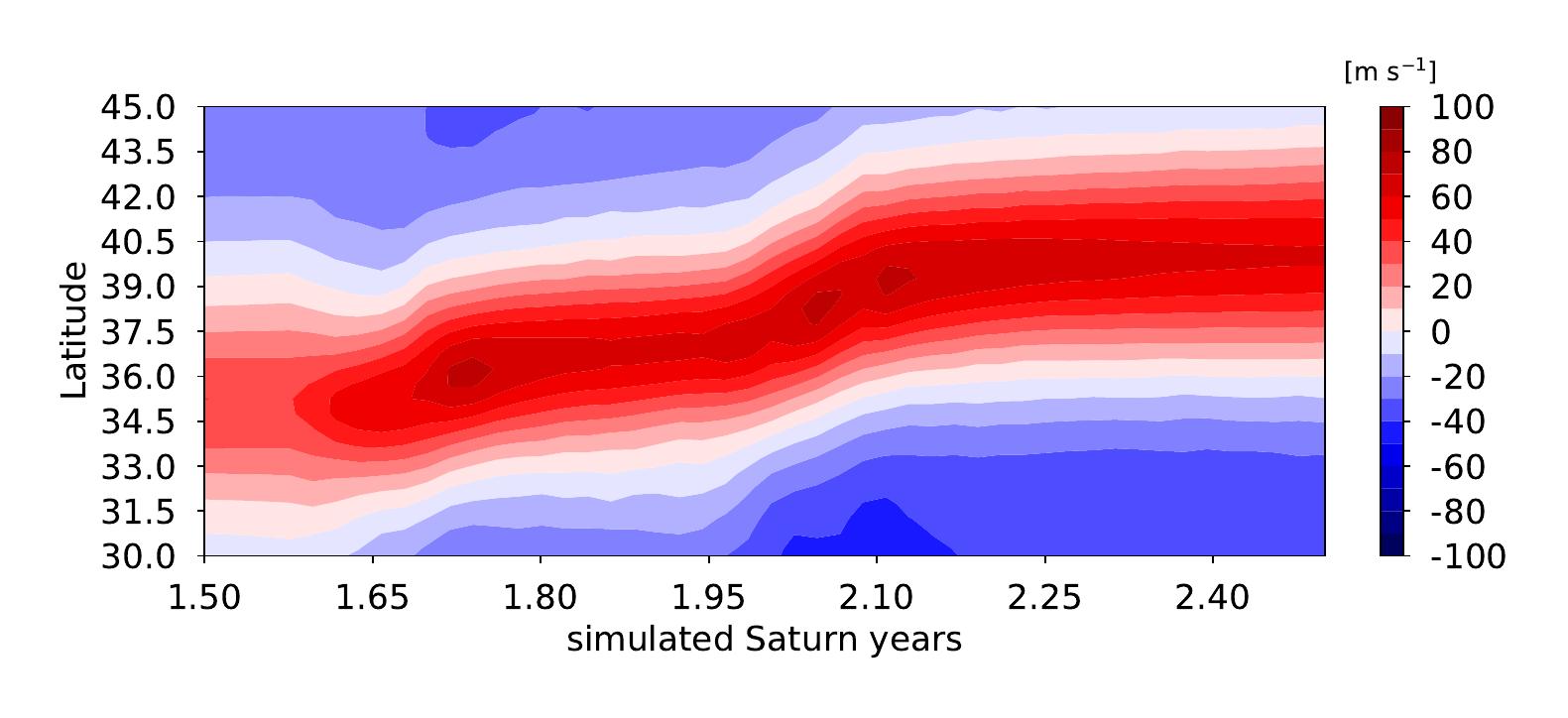}
\includegraphics[width=0.48\textwidth]{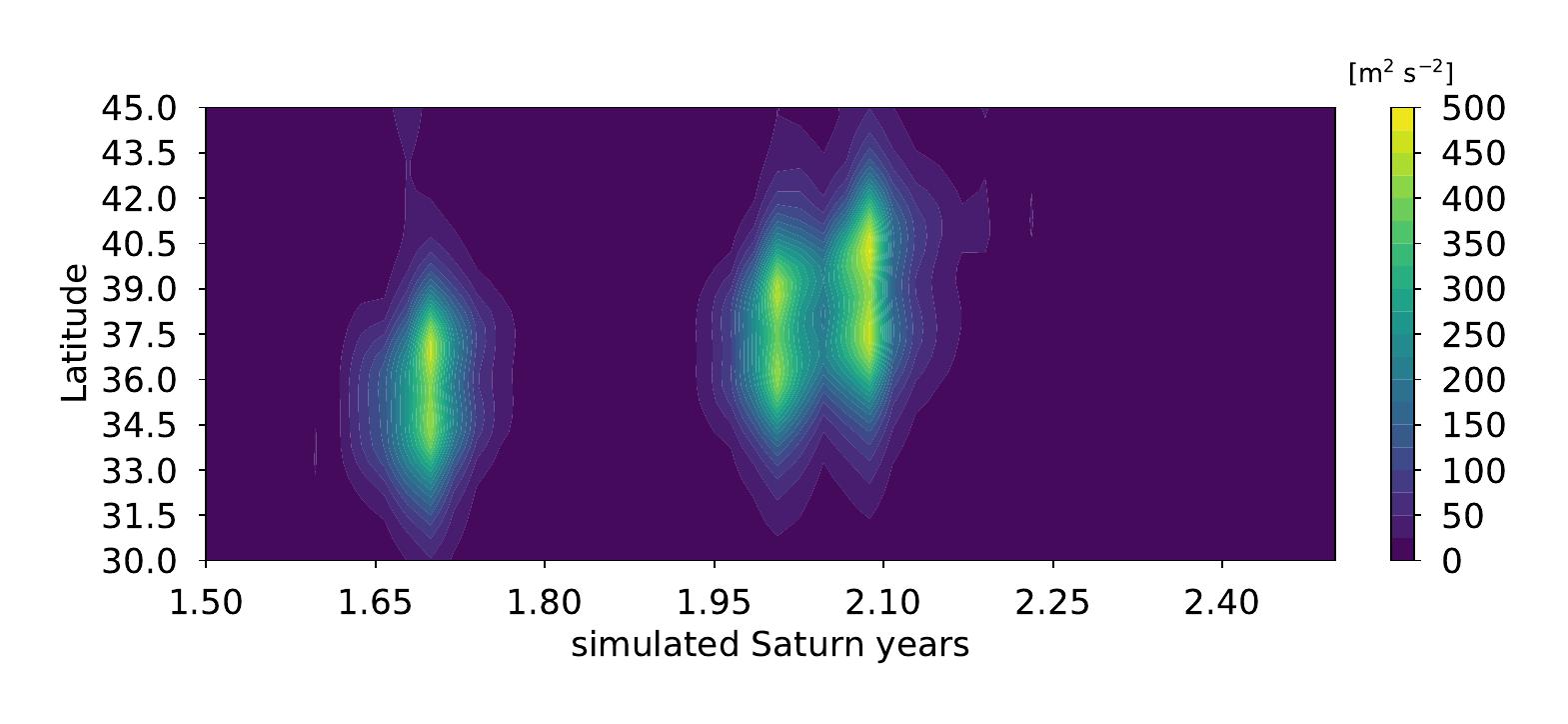}
\includegraphics[width=0.65\textwidth]{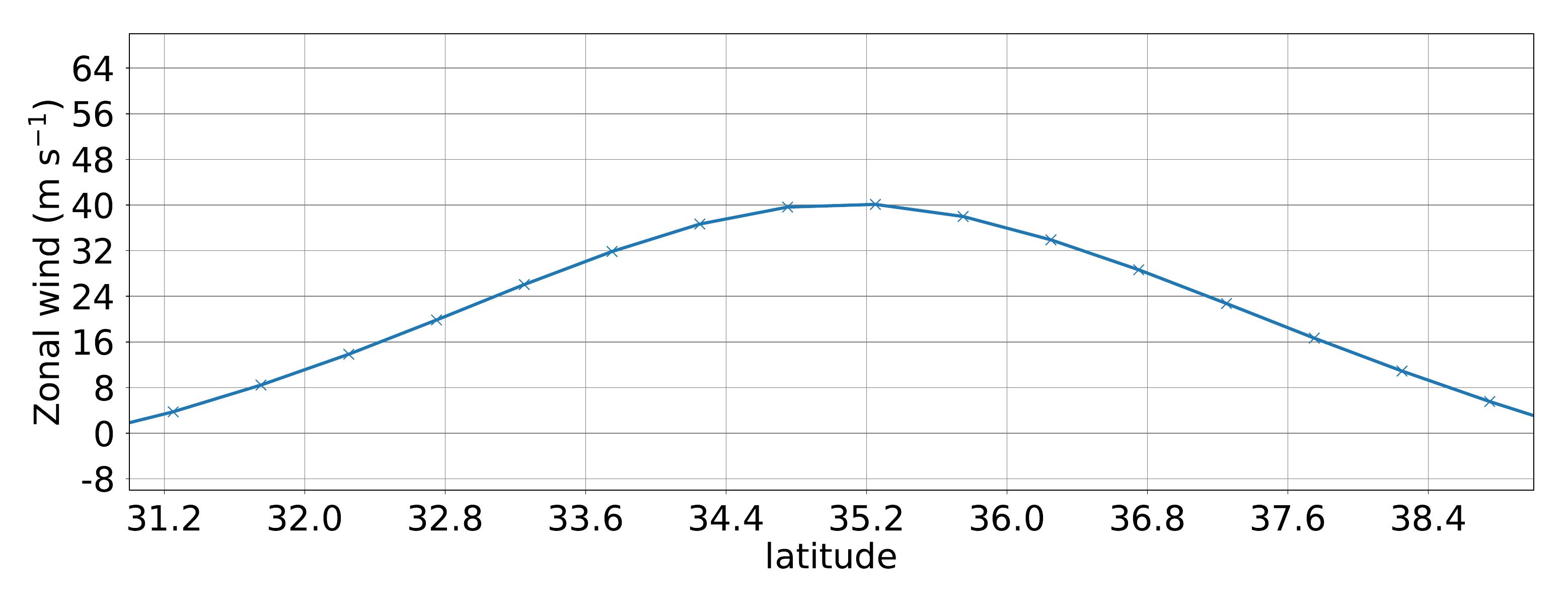}
\includegraphics[width=0.65\textwidth]{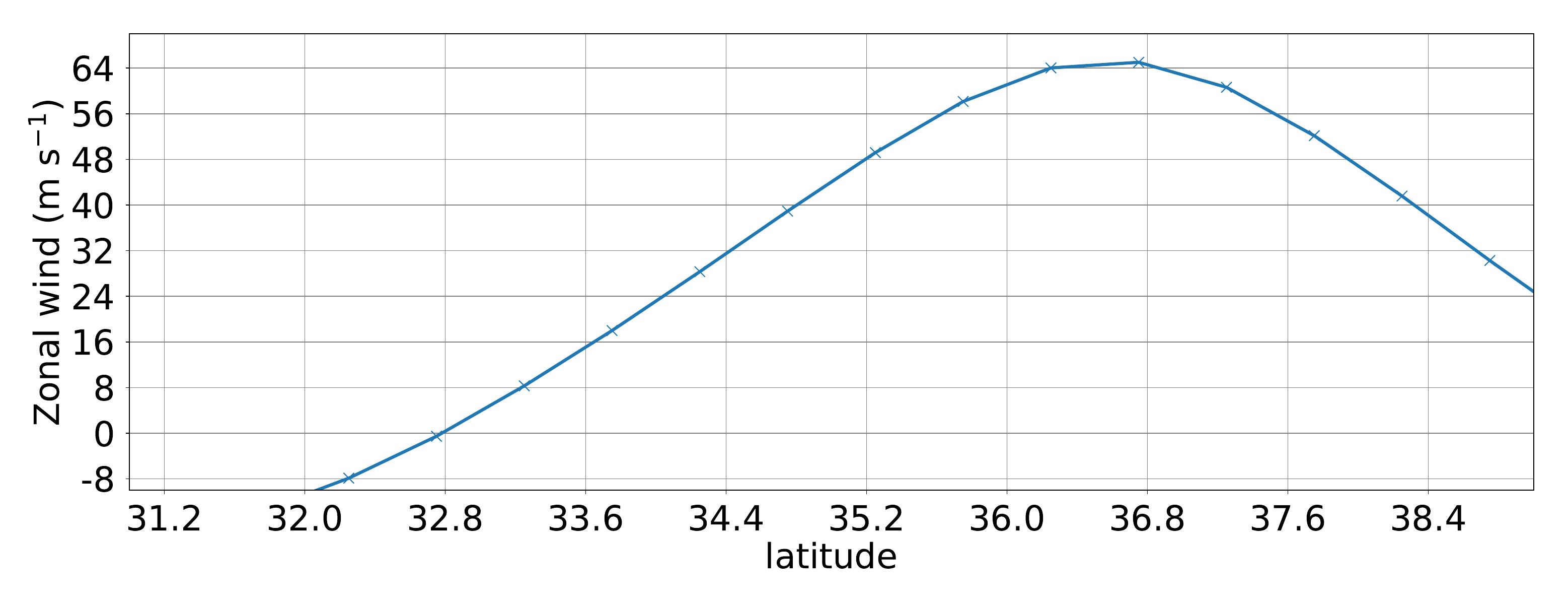}
\includegraphics[width=0.65\textwidth]{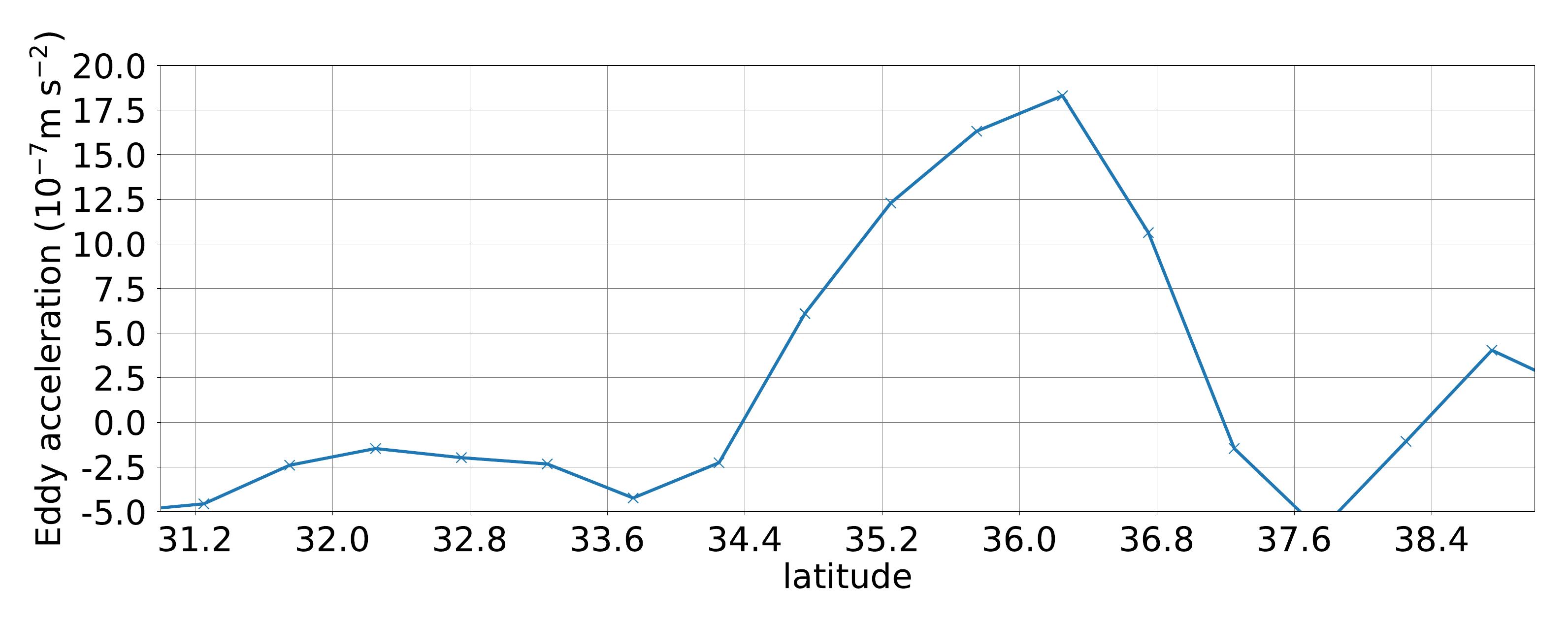}
\caption{\label{fig:migrationanalysis}
\emph{An episode of the poleward migration of a mid-latitude jet
is shown on the top two figures offering 
a magnified view of a relevant portion of
Figure~\ref{fig:jetevolution}
showing an episode of jet migration.
The latitudinal profiles of the zonal-mean zonal jet
at the beginning and in the end of the
temporal interval spanned by the top plots
are shown as middle plots (pressure level is $200$~mbar).
The bottom plot features the acceleration term
obtained by equation~\ref{eq:eddyacc}
(computation of the divergence of the 
Eliassen-Palm flux).}}
\end{center}
\end{figure}

A complementary framework to study
the evolution of jets -- especially the
eddy-related acceleration -- is
the Transformed Eulerian Mean approach.
In this approach, the Eliassen-Palm (EP) 
flux~$\mathcal{F}$ \citep[e.g.,][]{Vall:06},
which meridional component writes 
in isobaric coordinates \citep[e.g. equation 2.7 in][]{Andr:83}
\begin{equation}\label{eq:epflux} 
\mathcal{F}_{\varphi} = a \, \cos\varphi \left( - \overline{u^\prime v^\prime} + \psi \, \Dp{\overline{u}}{p} \right)
\qquad \textrm{with} \qquad
\psi = - \overline{v^\prime T^\prime} / \left( \frac{R \, \overline{T}}{c_p \, p} - \Dp{\overline{T}}{p} \right) 
\end{equation}
provides a direct link between 
the convergence / divergence of eddy momentum 
and the resulting acceleration / deceleration of zonal jets.
The horizontal contribution of eddies to zonal-mean wind acceleration
is the divergence of the meridional component~$\mathcal{F}_{\varphi}$ of the EP flux
\begin{equation}\label{eq:eddyacc}
\left[ \Dp{\overline{u}}{t} \right]_{\textrm{\scriptsize eddies}} 
= \frac{1}{a^2 \, \cos^2\varphi} \Dp{\mathcal{F}_{\varphi} \, \cos\varphi}{\varphi} 
\qquad
\left( 
\simeq - \frac{1}{a \, \cos^2\varphi} \Dp{\overline{u^\prime v^\prime} \, \cos\varphi}{\varphi} 
\right)
\end{equation}
\newcommand{\auckland}{(The vertical contribution of
eddies to zonal-mean wind acceleration is omitted
in this equation because, in the specific context
of our analysis of eddy-driven jets, 
it was found to be negligible).}
\noindent \auckland
We use the expression in equation~\ref{eq:epflux} 
to diagnose the eddy-driven acceleration in our Saturn GCM simulations;
we note, however, that the approximate expression in parenthesis 
(used e.g. to interpret Figure~\ref{fig:section_eqsuper} 
in section~\ref{sec:eqjets})
is reasonable in a vertically-integrated quasi-geostrophic framework,
with zonal averaging making mean momentum flux convergence terms 
to be small compared to the eddy momentum flux convergence terms \citep{Hosk:83,Vall:06,Chem:15}.

As was found from analyzing Figure~\ref{fig:dynbudg},
the~$35^{\circ}$N eastward jet in
the end of the first simulated year 
typically
undergoes a poleward migration associated
with a burst of eddy activity; this
eddy-driven migration is continuing in the
beginning of the second simulated year.
Figure~\ref{fig:migrationanalysis} 
indicates
that the divergence of the Eliassen-Palm
flux associated with this eddy activity
indeed acts to slow down the jet core
and accelerate its flanks, with a larger
acceleration being experienced in the poleward
side.

\subsection{Barotropic vs. baroclinic instability of the jets \label{sec:baroinstab}}

The importance of barotropic and baroclinic instabilities
has been discussed in the existing literature 
as the source for gas giants' banded jets 
\citep{Dowl:95,Liu:10jets}
and the evolution thereof, notably migration 
\citep{Will:03,Chem:15}.
Just as the vertical gradient of potential temperature
enables to assess convective instability,
the meridional gradient of potential vorticity
enables to assess 
barotropic / baroclinic instability \citep{Dowl:95,Holt:04,Vall:06}.
The Rayleigh-Kuo [RK] necessary condition 
for barotropic instability
is that the meridional gradient of PV
\begin{equation}\label{eq:rk}
\left[ \Dp{ \overline{q} }{y} \right]_{BT} 
= \beta - \DDp{\overline{u}}{y} 
\end{equation}
\noindent changes sign in the domain interior.
The Charney-Stern-Pedlosky [CSP] necessary condition
for baroclinic instability is that the
full-baroclinic meridional gradient of PV
\begin{equation}\label{eq:csp}
\left[ \Dp{ \overline{q} }{y} \right]_{BC}
= \left[ \Dp{ \overline{q} }{y} \right]_{BT}
- \frac{1}{\rho_0} \, \Dp{}{\mathcal{Z}} \left[ \rho_0 \, \frac{f_0^2}{N^2} \Dp{\overline{u}}{\mathcal{Z}} \right] \qquad \qquad \mathcal{Z} = -H \, \ln (p/p_0)
\end{equation}
\noindent 
either: 
changes sign in the interior [CSP1, similar to~RK],
is the opposite sign to $\overline{u}_{\mathcal{Z}}$ at the upper boundary [CSP2],
is the same sign as $\overline{u}_{\mathcal{Z}}$ at the lower boundary [CSP3],
or is zero and $\overline{u}_{\mathcal{Z}}$ is the same sign at both boundaries [CSP4].
The CSP criterion is not defined in a neutral
layer (where~$N^2 \sim 0$) such as Saturn's troposphere, 
thus we carry out the analysis of 
the two necessary conditions in equations~\ref{eq:rk}
and~\ref{eq:csp} near the tropopause level in our simulations.

\begin{figure}[p]
\begin{center}
\includegraphics[width=0.48\textwidth]{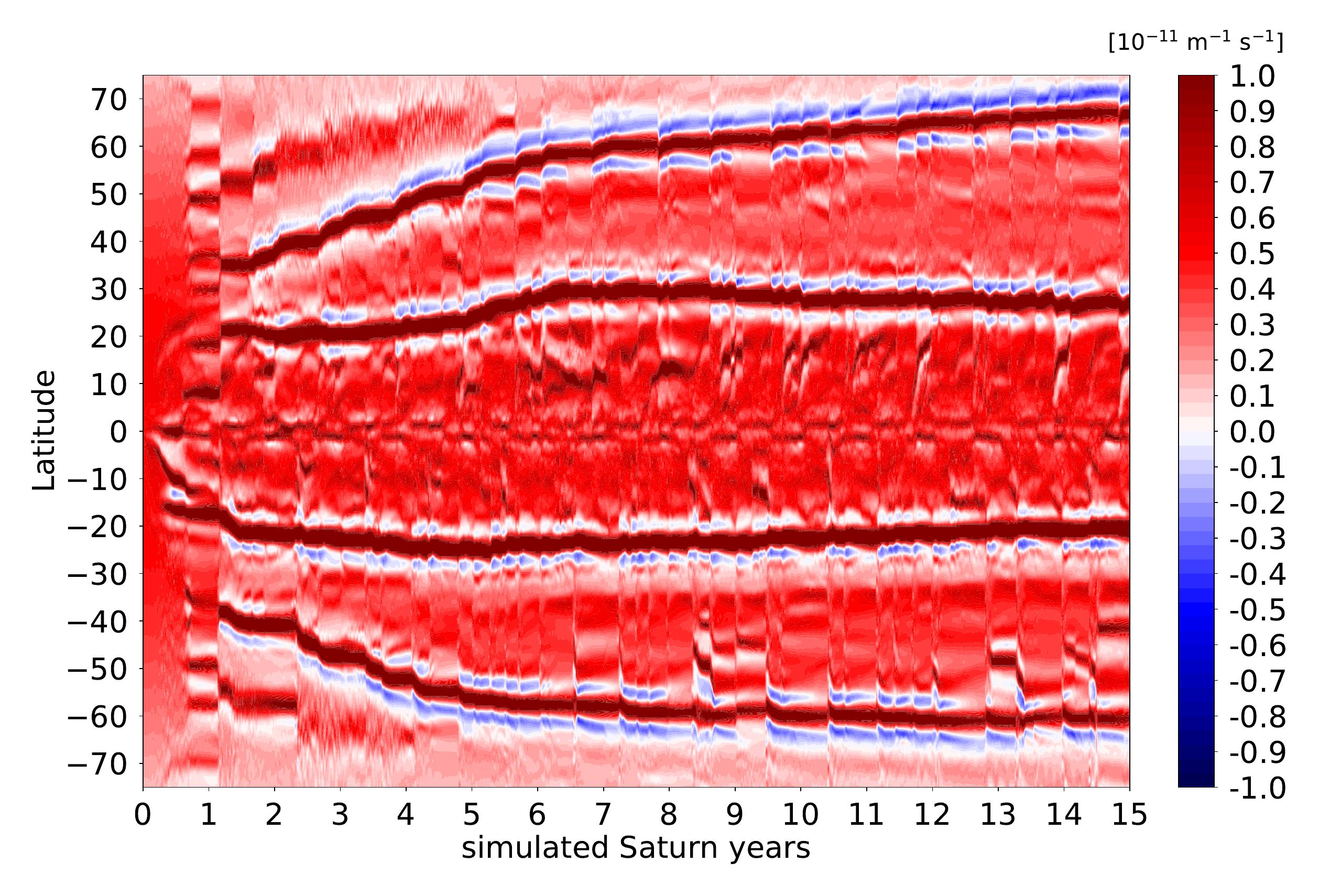}
\includegraphics[width=0.48\textwidth]{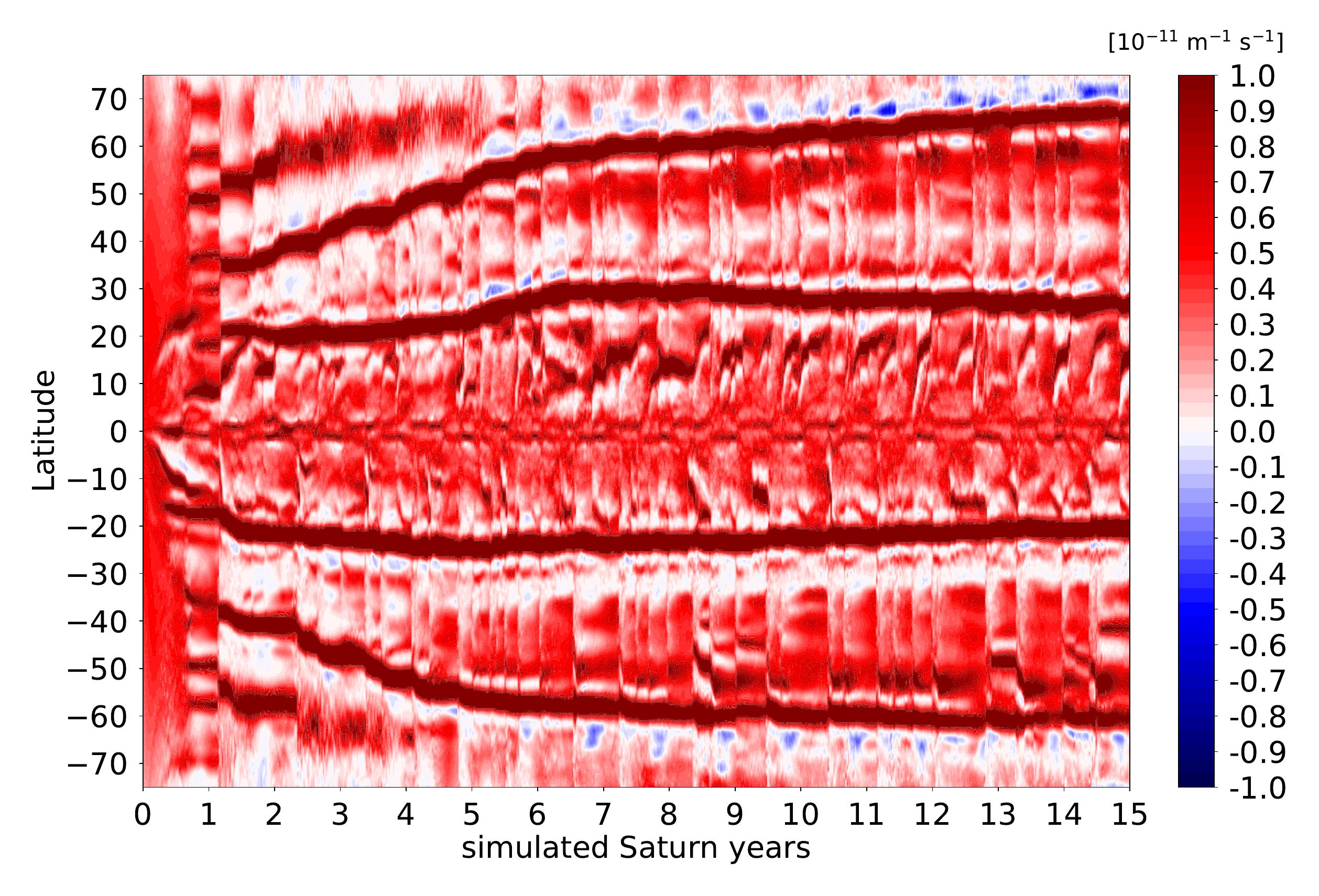}
\includegraphics[width=0.48\textwidth]{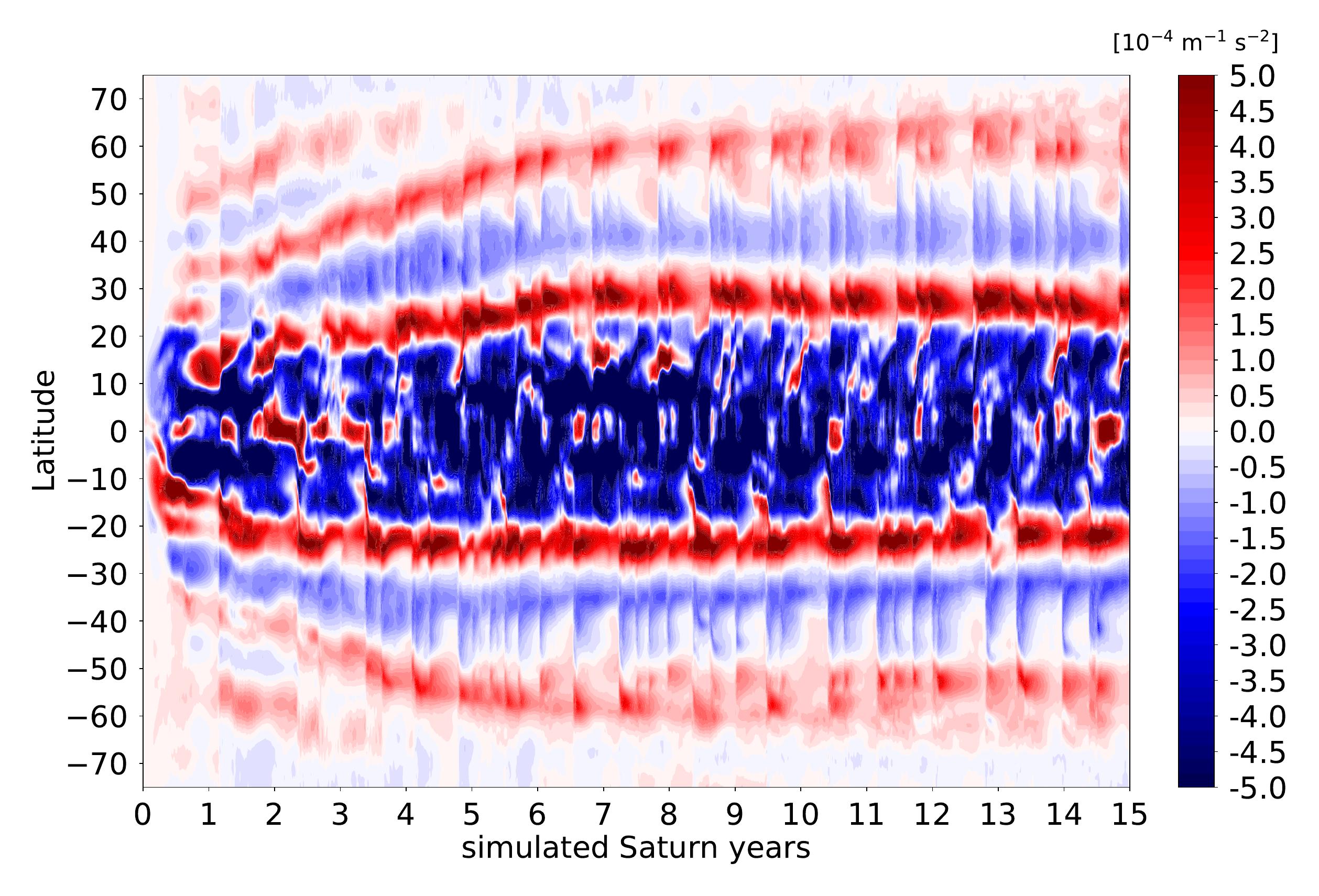}
\caption{\label{fig:baroinstab}
\emph{Evolution of the meridional gradients
of zonal-mean potential vorticity
(top left: barotropic, equation~\ref{eq:rk} for necessary condition RK,
top right: baroclinic, equation~\ref{eq:csp} for necessary condition CSP1)
and vertical shear of zonal-mean zonal wind (bottom)
at Saturn's tropopause (100 mbar)
within the whole 15-year duration of 
our Saturn DYNAMICO GCM simulation.
The diagnostic in equation~\ref{eq:rk}
cannot be computed in the troposphere
where the atmosphere is neutral ($N^2 \sim 0$)
and barotropic; the two other
diagnostics are shown at the tropopause for
consistency but are similar at deeper levels.}}
\end{center}
\end{figure}

The necessary condition RK for barotropic instability
can be assessed from Figure~\ref{fig:baroinstab}
by determining when the quantity
described by equation~\ref{eq:rk} changes sign.
In the first four simulated years,
the mid-latitude eastward jets migrating 
from~$30^{\circ}$N/S
to~$60^{\circ}$N/S
fulfil the RK condition on the
poleward flanks, but not on the equatorward flank.
In subsequent years (from year 5 to year 15)
in our Saturn DYNAMICO GCM simulations,
those eastward jets fulfil the RK on both flanks,
which is also the case for 
the weakly-migrating $20^{\circ}$N/S jets
throughout the 15-year simulation.
The simulated mid-latitude jets 
in our Saturn DYNAMICO GCM
are thus possibly impacted by barotropic instability;
this could be expected from the 
PV mapping in Figure~\ref{fig:instpv}
which clearly shows that strong inversions of
the meridional gradients of PV are found
at the location of the eastward jets.
This provides an explanation
for the extratropical eddies
found in the bulk of eastward jets
and discussed in section~\ref{sec:extraeddies}.
\newcommand{\helsinki}{It is worth reminding here 
that barotropic instability
acts to transfer momentum from jets to eddies, and not
the contrary.
Hence the positive conversion rate~$\mathcal{C}$ 
indicating an overall transfer from eddies to jets
(section~\ref{sec:global})
hints at baroclinic instability complementing
barotropic instability.} \helsinki

Referring to the CSP1 criterion,
Figure~\ref{fig:baroinstab} shows 
that the conditions for baroclinic instability
are met in polar regions -- this is also 
true for barotropic instability according 
to the RK criterion.
In mid-latitude eastward jets, 
the CSP1 necessary criterion 
for baroclinic instability
is however only verified
in the poleward flank of the jets;
the quantity~described in equation~\ref{eq:csp}
is mostly positive at all latitudes (outside polar regions).
\newcommand{\kinshasa}{The fact that the
CSP1 criterion for baroclinic instability
is fulfilled in the poleward flank of the mid-latitude jets,
and less so in the equatorward flank,
echoes the conclusions of \citet{Chem:15}
who demonstrated that the poleward migration of jets 
in idealized GCM simulations of high-rotation planets
is caused by a poleward bias in 
baroclinicity across the width of the jet.} \kinshasa

The condition CSP3 is fulfilled
in the simulated mid-latitude eastward jets
since the (baroclinic) meridional gradient of PV
is of the same sign as the 
vertical shear~$\overline{u}_{\mathcal{Z}}$,
i.e. positive as is shown in Figure~\ref{fig:baroinstab}
(see also section~\ref{sec:midlatjets}).
Those simulated jets could thus be
baroclinically unstable\footnote{CSP3 is often the decisive 
condition in the terrestrial environment
too, see~\citet{Vall:06}. An assessment of conditions
CSP~2 and CSP~4 in the case of our Saturn DYNAMICO GCM
simulations indicates that those are not fulfilled.}.
This is not the case for 
the equatorial eastward jet, 
and the broad westward jets, for
which vertical shear is negative.

We conclude that both barotropic
and baroclinic instabilities could
account for the 
maintenance, migration
and (in polar regions) disappearance
of eastward zonal jets in our
Saturn DYNAMICO GCM simulations.
The putative role of baroclinic instability
in driving the mid-latitude jets was also
highlighted by \citet{Liu:10jets};
the possibility that barotropic
and baroclinic instabilities exist
in Saturn's troposphere is also
argued in \citet{Stud:18}.
In that respect, the pole-to-equator 
meridional gradient
of temperature in the deep troposphere 
(e.g. at pressures greater and equal to 500 mbar)
is a key constraint for baroclinic instability,
owing to its link to the vertical shear of zonal wind
involved in the CSP3 criterion, through the 
thermal wind equation.

\newcommand{\tokyo}{The equator-to-pole 
temperature gradient of 4~K 
found at the 500-mbar level in Figures~\ref{fig:tsections} 
and~\ref{fig:bv} (bottom) translates into 
about 7~K at the 2.5-bar level,
in accordance with potential temperature
being constant in the neutral lower troposphere.
Both correspond to an equator-to-pole gradient
of potential temperature of~7~K.} \tokyo
\newcommand{\kyoto}{As is discussed in section~\ref{sec:thermal},
while a good agreement between our model and the observations
is obtained in the upper troposphere / lower stratosphere
(Figure~\ref{fig:compcirs}), the simulated baroclinicity 
in the lower troposphere is at least twice larger
than it is in the observations.
In addition to this, the shallowness of our Saturn GCM
\citep[as well as other models sharing the same strategy, e.g.][]{Liu:10jets}
prevents it from resolving the deep large-scale
circulations that could act to counteract
and further homogenize the meridional gradients
of potential temperature in the deeper troposphere \citep{Aurn:08}.
According to the CSP3 criterion,
an unrealistic baroclinicity at the bottom
boundary entails that part of the eddies resolved by
our Saturn DYNAMICO GCM could be spurious
and not occurring in the real Saturn -- contrary to the
baroclinicity associated with PV gradient reversals 
within the atmospheric fluid related to the CSP1 criterion.
This would impact 
the structure of the resolved zonal jets,
as well as their meandering and migration.} \kyoto
\newcommand{\ulanbator}{A similar word of caution
applies to the analysis of the conversion rate~$\mathcal{C}$
in Figure~\ref{fig:globalforcingcomp} 
and section~\ref{sec:global}, even if the order
of magnitude compared to observations is correct.} \ulanbator
\newcommand{\delhi}{For instance, the fact that 
baroclinic instabilities cause
unrealistically strong
jet meandering in polar regions
may be deemed an indication that 
meridional gradients
could be improved at the bottom of our model.
Deepening the model bottom 
and ensuring a more realistic
baroclinicity in the lower troposphere
is an area of future improvement of
our Saturn DYNAMICO GCM, as is the
case for most existing shallow-atmosphere models.} \delhi

\section{Conclusions \label{sec:discussion}}

The conclusions of our study can be summarized as follows.
\begin{enumerate}
\item The Cassini mission opened novel questions on tropospheric
and stratospheric circulations on Saturn, with new modeling challenges
(section~\ref{sec:intro} and challenges \ref{ch:radi} 
\ref{ch:dyna} \ref{ch:strat} \ref{ch:bound}).
\item The Global Climate Model (GCM) we built is named the
Saturn DYNAMICO GCM and couples the radiative transfer
of \cite{Guer:14} tailored for Saturn
with the icosahedral dynamical core 
DYNAMICO of \cite{Dubo:15}
tailored for massively-parallel computing resources
(section~\ref{sec:method}).
\item Care must be taken when developing
a GCM for gas giants in setting the subgrid-scale
dissipation and verifying the conservation of
global axial angular momentum (appendix~\ref{sec:sens}).
\item We reached the capability to simulate the dynamics of Saturn's atmosphere
from the troposphere to the lower stratosphere
during 15 Saturn years with an horizontal resolution of~$1/2^{\circ}$
longitude / latitude, which made our reference simulation 
(sections~\ref{sec:reference} and~\ref{sec:evolution});
our Saturn DYNAMICO GCM produces a satisfactory
thermal structure, and seasonal variability thereof,
compared to Cassini CIRS measurements
(section~\ref{sec:thermal}).
\item The number and intensity of mid-latitude eastward jets
(and broad westward jets)
reproduced by the reference GCM simulation
is compliant with observations, although slightly underestimated,
but no stable circumpolar jet (less so hexagonal-shaped) is reproduced;
those jets' intensities increase with altitude
and their latitudinal organization exhibits 
potential-vorticity staircases
(section~\ref{sec:midlatjets}).
\item The GCM simulation exhibits at the equator
both a superrotating zonal jet in the troposphere
and stacked alternating zonal jets
in the stratosphere;
nevertheless, the former is 
one order of magnitude less powerful than the observed jet,
and the latter are not downward-propagating with time
as would be expected from the observed equatorial oscillation
(section~\ref{sec:eqjets}).
\item Our model produces a wealth of 
Yanai (Rossby-gravity), Rossby and Kelvin waves
in the tropical channel,
part of them hinted at in available observations 
(section~\ref{sec:eqwave}).
\item Outside the tropics, 
the cores of eastward jets are perturbed
by Rossby waves reminiscent of Ribbon-like waves, 
while westward jets host an eddy activity
not especially organized in vortices,
which transitions in polar regions
into a predominance of large-scale vortices
(section~\ref{sec:extraeddies}).
\item In the 15-Saturn-year course of our Saturn DYNAMICO GCM
simulation, eastward jets undergo poleward migration
and perturbations by bursts of eddies
(section~\ref{sec:phenomevolution}).
\item The global kinetic energy conversion rate 
simulated in our Saturn DYNAMICO GCM,
albeit half the value of the observed estimates,
is positive and argues for a significant contribution of
eddy acceleration in driving the eastward jets
(section~\ref{sec:global}).
\item The acceleration (and, if applicable, migration)
of jets caused by eddy momentum transfers
is evidenced by local analysis with either
Eulerian-mean or transformed Eulerian-mean formalism
(section~\ref{sec:local}).
\item Eastward jets produced by our Saturn DYNAMICO GCM
are prone to both barotropic and baroclinic instabilities
(section~\ref{sec:baroinstab}).
\end{enumerate}

Based on the present study, and the comparison
between available observations
and our GCM simulations,
we can envision the following
improvements of our Saturn DYNAMICO GCM 
in the future:
\begin{enumerate}[label=$(\alph*)$]
\item to refine the horizontal resolution to~$1/4^{\circ}$;
\item to include a physically-based parameterization for
subgrid-scale dry and moist convection, and
to subsequently deepen the model bottom to 10-20 bars;
\item to extend the model top to the upper stratosphere
and to refine the vertical resolution in the troposphere 
and the stratosphere;
\item to implement a parameterization for 
the impact of unresolved mesoscale gravity waves
on the mean flow.
\end{enumerate}
This list is not exhaustive, but represents
the near-future evolution of our model.
Long-term developments would
be inspired by the model-coupling methodology
for the climate of telluric planets,
i.e. coupling our Saturn DYNAMICO GCM 
with
stratospheric photo-chemical models,
deep-interior convection models
and upper-atmosphere thermo-iono-spheric models.

\begin{figure}[ht]
\begin{center}
\includegraphics[width=0.5\textwidth]{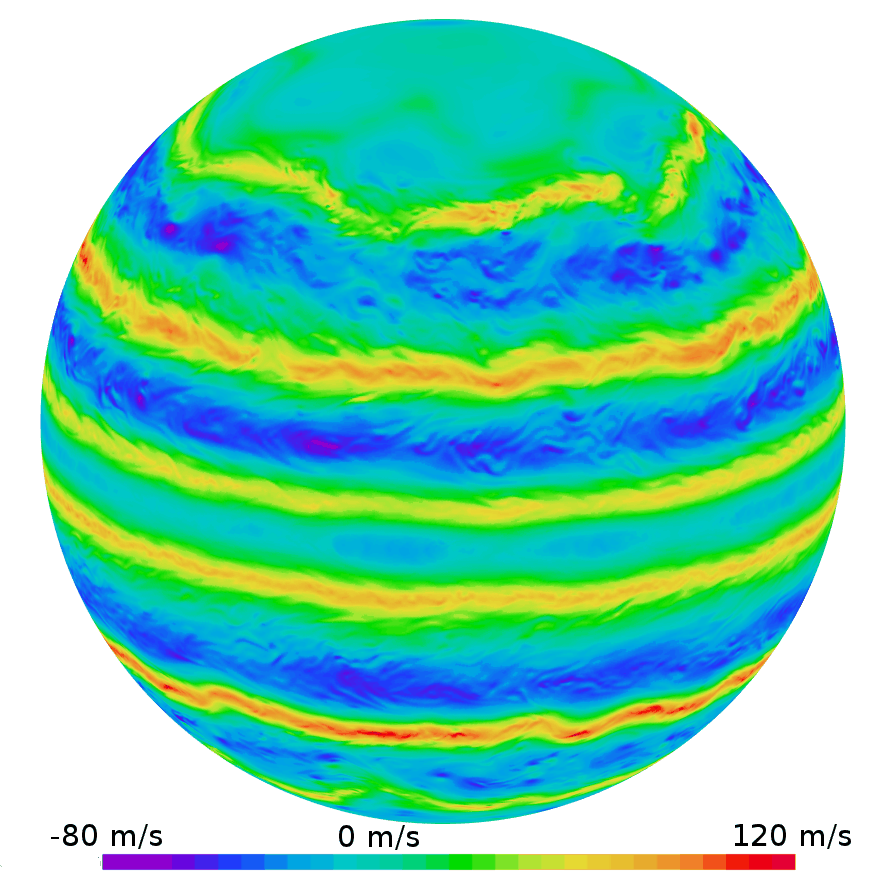}
\includegraphics[width=0.45\textwidth]{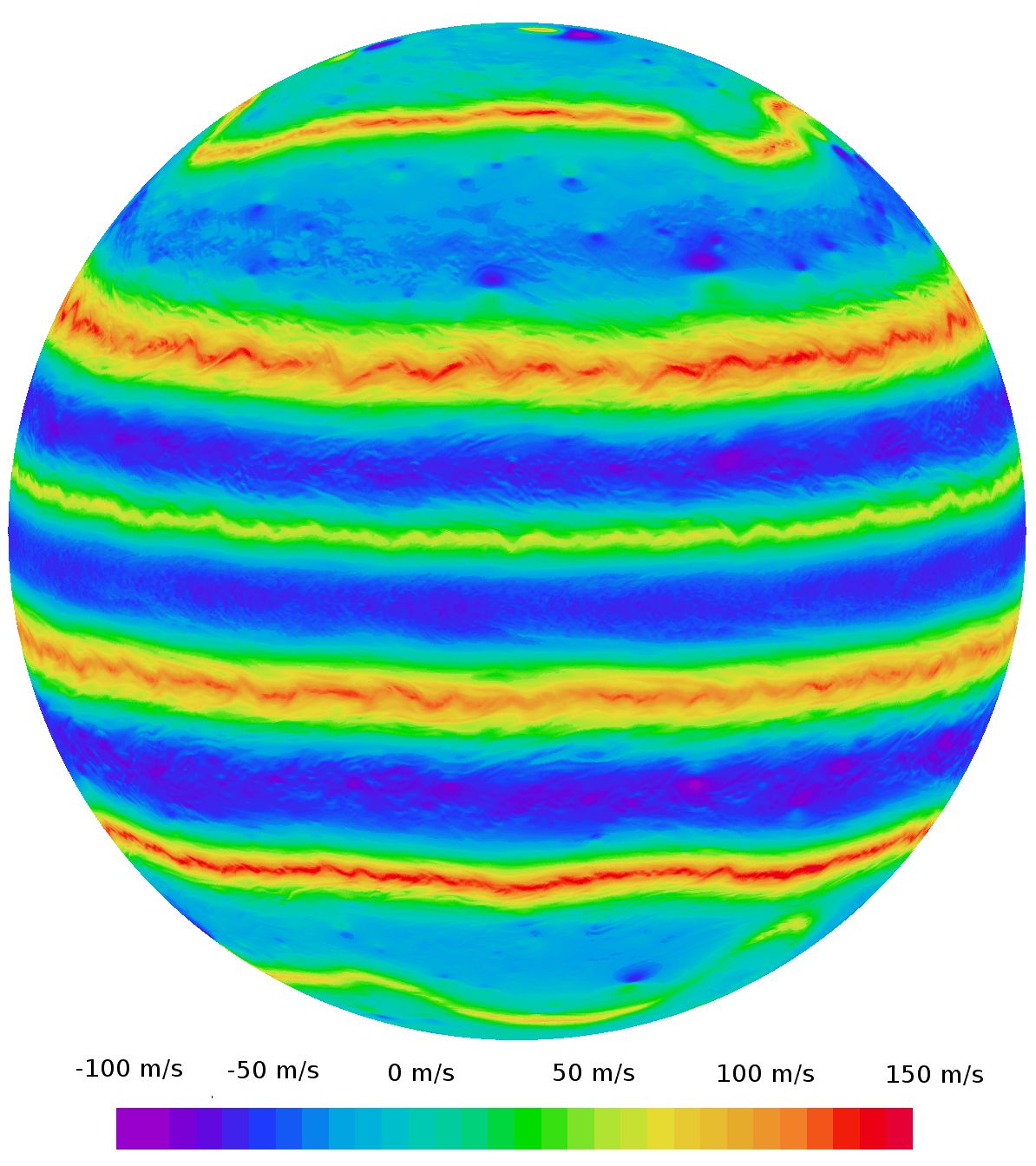}
\caption{\label{fig:granddefi}
\emph{Instantaneous zonal winds predicted at 0.5 bar (yellow/red: prograde jets, blue/violet: retrograde jets) 
after $500$ simulated Saturn days by our Saturn DYNAMICO GCM. 
Our simulations used a horizontal resolution of $1/4^{\circ}$ (left) 
and~$1/8^{\circ}$ (right) longitude, and extent from troposphere to the stratosphere.
In the left plot, jets' instabilities and filamentation can be noticed; 
in the right plot, the even finer horizontal resolution allows the model 
to reproduce the propagation of gravity waves on the flanks of the jets, 
as well as the possible emergence of traveling vortices (cf. blue/green spots in the northern hemisphere).
The objectives behind those simulations were more technical
(testing a massively-parallel computing cluster) than scientific:
the strong hyperdiffusion made the structures seen in the figure
to disappear after about a thousand simulated Saturn days.} }
\end{center}
\end{figure}

What the refinement of horizontal resolution
would bring to Saturn GCM studies
can be illustrated by the tests
of an earlier version of the
Saturn DYNAMICO GCM at 
the horizontal 
resolutions of~$1/4^{\circ}$ and~$1/8^{\circ}$.
We carried out overdissipated
GCM simulations at those resolutions,
using both a sponge layer and 
hyperdiffusion timescales of 
the order~$\tau_D = 500-1000$~s,
to ensure overly-conservative numerical stability
in order to test the performance of the model
on massively-parallel architectures
(up to~$60,000$~cores).
Those simulations are thus not optimized
for scientific return\footnote{This is 
all the more true since the outcome 
of a 2-Saturn-year simulation at~$1/2^{\circ}$ 
(hence not fully spun-up) was used
to initialize a~$1/4^{\circ}$ simulation, 
and similarly from the~$1/4^{\circ}$ to the~$1/8^{\circ}$ 
simulations} and
the approach
described in appendix~\ref{sec:sens}
will have to be carried out again
in the future for the~$1/4^{\circ}$ 
and~$1/8^{\circ}$ configuration of the model.
Yet, even those imperfect~$1/4^{\circ}$ and~$1/8^{\circ}$
simulations with the Saturn DYNAMICO GCM
demonstrate the potential of the model
to better resolve eddies, waves and vortices
with refined horizontal resolution, as is
shown in Figure~\ref{fig:granddefi}
and in the movie included as supplementary material,
showing wind amplitude and vorticity from
the~$1/8^{\circ}$ Saturn DYNAMICO simulation.

Finally, we note that the Saturn DYNAMICO GCM is only a first step 
towards a GCM system able to simulate the atmospheres of all giant planets,
ice giants Uranus and Neptune included.
The development of the Jupiter DYNAMICO GCM is currently ungoing \citep{Guer:16egu,Bois:18epsc},
along the lines drawn for Saturn by \citet{Guer:14} and this study.
The similarities and differences between Jupiter and Saturn
in their banded jets \citep{Inge:90,Dowl:95,Kasp:18}, 
eddy activity \citep{Saly:06}
equatorial oscillations \citep{Li:00qqo,Simo:07},
large-scale vortices \citep{Yous:03,Flet:10grs,Simo:14,Lega:05}
relate to fundamental research in geophysical fluid dynamics.
Employing GCMs for giant planets could help, along with observations,
to reach a detailed understanding of the big 
picture of giant planets' climate and meteorology.
This is all the more relevant to prepare
the next round of observations of the giant planets,
with either probes \citep{Mous:14},
telescopes \citep{Norw:14},
or orbiting spacecraft \citep{Cava:17}.

\section*{Acknowledgments} 
The authors thank the editor Darrell Strobel,
an anonynous reviewer and Adam Showman for
extremely constructive reviews which helped to
improve the manuscript.
We would like to thank
Tapio Schneider,
Yohai Kaspi,
Leigh Fletcher,
Glenn Orton,
Roland Young,
Peter Read,
Mike Flasar,
Fran\c cois Forget,
Michel Capderou,
Pierre Drossart,
Thibault Cavali{\'e},
Agustin Sanchez-Lavega,
Ricardo Hueso,
Th{\'e}r{\'e}se Encrenaz,
Emmanuel Lellouch,
Frédéric Hourdin,
Sébastien Fromang,
for useful discussions
and questions 
on preliminary versions of 
the work reported 
in this paper.

The authors 
acknowledge exceptional computing support from 
Grand Équipement National de Calcul Intensif (GENCI)
and
Centre Informatique National de l'Enseignement Supérieur (CINES).
All the simulations presented in this paper were carried out on the \emph{Occigen} cluster hosted at CINES.
This work was granted access to the High-Performance Computing (HPC) resources 
of CINES under the
allocations 2015-017357, 2016-017548, A001-0107548, A003-0107548, A004-0110391
made by GENCI.
We also thank CINES for a ``Grand Challenge''
exceptional allocation in early 2015 to test 
the performances of the Saturn DYNAMICO GCM
at various horizontal resolutions.

Dubos,
Cabanes,
Spiga,
Meurdesoif,
Millour
acknowledge funding from 
Agence Nationale de la Recherche (ANR),
project HEAT ANR-14-CE23-0010.
Guerlet, 
Indurain,
Spiga
acknowledge funding from 
Agence Nationale de la Recherche (ANR),
project OMAGE ANR-12-PDOC-0013.
Spiga,
Cabanes,
Guerlet
acknowledge funding from 
Agence Nationale de la Recherche (ANR),
project EMERGIANT ANR-17-CE31-0007.
Guerlet,
Spiga,
Lebonnois
acknowledge funding from
Centre National d'Études Spatiales (CNES)
project exploiting CIRS measurements onboard Cassini.
Boissinot,
Spiga
acknowledge funding from 
Region Île-de-France
DIM ACAV+
project JOVIEN.
Sylvestre,
Fouchet,
Spiga
acknowledge funding from 
Université Pierre et Marie Curie
(now Sorbonne Université)
Émergence program.
Leconte
acknowledges that this project 
has received funding from the European Research Council (ERC) 
under the European Union's Horizon 2020 research 
and innovation program (grant agreement No. 679030/WHIPLASH).

\appendix
\section{Impact of assumptions in the dynamical core \label{sec:sens}}

Numerical subgrid-scale dissipation in the model
was found to be a critical setting to deal with.
The sensitivity of simulated jets with horizontal dissipation
is a common issue in GCM: it was specifically discussed 
for the cases of 
Venus \citep{Lebo:12inter},
Titan \citep{Newm:11},
and hot Jupiters \citep{Thra:11}.
This is a particularly important issue
for a gas giant GCM,
and it is discussed in 
section~\ref{sec:dissip}.
Another important question
is to explore the behaviour of
global axial angular momentum in our
Saturn DYNAMICO GCM simulations,
which is done in section~\ref{sec:aam}.

As far as the sensitivity of the modeled jets 
to the settings adopted
for bottom drag~$(\tau_R,\varphi_R)$
is concerned, we rely on the work by
\citet{Liu:10jets} and \citet{Liu:11},
and adopt their 
settings~$\tau_R=100$~Earth days
and $\varphi_R = 33^{\circ}$
for our reference simulation. 
\newcommand{\alger}{Simulations 
with Saturn DYNAMICO
carried out with different values
of those parameters
(respectively~$\tau_R=10$~Earth days
and $\varphi_R = 10^{\circ}$)
confirm the conclusions of
\citet{Liu:10jets} and \citet{Liu:11}
that the bottom drag affects the
jets' width and speed.} \alger

\subsection{Exploring the impact of dissipation \label{sec:dissip}}

A subgrid-scale dissipation term is 
included in our Saturn DYNAMICO GCM
to prevent the accumulation of energy
at scales close to the grid resolution,
caused by the GCM not resolving the
turbulent scales at which this
energy is dissipated.
This hyperviscosity term is written 
in our Saturn DYNAMICO model
as an iterated Laplacian term on a given
variable~$\psi$
\begin{equation} \label{eq:dissip} 
\left[ \frac{\textrm{d}\psi}{\textrm{d}t} \right]_{\textrm{dissip}} 
= \frac{(-1)^{q+1} \, \ell_{\textrm{min}}^2}{\tau_D^{\psi}} \, \nabla^{2q} \psi 
\end{equation}
where~$q$ is the order of dissipation and~$\tau_D^{\psi}$ the damping timescale 
associated with the variable~$\psi$
at the smallest spatial scale~$\ell_{\textrm{min}}$ 
resolved by the model for a given horizontal discretization.
Large values of~$\tau_D^{\psi}$ means weaker dissipation:
$\tau_D^{\psi}$ is the times it takes to dissipate a perturbation
on variable~$\psi$ developing at the 
spatial scale~$\ell_{\textrm{min}}$.
The three variables denoted by~$\psi$ are 
vorticity, divergence, and potential temperature,
chosen to set
horizontal dissipation 
on respectively
the rotational component of the flow (e.g. Rossby waves),
the divergent component of the flow (e.g. gravity waves),
and the diabatic perturbations (e.g. coming from the physical packages).
In GCMs for telluric planets, 
strong variations 
from one grid point to one another
result from topography contrasts 
or mesoscale convective cells,
which calls for a preferential 
damping on the divergent flow \citep{Hour:12}.
In our Saturn DYNAMICO GCM simulations,
we adopt a simpler approach and
we set the same dissipation rate for all three variables, 
namely~$\tau_D^{\psi} \equiv \tau_D$ 
for divergence, vorticity and potential temperature.

In the GCM methodology, 
suitable values of~$(q,\tau_D)$ are determined empirically,
using a combination of past modeling experiences 
and trial-and-error approach using 
GCM simulations.
The ``right'' settings for numerical dissipation is 
a trade-off between 
ensuring model stability,
damping energy at the smallest resolved scales,
and minimizing impact on the large-scale flow.
A common practice is~$q$ ranging between~$1$ and~$4$, 
and~$\tau_D$ typically one-two terrestrial hours ($2000-5000$~s) 
for a~$1/2^{\circ}-1^{\circ}$ GCM simulation. 
We chose~$q=2$ (fourth-order dissipation) 
for the Saturn DYNAMICO GCM simulations,
since it is the setting adopted by our team 
for GCM for telluric bodies \citep{Hour:95b,Lebo:10},
and because $q=1$ is overly dissipative 
on circulations at large scales,
while $q=3$ led to results similar to $q=2$.
We then carried out several one-Saturn-year simulations 
with our Saturn DYNAMICO GCM 
to explore the sensitivity of 
the computed tropospheric jet structure to
the dissipation timescale~$\tau_D$.

\begin{figure}[p!]
\begin{center}
\includegraphics[width=0.85\columnwidth]{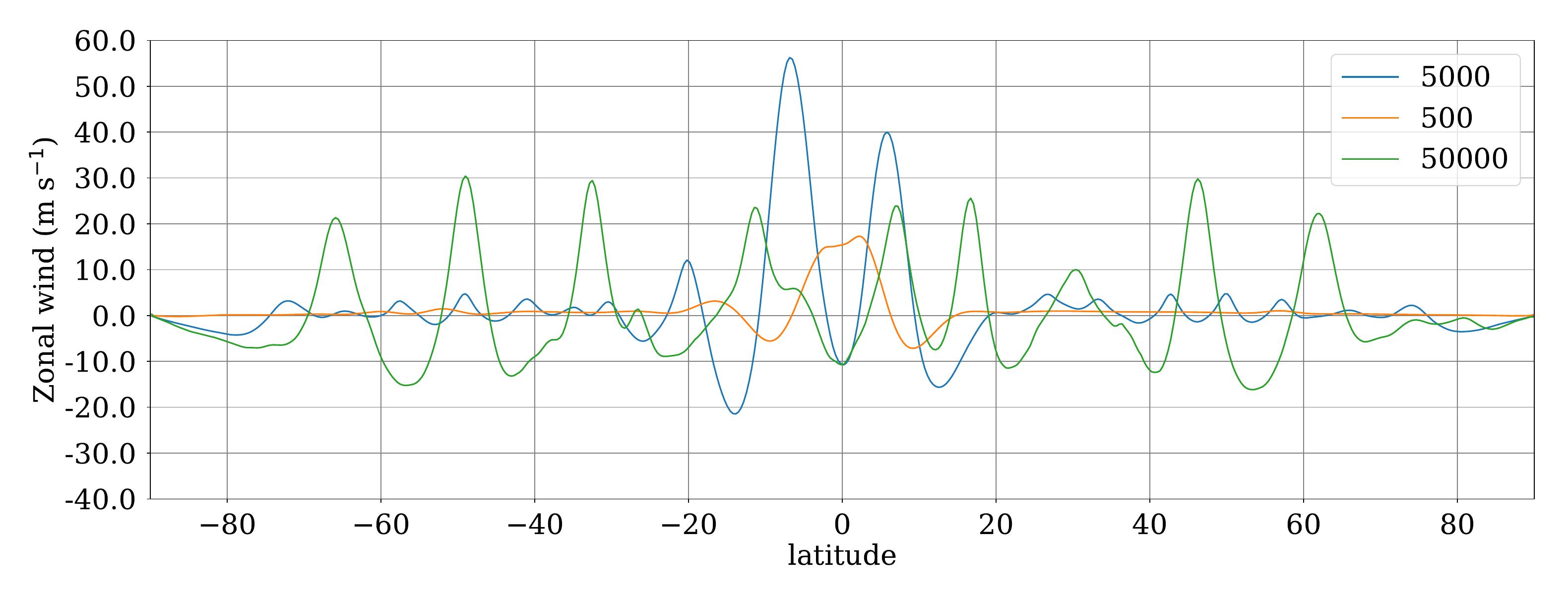}
\includegraphics[width=0.85\columnwidth]{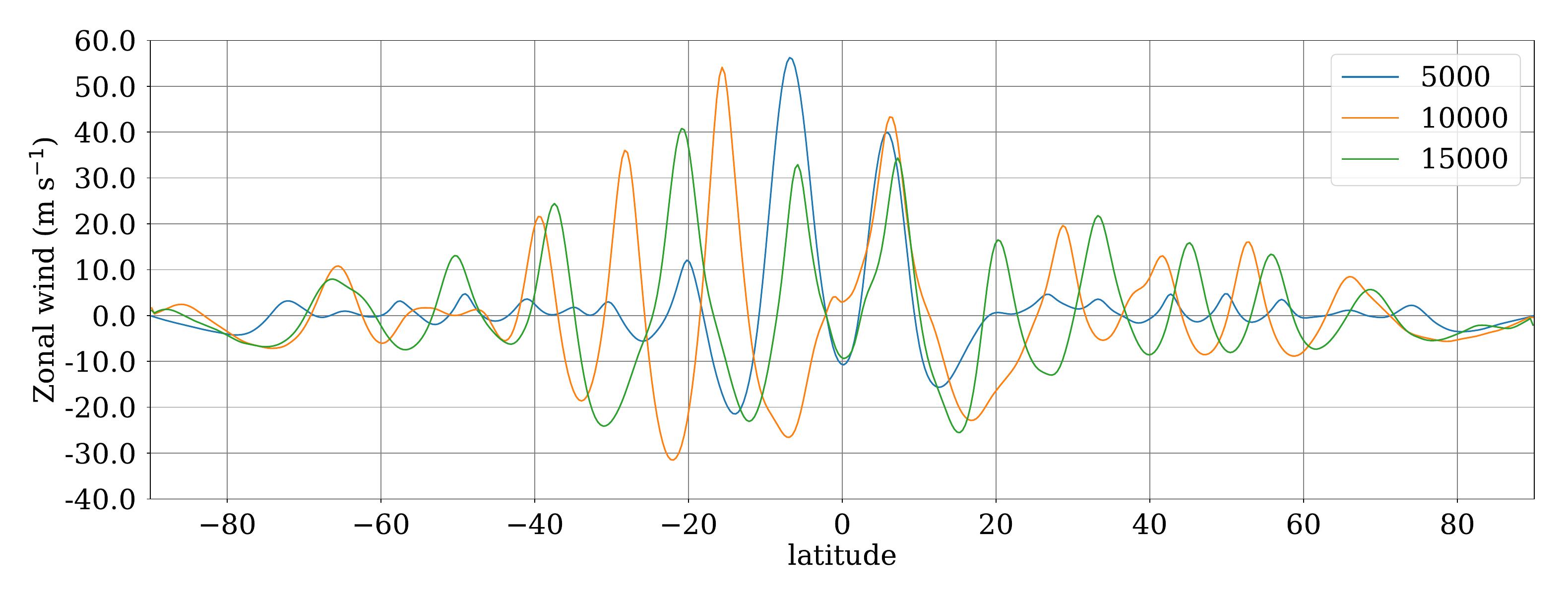}
\caption{\label{fig:dissip} 
\emph{Zonal-mean jets obtained at cloud level 
after~$20000$~integrated Saturn days 
(about one simulated Saturn year)
with our Saturn DYNAMICO GCM
run at a horizontal resolution of~$1/2^{\circ}$.
Those tests were carried out with a preceding, slightly different,
version of the DYNAMICO model compared to the one used
for the reference simulations of section~\ref{sec:reference}.
In the top plot, results are shown 
for three values of horizontal dissipation timescales
varying by one order of magnitude
(orange: strong $\tau_D = 500$~s,
blue: moderate $\tau_D = 5000$~s,
green: weak $\tau_D = 50000$~s).
In the bottom plot, results are shown 
for three values of horizontal dissipation timescales, two of them
(blue: $\tau_D = 5000$~s,
green: $\tau_D = 15000$~s)
enclosing the value chosen for 
the reference simulation discussed in
sections~\ref{sec:reference} and~\ref{sec:evolution}
(orange: $\tau_D = 10000$~s).} } 
\end{center}
\end{figure}

Figure~\ref{fig:dissip} shows that the simulated tropospheric jets in our~$1/2^{\circ}$ 
Saturn DYNAMICO GCM simulations are sensitive to the value assumed 
for dissipation timescale~$\tau_D$. 
A first order-of-magnitude sensitivity study (Figure~\ref{fig:dissip}, top) 
using extreme values disqualifies the strongest dissipation rate
($\tau_D=500$) which damps the mid-latitude jets out of existence.
Setting a weak dissipation ($\tau_D=50000$) 
lets jets significantly accelerate within one simulated year -- before 
the GCM simulation undergoes numerical instabilities in the second year of simulation.  
A refined sensitivity study (Figure~\ref{fig:dissip}, bottom) indicates 
that results obtained with~$\tau_D=10000$ or~$\tau_D=15000$ are essentially similar:
the choice of dissipation mainly impacts jets' meridional location.

The selective criterion to choose~$\tau_D$ for our
reference Saturn DYNAMICO GCM simulation
is then based on observations of Saturn's jets \citep{Porc:05,Garc:10,Stud:18}
from which we argue that, qualitatively, $\tau_D=10000$ sets 
a more realistic velocity profile. 
With $\tau_D=5000$, high and mid latitudes jets are inconsistently weak, 
while $\tau_D=15000$ smoothes the prevailing contrast 
that exists between equatorial and high-/mid-latitude jets.  
The value of~$\tau_D=10000$ is thus adopted for the 
15-year-long reference simulation discussed in 
sections~\ref{sec:reference} and~\ref{sec:evolution}.
It is important to note here that~$\tau_D=15000$
would have been an acceptable choice as well.
Figure~\ref{fig:dissip} (right) shows that,
even the intensity and location of jets
can be different between the Saturn DYNAMICO GCM simulations
with either~$\tau_D=10000$ or~$\tau_D=15000$,
the overall jet structure is qualitatively similar in both cases.

\newcommand{\madrid}{To understand why dissipation impacts the jets, 
it is important to keep in mind that eddies, which putatively drive the jets, 
have typical length scales 
of a couple degrees latitude for a grid spacing of~$0.5^{\circ}$
\citep[assuming the eddies are baroclinically driven
and taking the Rossby radius of deformation
as a first-order theoretical estimate 
of the eddy length scale, e.g.][]{Youn:17}. } \madrid
These eddies are thus potentially dampened by 
numerical dissipation that dominantly acts on 
the smallest-resolved scales. 
Undoubtedly, the next challenge is to refine GCM resolution
towards~$1/4^{\circ}$ (or better) to enhance simulations of eddies, 
and hence planetary jets.
Even at such refined horizontal resolution though, 
the chosen value of~$\tau_D$ will still be
impacting the jets' intensities and locations
(see also equation~\ref{eq:aambudgetred} 
in section~\ref{sec:aam}).
Ultimately, how dissipative the Saturn atmosphere is
for small-scale circulations remains an open question
and difficult to constrain with observations \citep{Inge:18}.

\subsection{Angular momentum \label{sec:aam}}

The atmospheric Axial Angular momentum (AAM)~$\mathcal{M}$
is defined by the sum of 
the planetary contribution~${\mathcal{M}}^m$ 
associated with 
the solid-body rotation of the planetary sphere 
and 
the relative contribution~${\mathcal{M}}^w$ 
associated with
the motions of the atmosphere with respect to the solid-body rotating reference
\begin{equation}
\mathcal{M} =
{\mathcal{M}}^m + {\mathcal{M}}^w =
\int_{V} \mu^m \, \textrm{d}m + 
\int_{V} \mu^w \, \textrm{d}m
\end{equation}
\noindent where ~$\int_V$
denotes global integration 
over the volume of the atmosphere
and~$\mu$ denotes the AAM terms per unit mass
for respectively the planetary and relative contributions
\begin{equation}\label{eq:specificaam}
\mu^m = \Omega \, a^2 \, \cos^2\varphi \qquad \mu^w = u \, a \, \cos\varphi
\end{equation}
Thus, the temporal evolution of AAM is 
either related to
atmospheric mass redistribution
(that is, the temporal evolution of surface pressure)
or
wind variability
(that is, the temporal evolution of zonal wind).
Assuming that hydrostatic primitive equations are considered, 
and in the absence of 
any surface torque and zonal mechanical forcing, 
the globally-integrated AAM
$\mathcal{M}$ is conserved
provided the top lid does not vary with longitude
\citep{Stan:03,Thub:08,Laur:14}, which is ensured by
the DYNAMICO formulation \citep{Dubo:15}.

\begin{figure}[ht]
\begin{center}
\includegraphics[width=0.9\textwidth]{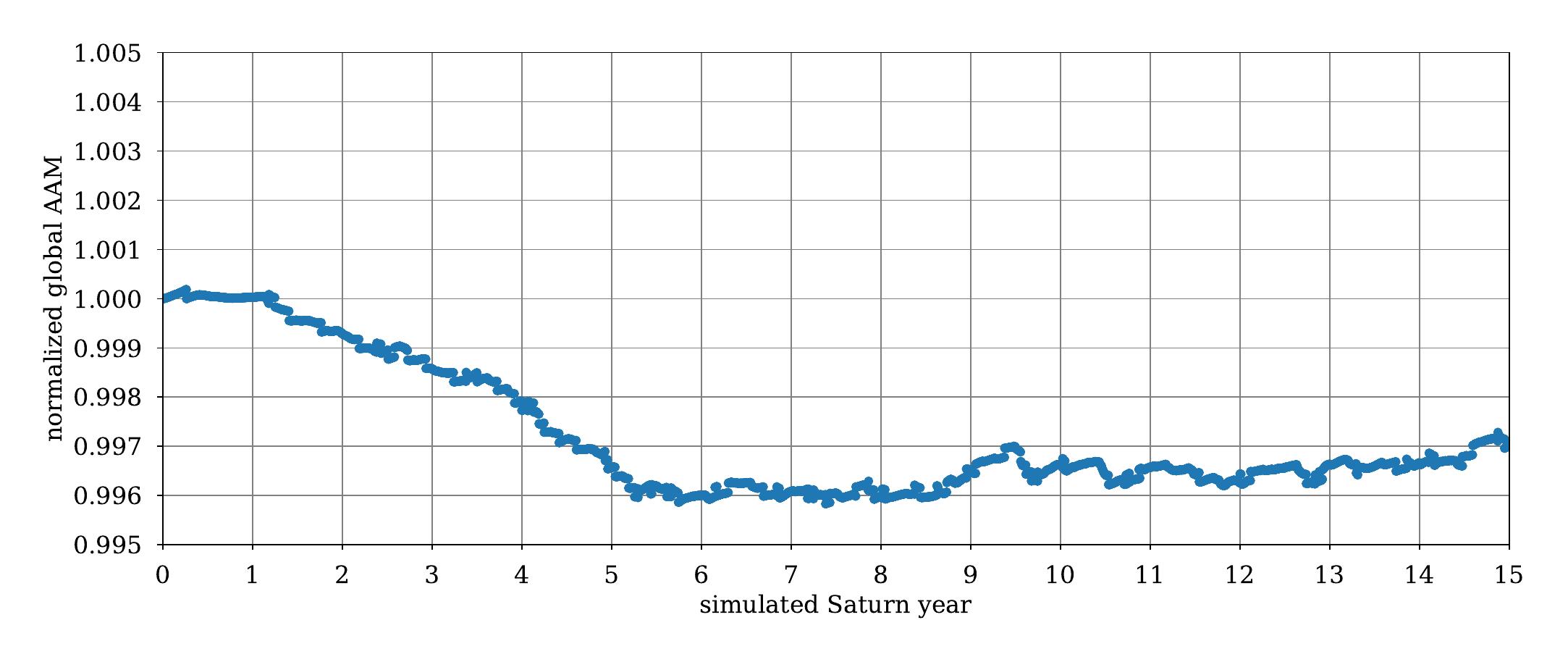}
\caption{\label{fig:aam} 
\emph{Temporal evolution of the global AAM~$\mathcal{M}$
(normalized to the initial value
$\mathcal{M}_0 = 5.244 \times 10^{32}$~kg m$^2$ s$^{-1}$)
for the complete duration of
our reference Saturn GCM simulation.}} 
\end{center}
\end{figure}

The conservation of global AAM $\mathcal{M}$ 
in our Saturn DYNAMICO model is satisfying (Figure~\ref{fig:aam}).
There is a very small decrease of global AAM 
($\sim 0.4\%$) in the first six simulated Saturn years,
which correspond to the dynamical spin-up of our model,
before the global AAM stabilizes at a final value
and only undergoes negligible temporal variations
of the order~$\sim 0.05\%$.
The properties of AAM conservation 
of our Saturn DYNAMICO model
compare very favorably to 
the performance of the dynamical cores 
described in \citet{Poli:14}, 
where most models exhibit typical AAM drifts of~$1-2 \%$
(reaching several tens of~\% for one particular model).

In addition to the conservation of global AAM,
negligible ``AAM noise'' must be ensured 
\citep{Lebo:12,Laur:14}. 
How the temporal evolution of global atmospheric AAM 
splits between each term of the GCM tendencies can be written
\begin{equation}\label{eq:aambudget}
\ddf{\mathcal{M}}{t} = \ddf{\mathcal{M}^m}{t} + \ddf{\mathcal{M}^w}{t} = F + T + S + D + \varepsilon,
\end{equation}
\noindent where
$F$ is the AAM tendency associated with subgrid-scale mixing in the physical packages 
(mostly boundary-layer effects),
$T$ is the AAM tendency associated with mountain torques,
$S$ is the AAM tendency associated with upper sponge layer
\citep[a reminder that upper-level sponge layer might alter significantly the AAM balance,][]{Shaw:07},
$D$ is the AAM tendency due to conservation errors in the
parameterization of horizontal dissipation,
and~$\varepsilon$ is a residual numerical rate of AAM variation due to conservation errors
(hereinafter named the ``AAM noise'' since it is a spurious source/sink of AAM in the model).
The AAM noise~$\varepsilon$ can be diagnosed in a GCM
by adding the temporal variations of 
${\mathcal{M}}^m$
and
${\mathcal{M}}^w$
computed by the primitive equations ($PE$) in the dynamical core,
excluding any term accounted for in~$F$, $T$, $S$ or $D$
(e.g. explicit diffusion operators are included in~$D$)
\begin{equation}
\varepsilon = \left[ \ddf{\mathcal{M}^m}{t} \right]_{PE} + \left[ \ddf{\mathcal{M}^w}{t} \right]_{PE}
\end{equation}
In the impossible perfect situation 
where no numerical approximations or errors are made
in the GCM dynamical core when solving the primitive equations, 
the AAM noise $\varepsilon$ should be zero
since AAM is exactly transferred from the solid-body rotation
to the atmospheric flow and vice versa;
a GCM simulation where~$\varepsilon$ is of similar
or larger magnitude than the other torques
would be questionable \citep{Laur:14}.

In our Saturn GCM simulations with DYNAMICO,
$D$ also includes the Rayleigh friction at the bottom of the atmosphere,
$S$ is zero because no upper-level sponge layer is used,
$T$ is zero since gas giants are devoid of surface,
and for a similar reason, in practice the term $F$
is one order of magnitude smaller than other terms.
This means that in the case of our Saturn GCM
(and more generally for any gas giant GCM)
equation~\ref{eq:aambudget} reduces to
\begin{equation}\label{eq:aambudgetred}
\ddf{\mathcal{M}}{t} = \ddf{\mathcal{M}^m}{t} + \ddf{\mathcal{M}^w}{t} = D + \varepsilon
\end{equation}
Hence the angular momentum noise~$\varepsilon$
has potentially a strong impact on 
the temporal AAM variability 
in a gas giant GCM.
A similar concern is raised by \citet{Lebo:12} 
in the case of idealized Venus simulations without topography.
The requirement of low AAM noise is less stringent
in realistic GCM simulations for Earth, Venus, or Titan
where the contribution of mountain torques~$T$ 
is large \citep{Lebo:12,Laur:14}.

\begin{figure}[htbp]
\begin{center}
\includegraphics[width=0.7\textwidth]{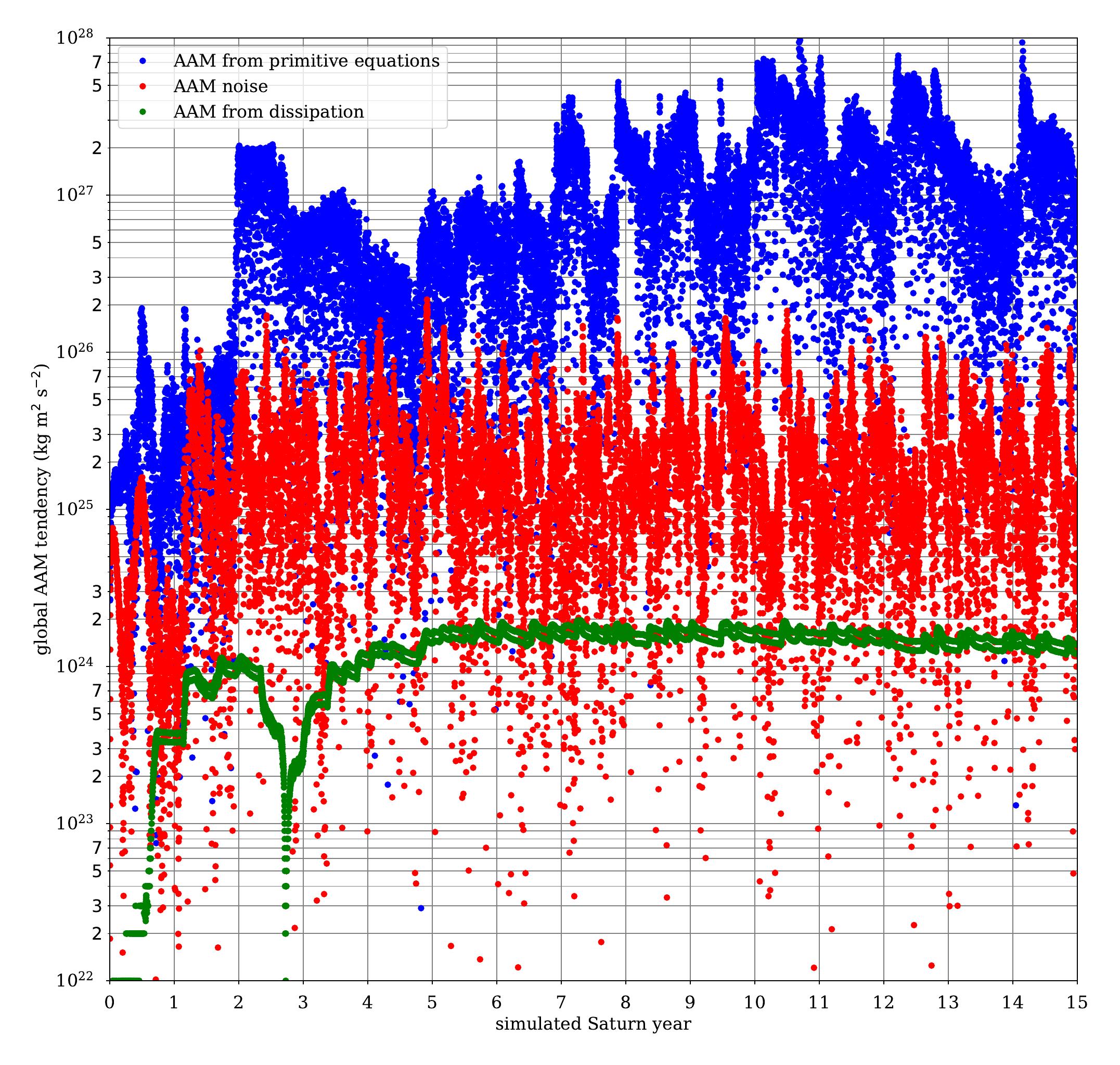}
\caption{\label{fig:aam_cond} 
\emph{Temporal evolution of 
the tendency~$\textrm{d}\mathcal{M} / \textrm{d}t$
of global AAM
for the complete duration of
our reference Saturn GCM simulation.
Blue points depict 
the term~$\left[ \textrm{d}\mathcal{M}^w / \textrm{d}t \right]_{PE}$
associated with wind tendencies computed in the primitive equations
resolved in the dynamical core of the Saturn GCM.
Red points depict
the term~$\varepsilon$ 
associated with AAM noise, i.e. 
residual numerical rate of AAM variation due to conservation errors.
Green points depict
the term~$D$ 
associated with AAM tendencies resulting in the dissipation
and Rayleigh drag scheme.
The log-scale figure 
is produced with the absolute values of
the various tendencies~$\textrm{d}\mathcal{M} / \textrm{d}t$.}}
\end{center}
\end{figure}

Compared to the analysis presented 
in \citet{Lebo:12} and \citet{Laur:14}
for telluric planets, the analysis
of AAM noise is more straightforward 
for gas giants.
The following condition\footnote{This condition is equivalent
to $$ \varepsilon + D \ll \ddf{\mathcal{M}^w}{t}, $$
i.e. the gas-giant (simplified) equivalent 
of the condition in \citet{Lebo:12}
$$ \varepsilon^* \ll \ddf{\mathcal{M}^w}{t} $$ }
must be ensured
\begin{equation}\label{eq:aamcondition}
\varepsilon \ll \left[ \ddf{\mathcal{M}^w}{t} \right]_{PE}
\end{equation}
Figure~\ref{fig:aam_cond}
shows that this condition~\ref{eq:aamcondition}
is fulfilled in our Saturn GCM,
with AAM noise being
one to two orders of magnitude
smaller than AAM tendencies
computed in 
the dynamical core's primitive equations
(except for a short duration
close to one and half simulated year).
This indicates that AAM noise
does not alter the dynamical
predictions for jets and eddies.
The same conclusion stands for
the AAM tendencies associated with dissipation and Rayleigh drag
(Figure~\ref{fig:aam_cond}),
which are even smaller than 
the AAM noise.
This indicates that the 
horizontal dissipation
set in our GCM (section~\ref{sec:dissip})
has a negligible impact
on the AAM budget -- this
is consistent with the
overall jet structure
being similar in simulations using
either~$\tau_D=10000$ or~$\tau_D=15000$.

In an attempt to explore
the model settings that could
influence the AAM noise $\varepsilon$,
we found that it is basically
insensitive to
time step, 
vertical discretization,
horizontal dissipation
(only a very strong horizontal dissipation
significantly lowers AAM noise,
but adversely impacts the intensity of jets).
This is consistent with
\citet{Laur:14} conclusions
that AAM noise is likely
related to small conservation
errors associated with
the horizontal discretization
of primitive equations.
The low AAM noise in our Saturn DYNAMICO GCM
simulation level ensures that 
dynamical spin-up of the jets is not 
due to convergence of numerically-spurious 
angular momentum.


\end{document}